\pdfoutput=1
\RequirePackage{ifpdf}
\documentclass[12pt,a4paper]{article}
\usepackage[utf8]{inputenc}
\usepackage[english]{babel}
\usepackage{amsmath}
\usepackage{jinstpub}
\usepackage{float}
\usepackage{mathtools}
\usepackage{graphicx}
\usepackage{multirow} 
\usepackage{booktabs}
\usepackage{bm}
\usepackage{fancyhdr}
\title{Convolutional Neural Networks Applied to Neutrino Events in a Liquid Argon Time Projection Chamber }
\collaboration{MicroBooNE Collaboration}

\abstract{
We present several studies of convolutional neural networks applied to data coming from the MicroBooNE detector, a liquid argon time projection chamber (LArTPC). The algorithms studied include the classification of single particle images, the localization of single particle and neutrino interactions in an image, and the detection of a simulated neutrino event overlaid with cosmic ray backgrounds taken from real detector data. These studies demonstrate the potential of convolutional neural networks for particle identification or event detection on simulated neutrino interactions. We also address technical issues that arise when applying this technique to data from a large LArTPC at or near ground level.  
}

\begin{document} 

\author[g]{R.~Acciarri}
\author[z]{C.~Adams}
\author[h]{R.~An}
\author[w]{J.~Asaadi}
\author[a]{M.~Auger}
\author[g]{L.~Bagby}
\author[g]{B.~Baller}
\author[q]{G.~Barr}
\author[q]{M.~Bass}
\author[x]{F.~Bay}
% ...

\affiliation[a]{Universit{\"a}t Bern, Bern CH-3012, Switzerland}
% ...
\affiliation[g]{Fermi National Accelerator Laboratory (FNAL), Batavia, IL 60510, USA}
\affiliation[h]{Illinois Institute of Technology (IIT), Chicago, IL 60616, USA}
% ...
\affiliation[q]{University of Oxford, Oxford OX1 3RH, United Kingdom}
% ...
\affiliation[w]{University of Texas, Arlington, TX, 76019, USA}
\affiliation[x]{TUBITAK Space Technologies Research Institute, METU Campus, TR-06800, Ankara, Turkey}
% ...
\affiliation[z]{Yale University, New Haven, CT, 06520, USA}

\maketitle

\newpage

\newpage

\section{Introduction}

Liquid argon time-projection chambers (LArTPCs) are playing an increasingly important role in neutrino experiments.  LArTPCs, a type of particle detector, produce high resolution images of neutrino interactions, opening the door to new, high-precision measurements.  For example, the efficient differentiation of electrons from photons allowed by a LArTPC enables precision measurement of CP violation and sensitive searches for sterile neutrinos.   This work studies the application of a machine learning algorithm, deep convolutional neural networks (CNNs)~\cite{AlexNet}, to the reconstruction of neutrino scattering interactions in LArTPCs. 

CNNs are a type of artificial neural network developed for image analysis, e.g. classifying the subject of an image or locating the position of different objects within an image. CNNs are capable of learning patterns and features from images through a training procedure where they process many examples.  CNNs can be adapted from one application to another by simply changing the set of training images. CNN-based techniques represent a widely-used solution for image analysis and interpretation.  Many of the  problems these algorithms face are analogous to the ones found in the analysis of images produced by LArTPCs.

In this work, we apply CNNs to events in the MicroBooNE detector.  This detector, located at the Fermi National Accelerator Laboratory (FNAL), is the first large-scale LArTPC to collect data in a neutrino beam in the United States. This is the first detector in a series of LArTPCs planned for future neutrino research~\cite{SBN, DUNE}.   In each case, the neutrino interacts and produces charged particles which traverse the detector and leave an ionization trail, or trajectory.  Performing physics analyses with LArTPC data depends upon accurate categorization of the trajectories. Categorization includes sorting trajectories by particle type, deposited energy, and angle with respect to the known incoming neutrino direction.   
Finding instances of neutrino interactions or individual particles within a LArTPC image and then classifying them maps directly to the types of tasks for which CNNs have been proven effective, tasks known as object detection and image classification, respectively.

This work addresses the following technical questions related to applying CNNs to the study of LArTPC images:
\begin{itemize}
\item Can CNNs, which have been developed for everyday, photographs which are information-dense, be successfully applied to LArTPC event images, which are more information sparse?  In contrast to natural photographs, which contain a wide range of textures and patterns throughout the image, LArTPC images consist of lines from particle trajectories that are spaced far apart from one another.  Much of the LArTPC image is empty.
\item Current networks are designed to work with images that are smaller (approximately 300$\times$300 pixels) than those produced by a large LArTPC (approximately 3000$\times$9600 pixels), such as MicroBooNE.  How does one balance the desire to use as high a resolution as possible with the constraints on image size imposed by current hardware limitations?  What is a possible scheme to downsize the image?
\item With a downsizing scheme in place, what is its effect on the network performance?  
\item How does one coordinate the multiple views of an event that LArTPC detectors produce?  
\end{itemize}
We document a number of strategies to overcome these challenges.

In what follows, we first briefly describe the MicroBooNE detector and the images it produces.  We then describe CNNs and their application to LArTPCs in general.   This is followed by three demonstrations that make use of the MicroBooNE simulation and cosmic ray data.  The demonstrations use publicly available, open-source software packages.


\section{The MicroBooNE Detector}


\begin{figure}[t]
 \centering
 \includegraphics[width=1.0\textwidth]{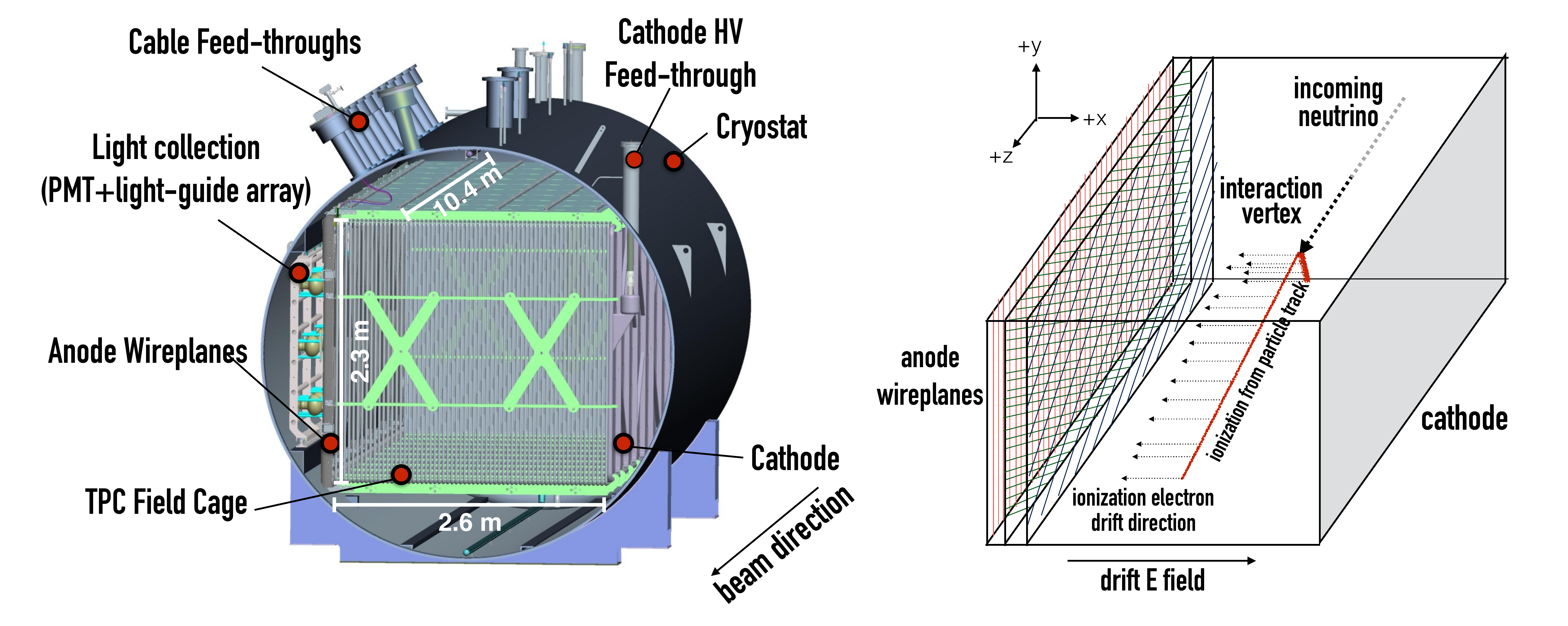}
  \caption{{\it Left.} An illustration of the MicroBooNE detector. Various components of the detector are labeled. {\it Right.} Schematic illustrating the operation of the TPC.  The TPC consists of a cathode plane and anode wire planes. The cathode is biased to $-70$ kV which places a 273 V/cm uniform electric field between the cathode and anode.  Ionization electrons from charged particles, produced in neutrino interactions, drift to the anode planes, which provide measurements of the drift time and ionization charge.}
  \label{fig:uBimage}
\end{figure}

MicroBooNE employs as its primary detector a large liquid argon time projection chamber with an active mass of 90 tons.  
Figure~\ref{fig:uBimage} shows a diagram of the MicroBooNE LArTPC.
The detector operates within a cryogenic vessel, which houses the TPC and the light collection system, the latter consists of an array of 32 photomutiplier tubes (PMTs) and four acrylic light-guides.  The TPC structure is composed of a rectangular, stainless-steel cathode plate set parallel to three sets of anode wire planes. The distance between the cathode plate and the closest wire plane is 2.6 m.  Both the cathode and anode wire planes are 2.3 m high and 10.4 m wide.  The space between them define the box-shaped, active volume of the detector.  We describe the MicroBooNE TPC with a right-handed coordinate system where the $Z$-axis is aligned with the long dimension of the TPC and the direction of the neutrino beam. The beam travels in the $+Z$ direction.  The $X$-axis runs normal to the wire planes, beginning at the anode and moving in the direction of the cathode. The $Y$-axis runs parallel to the anode and cathode, in the vertical direction.  The PMT array is mounted inside the cryostat and  to the wall closest to the anode wire planes.  The PMTs are set behind the wire planes, outside the space between the cathode and anode, and face towards the center of the TPC volume in the $+X$ direction.  The TPC and PMT system together provide the information needed to reconstruct the trajectories of charged particles traveling through the TPC. 

Figure~\ref{fig:uBimage} contains a schematic illustrating the basic operating principles of the TPC, which we describe here.  Charged particles traversing the liquid argon produce ionization and scintillation light. The ionization electrons drift to the TPC wires over a time period of milliseconds due to a 273 V/cm field between the cathode and anode planes produced by applying $-70$ kV on the cathode.  The TPC wires that the electrons encounter consist of three planes of parallel wires each oriented at different angles.  For all three planes, the wires lay parallel to the $YZ$ plane. The wires of the plane furthest from the TPC volume, referred to as the $Y$-plane, are oriented vertically along the $Y-$direction. The electric field lines terminate on these wires causing the ionization electrons to collect on them.  The other two planes, called $U$ and $V$, observe signals by induction as the electrons travel past them before being collected on the $Y$ plane. The wires of the $U$ and $V$ planes are oriented $\pm$60 degrees, respectively, with respect to the $Y$-axis.  There are 2400 wires that make up each of the $U$ and $V$ planes, and 3456 wires that make up the $Y$ plane. 
Electronics on each wire record the charge and readout time associated with signals on a wire. 



\subsection{Images in the MicroBooNE LArTPC}

\begin{figure}[t]
  \centering  
\includegraphics[width=0.495\textwidth]{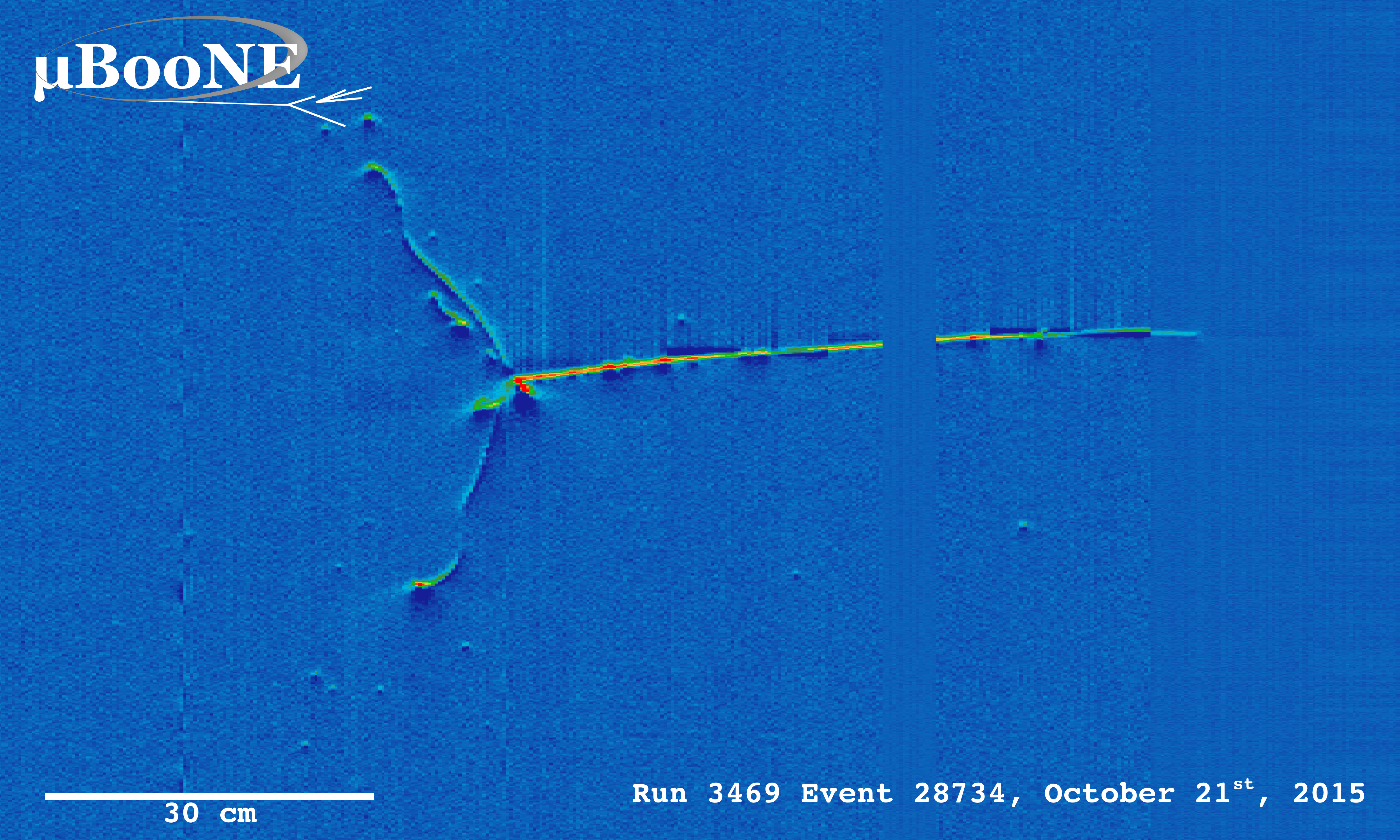}\\
\includegraphics[width=0.495\textwidth]{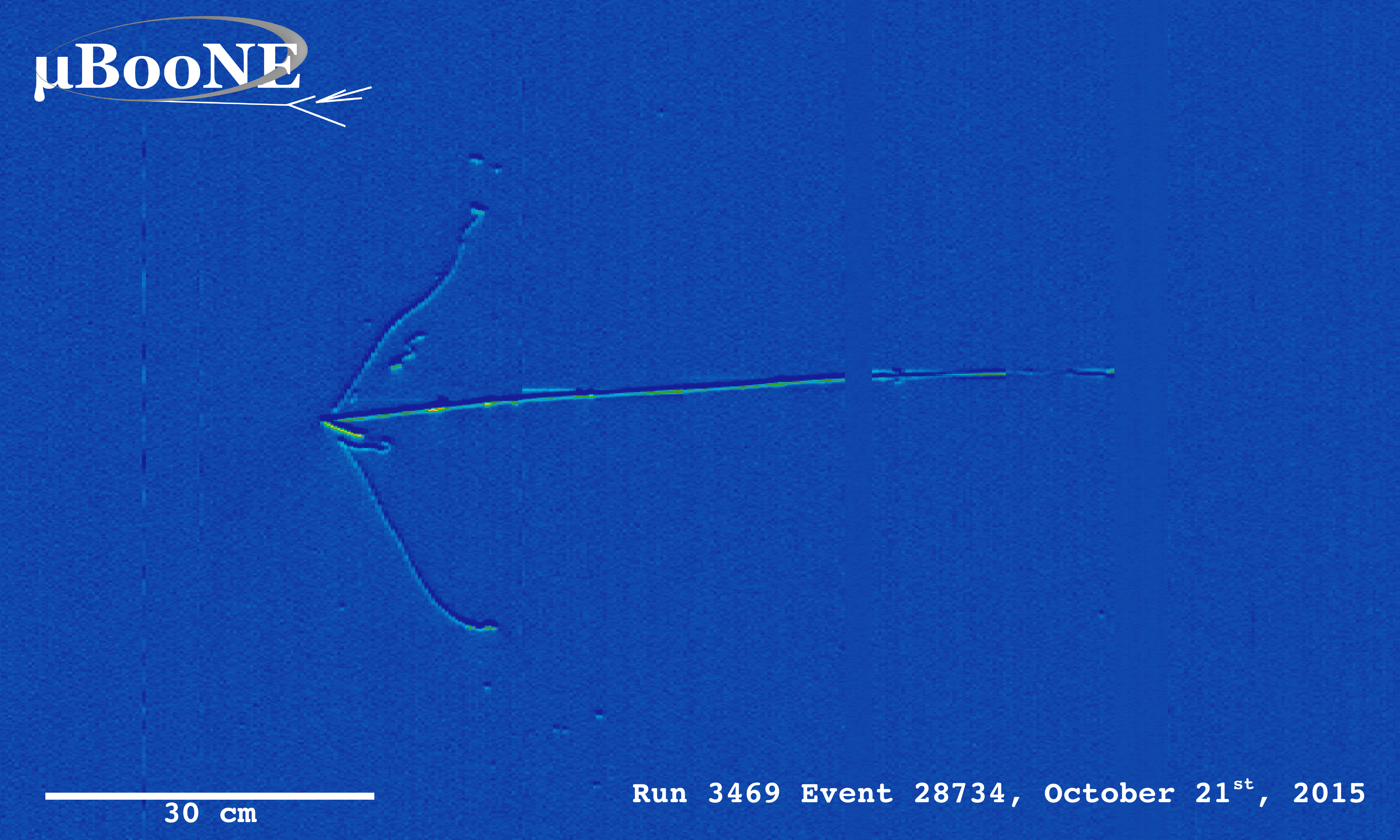}\\
\includegraphics[width=0.495\textwidth]{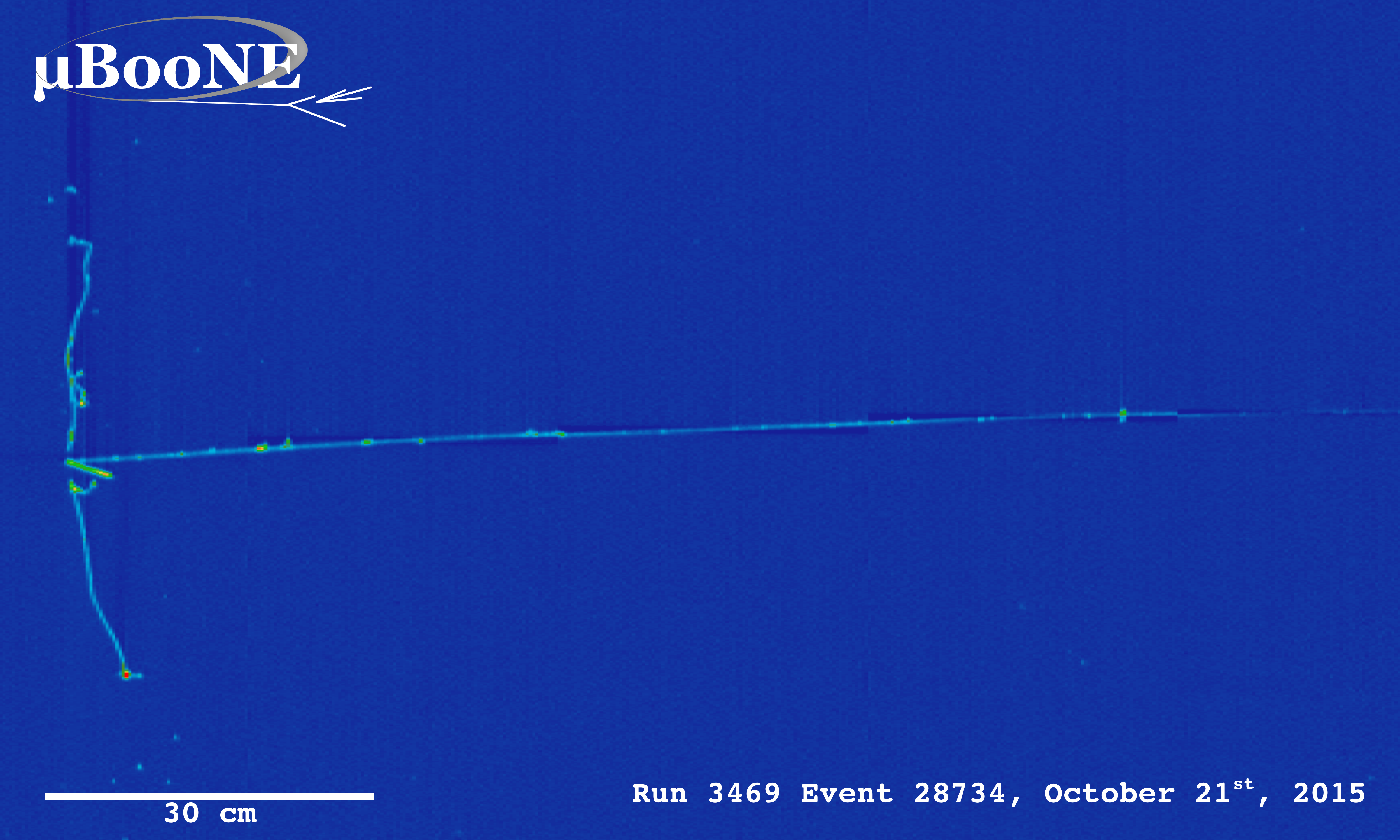}\\
\caption{Example neutrino candidate observed in the MicroBooNE detector.  The same interaction is shown in all three wire plane views.  The top image is from wire plane $U$; the middle from wire plane $V$; and the bottom from wire plane $Y$. The image is at full resolution and is only from a portion of the full event view.}
  \label{fig:UBEVD}
\end{figure}

Application of a CNN is straightforward for LArTPCs, because the data they produce are a set of images containing charged particle trajectories projected on a 2D plane, with each wire plane providing a different view of the event. The two axes of an image are the wire number and readout time. The first dimension moves across the detector, upstream to downstream in the neutrino beam direction, while the second dimension is a proxy for the $X$ direction axis\footnote{This assumes a constant electron drift velocity. Space charge (build up of slow moving positive ions in the detector) leads to distortions of this ideal and must be corrected for.}. In MicroBooNE, detector wires are separated by 3 mm and carry a signal waveform digitized at 2 MHz frequency with a 2 $\mu s$ shaping time. We form one image per wire plane by filling each column with the digitized waveforms from each wire. In such images, one pixel corresponds to 0.55 mm along the time axis given the measured drift velocity of 0.11 cm/$\mu s$. The pixel values of the image represent the analog-to-digital-converted (ADC) charge on the wire at the given time tick.  ADC counts are encoded over a 12-bit range. This scheme can produce high resolution images for each wire plane with full charge resolution.  In figure~\ref{fig:UBEVD}, we show three images (in false color) of a neutrino interaction candidate viewed by all three planes as examples of the high quality information obtained by a LArTPC. The task of reconstruction is to parse these images and determine information about the neutrino interaction that produced the tracks observed.

The time at which the detector records an image is initiated by external triggers.  Each trigger defines an {\it event}, which is defined as a 4.8 ms period, during which waveforms from the TPC wires and the PMTs are recorded. The situation during which the detector is triggered defines the types of events recorded, two of which will be used in our analyses. The first type of event, {\it on-beam} events, occurs in synchrony with the arrival of the neutrino beam at the detector. These events might contain a neutrino interaction.  The second type, {\it off-beam} events, occur out-of-synchrony with the neutrino beam and should contain only cosmic-ray trajectories.


For most on-beam events, neutrinos from the beam will pass through the detector without interacting.  This means that the vast majority of events collected are uninteresting in the context of a neutrino physics analysis and must be filtered out. The light collection system plays a crucial role in this. Along with ionization, charged particles traversing the detector will also produce scintillation light coming from the argon.  This light is observed by the PMTs within tens of nanoseconds of the interaction time.   Thus, the light tells us the time when the particle passes through the detector. By selecting only those events which have scintillation light occurring in time with the beam, most of the events without an interaction are removed. This cut is applied to the event images we analyze.
Those that pass will be referred to as {\it PMT-triggered} events.

\section{Convolutional Neural Networks}
\label{sec:cnnbg}

CNNs can be thought of as a modern advancement on feed-forward neural networks (FFNNs), which are commonly employed technique in high energy particle physics analyses. 
CNNs can be seen as a special case of FFNNs, one designed to deal with spatially localized data, such as those found in images. 
The network architectures of CNNs can be more complex than those used by FFNNs and include operations that go beyond those performed by the networks’ individual neurons. These extensions have allowed CNNs to become the state-of-the-art solution to many different kinds of problems, most notably photographic image classification. CNNs have even begun to find uses in other neutrino experiments~\cite{dayabay,nova,NEXT} and other LArTPC experiments have ongoing development of deep learning techniques.

Consider event sample classification, a typical analysis performed by FFNNs in high-energy physics.  Here the goal is to assemble a set of $N$ features, calculated for each data event, and use them to separate the events in one's data into $C$ classes. For example, one might aim to separate the data into a signal-enriched sample and a background-enriched sample. For many experiments, a common approach to reconstruction is to take raw data from various instruments and distill it into several quantities, often focused around characterizing observed particles, that can be used to construct these $N$ features. This information can then be used to train an FFNN.  However, the process of developing reconstruction algorithms can often be a long, complicated task. In the case of LArTPCs, we can employ CNNs, which can do much of this work automatically by learning its own set of features from data. This ability of CNNs is due to their architecture, which differs substantially from that of FFNNs.

\begin{figure}[t]
 \centering
 \includegraphics[width=0.4\textwidth]{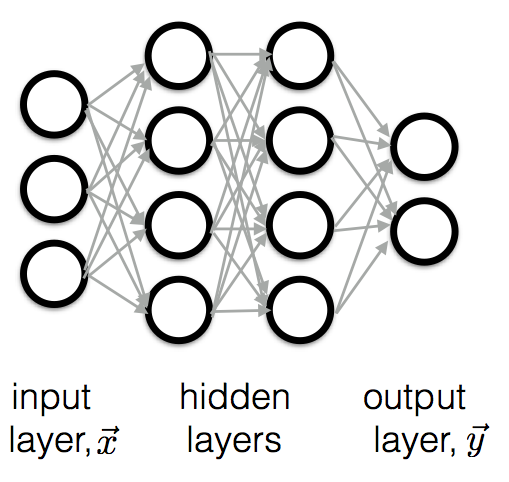}
  \caption{Feed-forward neural network (FFNN) architecture. The network is arranged in layers. There is an input layer consisting of an $N$-dimensional vector, $\vec{x}$, of features. In this case we have $N=3$.  There are one or more hidden layers. In this diagram, two are shown.  Finally, there is the output layer, a $C$-dimensional vector, $\vec{y}$, with $C$ commonly chosen to be the number output classes. In this case we have $C=2$.}
	\label{fig:feed_forward}
\end{figure}

In FFNNs, the input feature vector, $\vec{x}$ is passed into one or more hidden layers of neurons. These hidden layers eventually feed into an output layer containing $C$ neurons, which together produce a $C$-dimensional output vector, $\vec{y}$.  Figure~\ref{fig:feed_forward} contains a diagram illustrating a possible network architecture for a FFNN. One characteristic feature of these networks is that each neuron receives input from every neuron in the preceding layer. In principle, one has the freedom to define what each neuron does with this input. In practice, however, it is common for the neuron to perform a weighted sum of the inputs and then pass this sum into an activation function that models the neuron's response. In other words, for a given neuron, whose label is $i$, and an input vector, $\vec{x}$, the output, $f_i(\vec{x})$, of the neuron is 
\begin{equation}
f_i(\vec{x}) = \sigma\left( \vec{w}_i\cdot\vec{x} + b_i \right),
\end{equation}
where $w_i$ are known as the weights of neuron $i$, $b_i$ is the bias of neuron $i$, and $\sigma$ is some choice of activation function.  Common choices for $\sigma$ include the sigmoid function or hyperbolic tangent.  Currently, most CNN models use an activation function known as the Rectified Linear Unit, or ReLU.  Here $\sigma_{\textrm{ReLU}}$ is defined as
\begin{equation}
\sigma_{\textrm{ReLU}}(\vec{x}) =  
\left\{
  \begin{array}{ll}
      \vec{w}_i\cdot\vec{x} + b_i &\textrm{\hspace{1cm}} \vec{w}_i\cdot\vec{x} + b_i \geq 0 \\
       0 &\textrm{\hspace{1cm}} \vec{w}_i\cdot\vec{x} + b_i<0.
  \end{array}
\right.
\end{equation}
The way these networks learn is that one chooses values of the weights and biases for all neurons such that, for a given $\vec{x}$, the network produces some desired, $\vec{y}$. Going back to the signal vs. background example, the goal is to choose values for the network parameters such that, for a given event, the FFNN is able to map an $N$-dimensional feature vector of reconstruction quantities, $\vec{x}$, to $\vec{y}=(1,0)$ if the event is a signal event, or $\vec{y}=(0,1)$ if the event is a background event.  

\begin{figure}[t]
 \centering
 \includegraphics[width=0.5\textwidth]{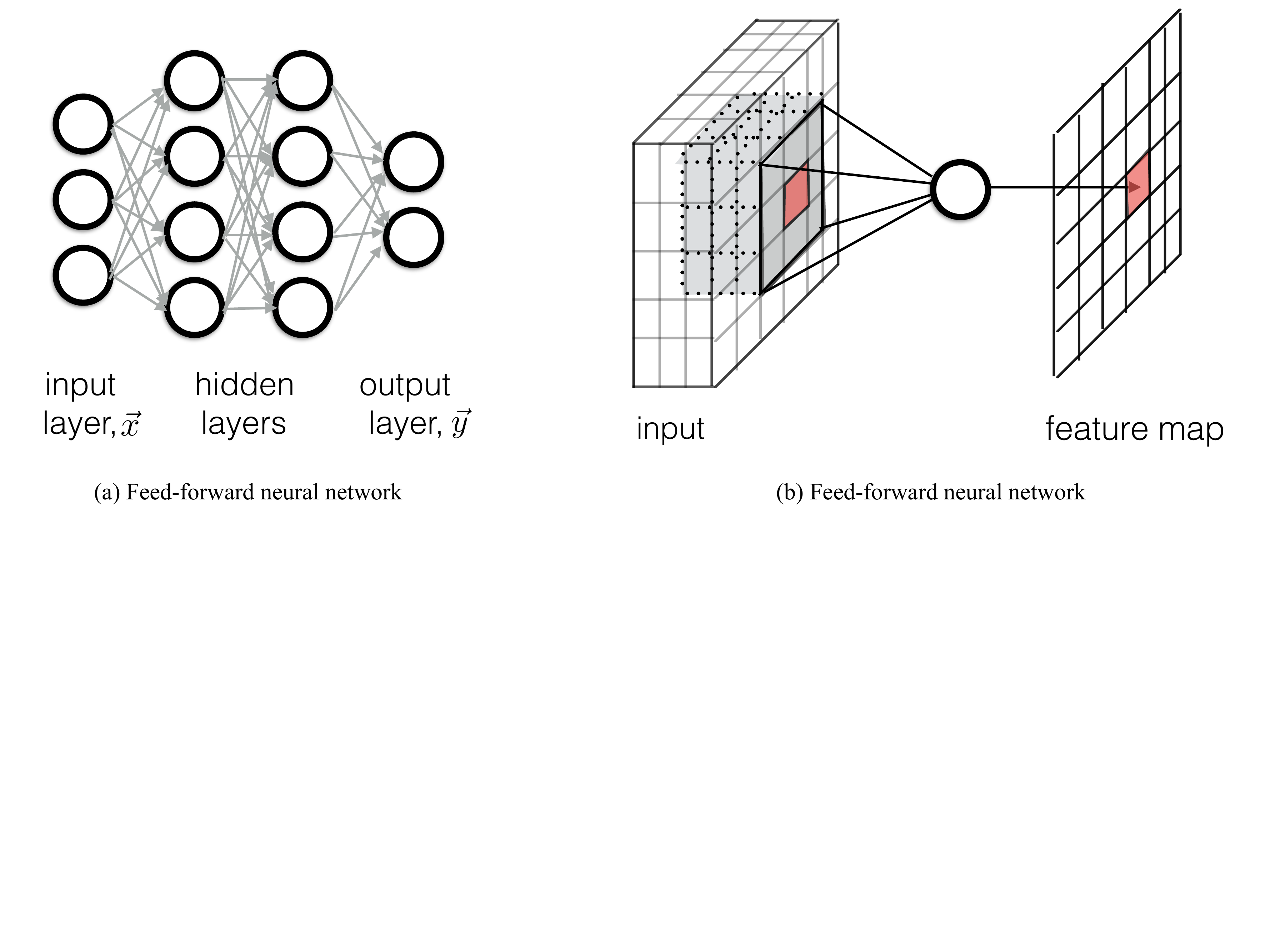}
  \caption{
 The input and output connections of one neuron in a convolutional neural network (CNN). The neuron, shown as a circle in the diagram, looks at a local 3D volume of pixel values in the input tensor, and is scanned over the entire input tensor to create the output feature map.}
	\label{fig:cnn_neuron}
\end{figure}

For CNNs, the way neurons are connected is different. In this case, sets of hidden neurons are still grouped together in layers.  However, a neuron in a given layer receives only the inputs of a local region of neurons in the preceding layer. Instead of an $N$-dimensional input vector, the input for a given neuron is laid out as an $M\times N \times L$ array of values.  This is a natural representation of the pixel values in an image.  For an image with height, $M$, and width, $N$, the third dimension, $L$, is the number of color channels, e.g. three for red-green-blue (RGB) images. 
The output of a given neuron, $i$, with a 3D array of weights, $W_i$, looking at a local volume of pixel values centered around pixel, ($m_j$,$n_j$), of input array, $X_j$, is given by
\begin{equation}
f_{i,j}(X) = \sigma\left( W_i \cdot X_j + b_i \right),
\end{equation}
where the width and height of $W_i$ are network parameters and referred to as the receptive field of a neuron.  The product $W_i\cdot X_j$ is an element-wise multiplication of the weight tensor and the input local volume $X_j$.
Figure~\ref{fig:cnn_neuron} illustrates this connection structure.  In the figure, a neuron receives input from a local volume, highlighted in gray, of the input tensor, represented by the 3D grid on the left. The volume is centered around the input pixel, $(m_j,n_j)$ highlighted in red.  In a CNN, this neuron acts only on one sub-region at a time, but, in turn, it acts on all sub-regions of the input as it is scanned over the height and width dimensions.  During this scan, the output values of the neuron are recorded and arranged into a 2D grid. The grid of output values, therefore, forms another image known as a feature map. In figure~\ref{fig:cnn_neuron}, this feature map is represented by the rightmost grid.

The neuron in this architecture acts like an image filter.  The value of the weights of the neuron can be thought as representing some pattern. For example, patterns might include different orientation of edges or small patches of color.  If the local region of the input matches that pattern, the activation is high, otherwise the activation is low.  Having scanned the feature across the input, the resulting output is an image indicating where that pattern can be located. In other words, the pattern has been filtered into the output image. By allowing $F$ such neurons, or filters, to process the input, one gets $F$ feature maps which can be arranged to make another 3D array of values.  Also, note that the operation of scanning the neuron across the input is a convolution between the input and the filter. The collection of filters that operate on an input array and produce a set of feature maps is known as a convolutional layer. 

Like FFNNs, one can arrange a CNN as a series of layers where the output of one convolutional layer is passed as input to another convolutional layer. With such a stack of layers, CNNs learn to identify high-level objects through combinations of low-level patterns. Furthermore, because a given pattern is scanned across the image, translation-invariant features are identified.

This architecture has a couple of advantages. First, the parameters of the neurons, or filters, are learned. Like FFNNs, training the network involves an algorithm which chooses the parameters of the neurons such that the network learns to map an input image to some number or vector. This ability to learn its own features is one of the reasons CNNs are so powerful and have seen such widespread use. When approaching a new problem, instead of having to come up with a set of features to extract from the data, with CNNs one can simply collect as much data as possible and start training the network. Second, the number of parameters per layer is much smaller than the number of parameters per layer of an FFNN.  In an FFNN, each neuron has a weight for all input neurons in the preceding layer.  This is large number of parameters, if one used every pixel of an image as input to the network.  In contrast, the parameters per layer for CNNs is only the volume of local input region for each filter times the number of filters. This efficient use of parameters allows CNNs to be composed of many, successive convolutional layers.  The more layers one uses, the more complex an object can be represented. And currently, the understanding is that ``deep'' networks with a very large number of layers (on the order of one hundred or more) tend to work better than shallow networks.


The attractiveness of a machine learning approach is that it might reduce the number of pattern recognition algorithms that physicists must develop.  Likely, different algorithms will be required for specific types of physics processes. Such algorithms can take many person-hours to develop and then optimize for efficient computation.  Instead, CNNs potentially offer a different strategy, one in which networks are left to discover their own patterns to recognize the various processes and other physical quantities of interest. Furthermore, processing an image takes on the order of tens of milliseconds on a graphics processing unit (GPU), which should be just as fast, if not faster, than other pattern recognition algorithms.  However, this approach is still relatively unexplored, and so in this work we describe three studies, or demonstrations, applying CNNs to LArTPC images. 


\subsection{Demonstration Steps}
We demonstrate the applicability of CNNs through the following tests:
\begin{itemize}
\item{Demonstration 1--}
Classification and detection of a simulated single particle within a single-plane image;
\item{Demonstration 2--}
Neutrino event classification and interaction localization within a single-plane image;
\item{Demonstration 3--}
Neutrino event classification with a full 3 wire-plane model.
\end{itemize}
The first study shows that a CNN can be applied to LArTPC images despite the fact that their content is quite different from the photographic images for which the technique was developed. 
In this study we select a sub-region of a whole event image, then we further downsize the image by summing neighboring pixel values.  These operations are referred to as image cropping and downsizing respectively, and are important steps for a large LArTPC detector such as MicroBooNE due to computing resource limitations.
The second study is an initial trial for localizing neutrino interactions within a 2D event image using only one plane of the three views. In the last case, we construct a network model that combines full detector information including all three views and optical data, using data augmentation techniques.   

\subsection{Network Models and Software Used}

\begin{table}[t]
\begin{center}
\caption{A summary of the publicly-available code used in each of the demonstration exercises.}
\begin{tabular}{clll}
Software & ref. & Purpose & Used in Demonstrations\\
\hline 
\vspace{0.05in}
LArSoft  & \cite{larsoft050800} & Simulation and Reconstruction & 1-3 \\
uboonecode & \cite{uboonecode050800} & Simulation and Reconstruction & 1-3 \\
LArCV & \cite{LArCV} & Image Processing and Analysis & 1-3 \\
Caffe & \cite{caffe} & CNN Training and Analysis & 1-3 \\
AlexNet  & \cite{AlexNet} & Network Model & 1,2 \\ 
GoogLeNet & \cite{GoogLeNet} & Network Model & 1 \\
Faster-RCNN & \cite{fasterrcnn} & Network Model & 1,2 \\
Inception-ResNet-v2 & \cite{Inceptionv4} & Network Model & 2 \\
ResNet & \cite{ResNet} & Network Model & 3 \\
\bottomrule
\end{tabular}
\label{Tab:codeused}
\end{center}
\end{table}

For the demonstrations performed in this work, we use several prevalent networks that have been shown to perform well at their given image processing task, typically measured by having nearly the best performance in some computer vision competition at the time of publication. Our implementation of the networks and the analysis of their outputs uses open-source software along with some custom code that is publicly available. We summarize the networks and software we use in Table~\ref{Tab:codeused}. For demonstration 1 we use AlexNet~\cite{AlexNet} and  GoogLeNet~\cite{GoogLeNet} for image classification task. Demonstration 2 introduces a simplified Inception-ResNet-v2~\cite{Inceptionv4}. For demonstration 3, a network, based on ResNet~\cite{ResNet}, is employed. Demonstrations 1 and 2 involve a network that can locate an object within a sub-region of an image ({\it Region of Interest}, or {\it ROI}, finding). For this task we use a network known as the Faster-region convolutional neural network, or Faster-RCNN~\cite{fasterrcnn}.

All of the above models are CNNs with various depths of layers. The choice of AlexNet for demonstrations 1 and 2 is motivated by the fact that this relatively shallow model is often used as a benchmark in the field of computer vision ever since it was first introduced for the annual Large Scale Vision Recognition Challenge (LSVRC) in 2012. GoogLeNet, which has 22 layers compared to 8 layers in AlexNet, is another popular model which we employed to compare to AlexNet in demonstration 1. 

Faster-RCNN, the state-of-the-art object detection network used in demonstrations 1 and 2, can identify multiple instances of different object classes within the same image. In this paper we combine this architecture with AlexNet for object detection networks tasked with locating various particle types or a neutrino interaction in a given event image.


We also use truncated versions of two networks, ResNet~\cite{ResNet} and Inception-ResNet-v2~\cite{Inceptionv4}, to perform neutrino event classification in demonstrations 2 and 3.  Both networks use a type of sub-network architecture know as residual convolutional modules introduced in ref.~\cite{ResNet}. These modules are believed to help the network achieve higher accuracy and learn faster.  (For a description of a residual model, see ref.~\cite{ResNet} or appendix~\ref{app:3planenuid}.)

Throughout all studies in this work, we use one of the most popular open-source CNN software frameworks, {\sc{Caffe}}~\cite{caffe}, for CNN training and analysis. Input data is in a ROOT file format~\cite{ROOT} created by the {\sc{LArCV}} software framework~\cite{LArCV}, which we developed to act as the interface between {\sc{LArSoft}} and {\sc{Caffe}}. {\sc{LArCV}} is also an image data processing framework and is used to further process and analyze images as described in the following sections. One can find our custom version of {\sc{Caffe}} that utilizes the {\sc{LArCV}} framework in \cite{LArCV}.  

The computing hardware used in this study involves a server machine equipped with two NVIDIA Titan X GPU cards~\cite{titanx}, chosen in part due to their large amounts of on-board memory (12 GB). We have two such servers of similar specifications at the Massachusetts Institute of Technology and at Columbia University, which are used for training and analysis tasks.

%
%

\section{Demonstration 1: Single Particle Classification and Detection}

In this demonstration, we investigate the ability of CNNs to classify images of single particles located in various regions of the MicroBooNE detector.  We show that CNNs can locate particles by specifying a bounding box that contains the particle.  This demonstration also proposes a strategy for working with the large image sizes from large LArTPCs.  The detector creates three images per event, one from each plane, which at full resolution have dimensions of either 2400 wires $\times$ 9600 digitized time ticks for the induction planes, $U$ and $V$, or 3456 wires $\times$ 9600 time ticks for the collection plane, $Y$.  Future LArTPC experiments will produce similar size or even larger images. Such images are too large to use for training a CNN of reasonable size on any GPU card available on the market. To mitigate this problem one must crop or downsize the image, or possibly both.

Therefore, in demonstration 1, we study the following things.
\begin{enumerate}
\item The performance of five particle classification ($e^-$, $\gamma$, $\mu^-$, $\pi^-$, proton) CNNs, using AlexNet and GoogLeNet as our nework models, applied to cropped, high-resolution images.
\item A comparison of the above with low-resolution images (that have been downsized by a factor of two in both pixel axes).
\item Two-particle separation for $e^-/\gamma$ and $\mu^-/\pi^-$.
\item The performance of a CNN to draw a bounding box around the single particles.
\end{enumerate}
The first task serves as a proof-of-principle that CNNs have the capability to interpret LArTPC images. The second is a comparison of how the downsizing factor affects our classification capability. The third focuses on particular cases interesting to the LArTPC community. Finally, the fourth is a simple test case of applying an object detection network, Faster-RCNN~\cite{fasterrcnn}, to LArTPC images.

\subsection{Sample Preparation}
\label{sec:5PSamplePrep}

\subsubsection{Producing Images}

For this study, we generated images using a single particle simulation built upon {\sc{LArSoft}}~\cite{larsoft050800}, a simulation and analysis framework written specifically for LArTPC experiments.  {\sc{LArSoft}} uses {\sc{geant4}}~\cite{geant4} for particle tracking. {\sc{LArSoft}} also implements a model of the electron drift inside the liquid argon medium.  Each event for this demonstration simulates only one particle: $e^-$, $\gamma$, $\mu^-$, $\pi^-$, or proton. 
Particles are generated uniformly inside the TPC active volume. Their momenta are distributed uniformly between 100 MeV and 1 GeV, except for protons which were generated with kinetic energy distributed uniformly between 100 MeV and 788 MeV where the upper bound is set to cover about the same momentum range, 1 GeV/c, of the other particles. The particles' initial directions are generated in an isotropic manner.  After their location and kinematics are determined, the particles are passed to {\sc{geant4}} for tracking.

After the events are tracked by {\sc{geant4}}, the MicroBooNE detector simulation, also built upon {\sc{LArSoft}}, is used to generate signals on all the MicroBooNE wires.  This is done with the 3D pattern of ionization deposited by charged particles within the TPC.  The detector simulation uses the deposited charge to simulate the signals on the wires using a model for the electron drift that includes diffusion and ionization lost to impurities. This model provides a distribution of charge at the wire planes which then goes into simulating the expected raw, digitized signals on the wires.   These raw signals are based on 2D simulations of charge propagation and electronics response on the induction and collection planes. We then convert these raw wire signals into an image.

The conversion of wire signals to an image, both for the simulation and for detector data used in later demonstrations, proceeds through the following steps.  First, the digitized, raw wire signals are sent through a noise filter~\cite{ubnoisefiltering}. 
Then, they are passed through an algorithm which converts the filtered, raw signals into calibrated signals~\cite{ubsignalprocessing}.  
The values of the calibrated, digitized waveforms then populate a 1-channel image, i.e. a 2D array, with one axis being time and the other being a sequence of wires.  We form an image for each of the three wire planes. Because the $Y$ plane has more wires than the $U$ and $V$ planes, a full event image for each plane has a width of 3456 pixels. For the $U$ and $V$ planes, which only have 2400 wires, the first 2400 pixels are filled, while the rest are set to zero. As a technical aside, we note that the value of each pixel is stored as a float as that is one of the input choices for {\sc{Caffe}}. We do not convert the ADC values into 8-bit values typical of natural images.

A full MicroBooNE event has a total of 9600 time ticks.  For this and the other demonstrations, we use a 6048 time tick subset ranging from tick 2400 to 8448 with respect to the first tick of the full event.  This subset includes time before and after the expected signal region. A reason to include these extra time ranges is to demonstrate that the analysis methodology employed in this study can be used for neutrino analysis in the future with the same hardware resources we have today. Particles are generated at a time which corresponds to the time tick 800 in the recorded readout waveform, and all associated charge deposition is expected to reach the readout wire plane by time tick 5855 given the by drift velocity. Finally, this simulation includes a list of unresponsive wires, which carry no waveform.  In total, about 830 (or 10\%) of the wires in the detector are labeled as unresponsive~\cite{ubnoisefiltering}. The number of such wires on the $U$ plane is about 400, on the $V$ plane about 100, and on the $Y$ plane about 330.  This list of wires is created based on real detector data.

Following a sample generation of images composed of 3456 wires and 6048 time ticks, we downsize by a factor of two in wire and six in time to produce a resulting image size of 1728$\times$1008 pixels. We downsize by merging pixels through a simple sum of pixel values. Although this causes a loss of resolution and, therefore, a loss of information, particularly along the wire axis, it is necessary due to hardware memory limitation. However the loss of information is smaller than one would naively expected in the time direction because the digitization period is shorter than the electronics shaping time, which is 2 microseconds, or 4 time ticks. 

\begin{figure}[tb]
  \centering  
  \vspace{-0.5in}
  \includegraphics[width=0.4\textwidth]{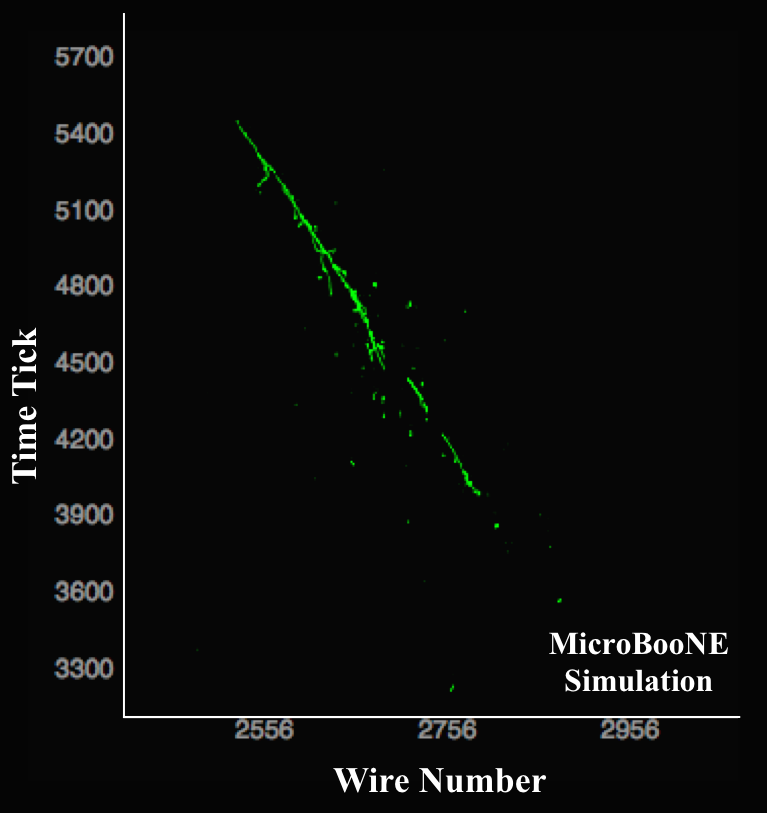}
  \includegraphics[width=0.4\textwidth]{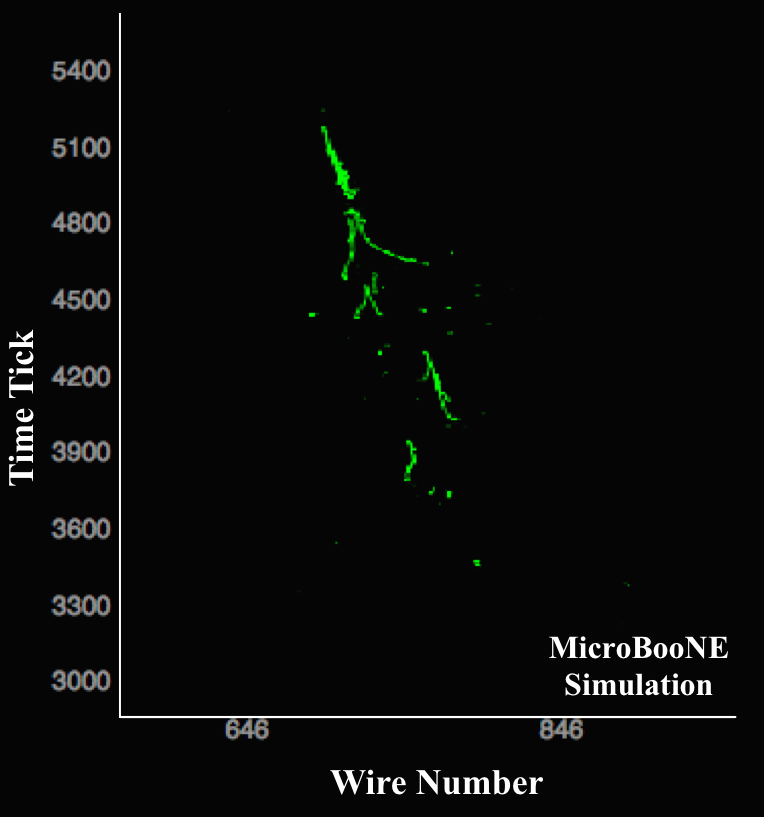}
  \\
  \vspace{0.1in}
  \includegraphics[width=0.4\textwidth]{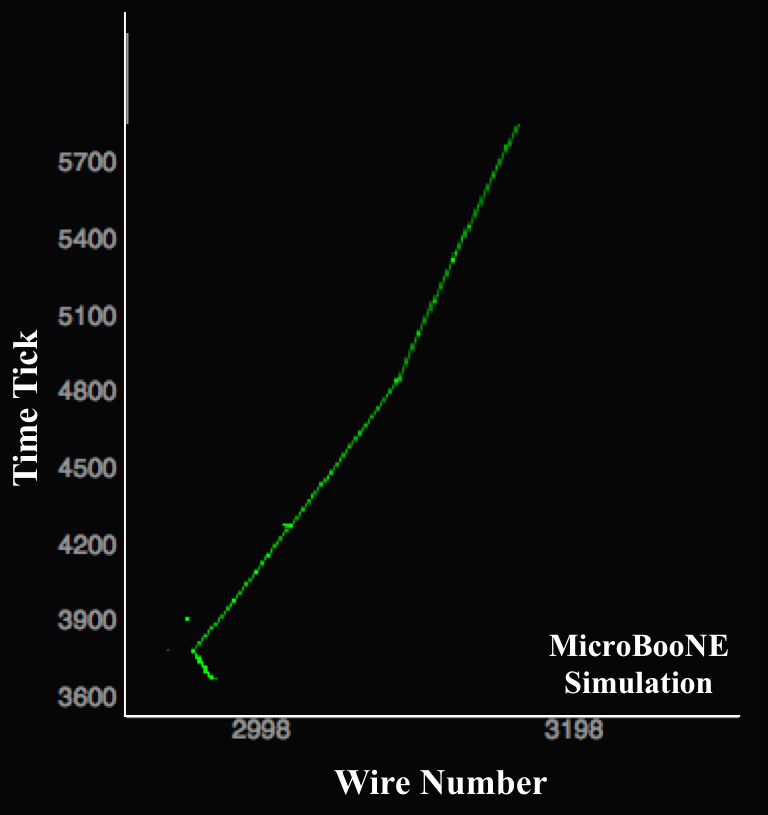}
  \includegraphics[width=0.4\textwidth]{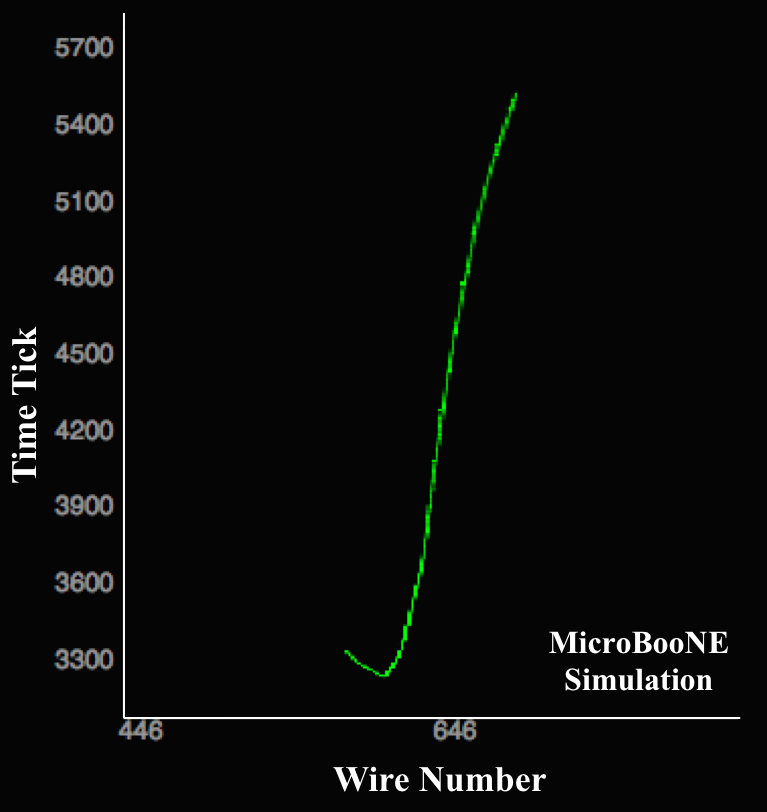}
  \\
  \vspace{0.1in}
  \includegraphics[width=0.4\textwidth]{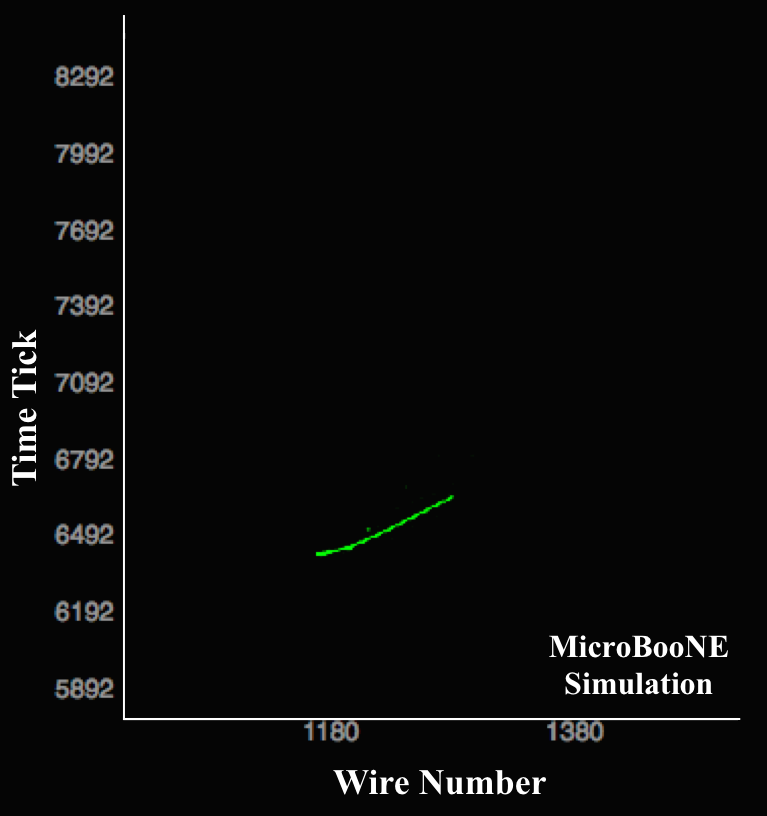}
\caption{Example event display for 3D box projection images for each particle type on the collection plane. From left top in clockwise direction, $e^-$, $\gamma$, $\mu^-$, proton, and $\pi^-$.}
  \label{fig:HiResEVD}
\end{figure}

Next, we run an algorithm in the LArCV package, referred to as \texttt{HiResDivider}, whose purpose is to divide the three, whole-event images into sets of three-plane sub-images where the set of sub-images views the same region of the detector over the same time period.  The algorithm is configured to produce sets of sub-images that are 576$\times$576 pixels.  Note that while this demonstration only uses images from the $Y$ plane, such a method will be useful for future work with large LArTPC images as it is a way to work with smaller images.  The algorithm first sub-divides the whole detector active volume into a number of equally sized 3D boxes, and then crops around the wires that read out that 3D region.  For our study, we find the 3D box that contains the start of a particle trajectory and extract the corresponding sub-images. While we use simulation-based information in our application, for future algorithms running on data, the same technique developed here can be applied following the identification of a neutrino vertex by another CNN or other pattern recognition algorithms. Example event displays of the $Y$ plane view are shown~in~figure~\ref{fig:HiResEVD}. 

\begin{figure}[t]
  \centering  
  \includegraphics[width=0.55\textwidth]{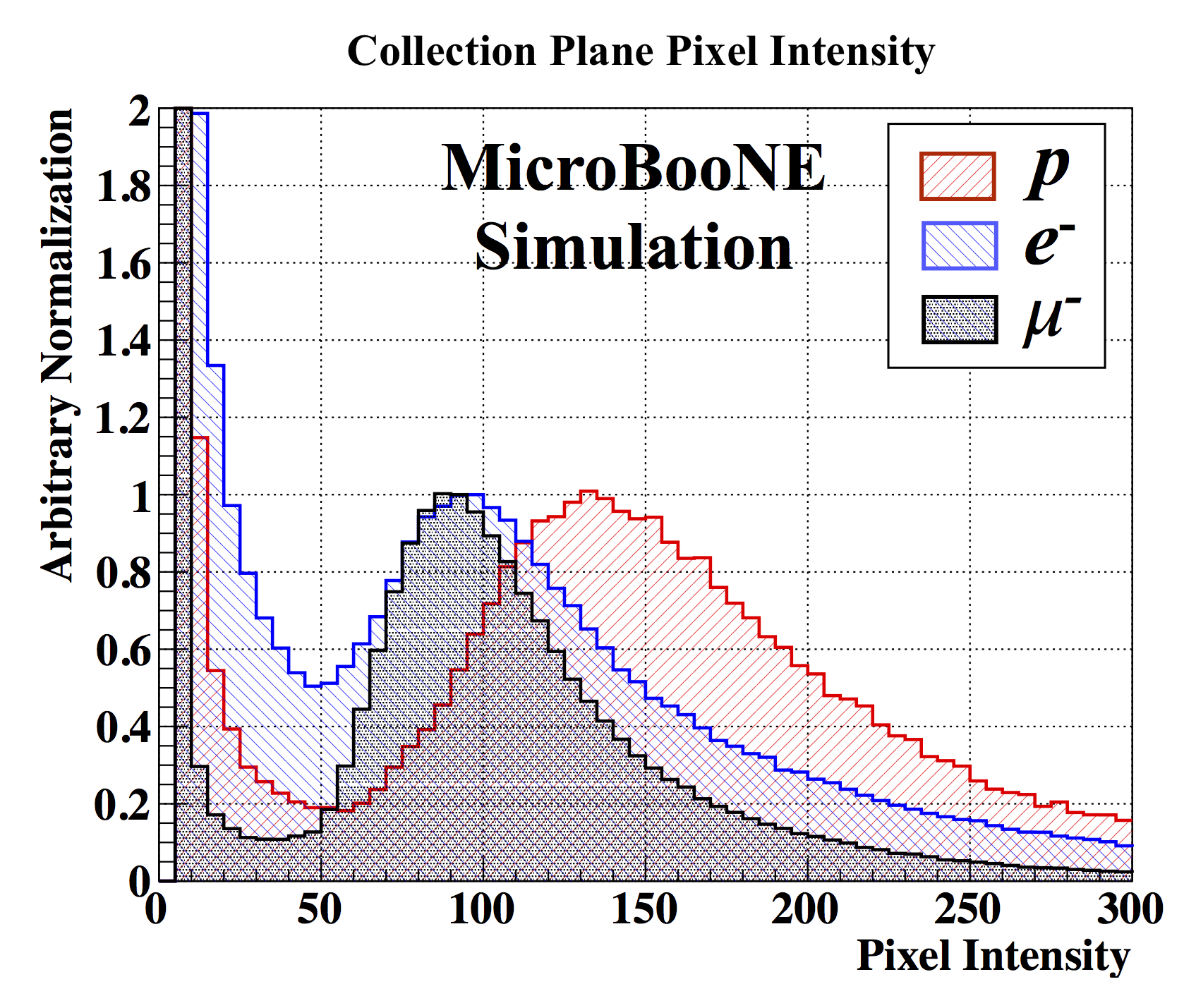}
\caption{Pixel intensity distribution (pixel value resulting from amplitude after merging operations of source waveform) for the collection plane. Red, blue, and black curves correspond to proton, $e^-$, and $\mu^-$ respectively. The vertical axis represents a relative population of pixels but normalized with an arbitrary unit to match the signal peak amplitude. }
  \label{fig:HiResPI}
\end{figure}

Finally, we apply a filter to remove images that contain almost no charge deposition information within the image. This is typically due to a particle immediately exiting the 3D box identified by \texttt{HiResDivider} algorithm or due to the particle traversing a region of unresponsive wires. An image was considered empty based on the number of pixels that carry a pixel intensity (PI) value above some threshold. 
 Figure~\ref{fig:HiResPI} shows the PI distribution on the $Y$ plane view for protons, $e^-$ and $\mu^-$ after running a simple, threshold-based signal-peak-finding algorithm on each column of pixels. A threshold on the filled pixel count is set to 40 for all particles except for protons which is set to 20, due to the short distance they travel.

Finally, in preparing the single-particle training sample, we also discard some proton events based on simulation information where a nuclear inelastic scattering process occurred which generated a neutron that in turn hit another proton. Such images contain multiple protons with space between them. Discarding these images helped the network to learn about a single proton image, which is aligned with our intention in this single-particle study. In the end, 22,000 events per particle type for training and about 3,800 per particle type for training validation were produced.

\subsubsection{Bounding Boxes for the Particle Detection Networks}

For the object detection networks, we must save a set of 2D ROIs or bounding boxes on each plane that is derived based on simulated particle energy deposition profiles. Using truth information from the simulation, a truth bounding box is stored for each particle that encapsulates where a particle deposited energy in the TPC. We then convert from detector coordinates into the image coordinates.  

\subsection{Network Training}	

\subsubsection{Classification}

For the particle classification studies in this demonstration, we train the networks, AlexNet and GoogLeNet, to perform this task. Training a network in essence involves adjusting the parameters of the network such that it outputs the right answer, however defined by the user, when given an image.  In practice, training a network involves sending a small collection of images, called a {\it batch}, along with a label for each image to the GPU card where images are passed through the network for classification. For a given input image, a network makes a prediction for its label. Both the label and the prediction typically take the form of a vector of positive real numbers where each element of the vector represents each class of object to be identified.  For an image which contains only one example, the label vector will have only one element with a value of one, and the rest being zero.  For the training images, the provided label is regarded as {\it truth} information.  The network outputs the predicted label with each element filled with numbers between 0 and 1 based on how confident the network is that the image contains each class. This prediction is often normalized such that the sum over all labels is set to 1. In this work, we refer to the elements of this normalized, predicted vector simply by {\it score}. Based on a provided label of an image and computed scores, a measure of error, called {\it loss}, is computed and used to guide the training process. 

The goal of training is to adjust the parameters of the network such that the loss is minimized. The loss we use is the natural logarithm of the squared-magnitude of the difference vector between the truth label and the prediction. Minimization of the loss is performed using stochastic gradient descent~\cite{backprop} (SGD). We attempt to minimize loss and maximize accuracy of the network during training, where accuracy is defined as the fraction of images for which the predicted label with the highest score matches the truth label. In order to avoid the network becoming biased towards recognizing just the images in the training sample, a situation known as over-training, we monitor the accuracy computed for a test or ``validation'' sample which does not share any events with a training set. 

\begin{figure}[t]
  \centering  
  \includegraphics[width=0.49\textwidth]{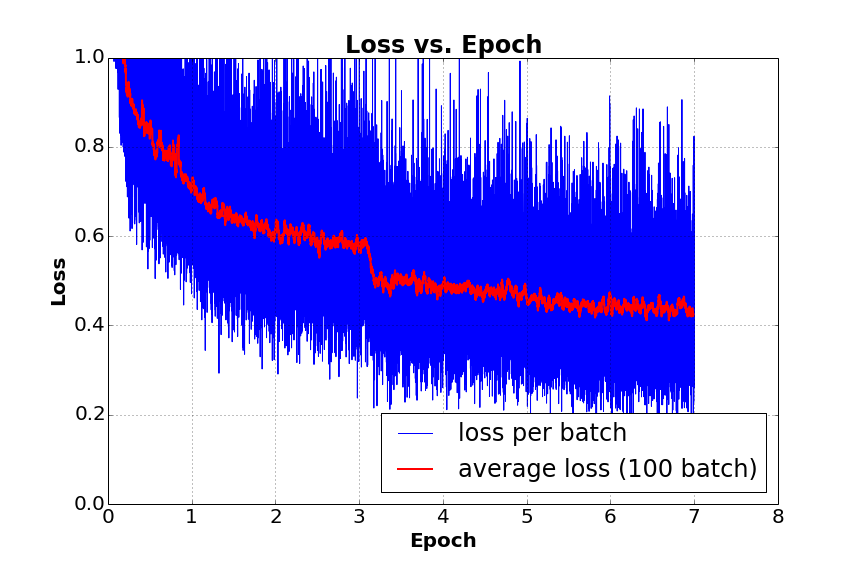}
  \includegraphics[width=0.49\textwidth]{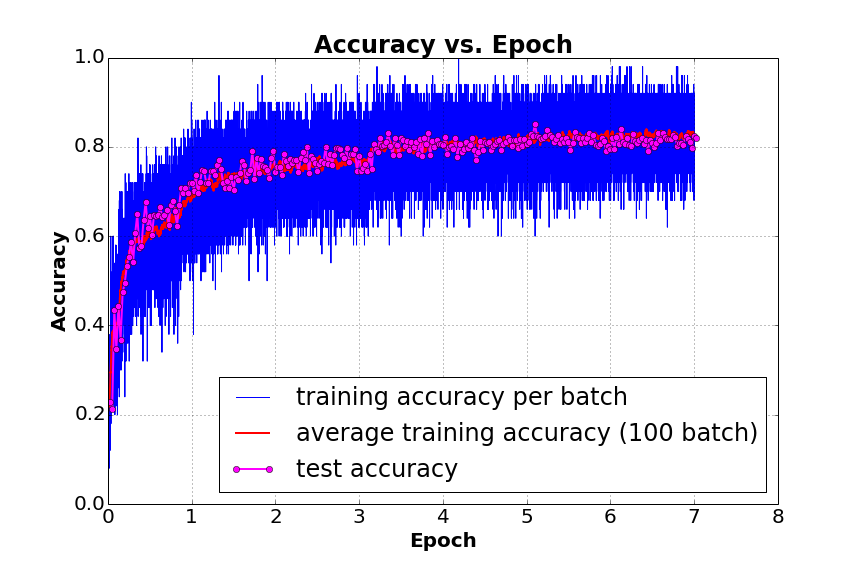}\\
  \includegraphics[width=0.49\textwidth]{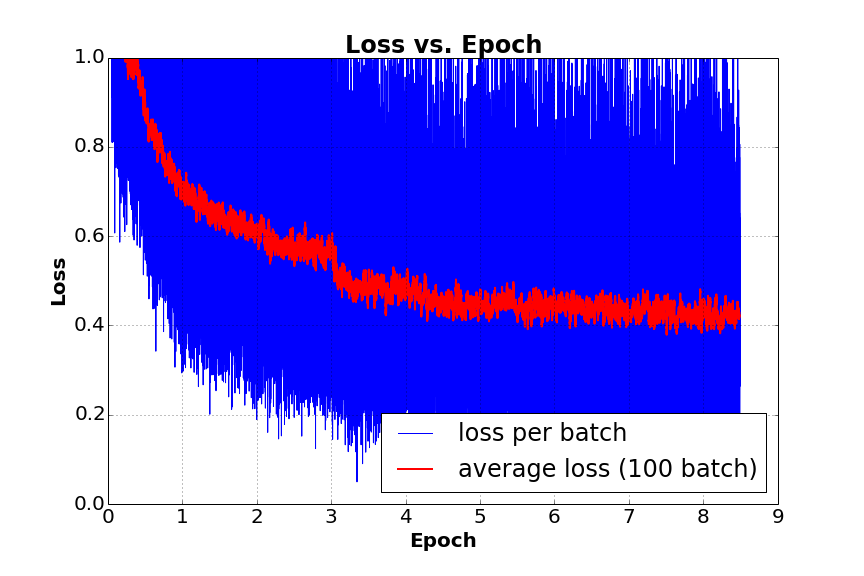}
  \includegraphics[width=0.49\textwidth]{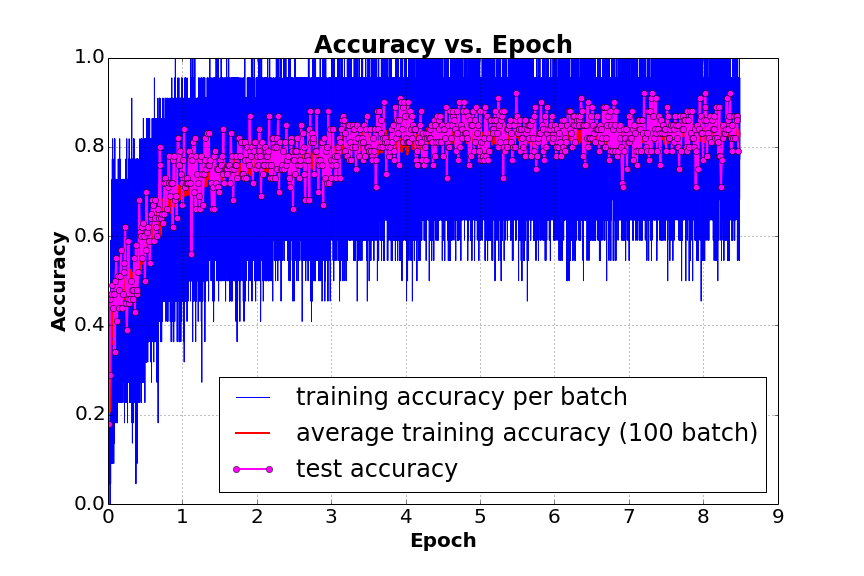}
\caption{Network training loss (left column) and accuracy (right column) for AlexNet (top row) and GoogLeNet (bottom row). The horizontal axis denotes elapsed time since the start of training. Blue data points are computed within the training process per batch of images. Red data points are computed by taking an average of 100 consecutive blue data points.  The magenta data points on the accuracy curve are computed using a validation sample to identify over-training. At around epoch 3 the learning rate of the network has been reduced for fine tuning. This results in a step of decrease in the loss and increase in the network accuracy.}
  \label{fig:5PHiResAccuracy}
\end{figure}

During the course of properly training a network, we monitor the accuracy computed for the test sample and watch to see if it follows the accuracy curve of the training sample. Both accuracy and loss are plotted against a standard unit of time, referred to as an epoch, which is the ratio of the number of images processed by the network for training to the total number of images in the whole training set. In other words, the epoch indicates how many times, on average, the network has seen each training example.  It is standard to train over many epochs.  During each iteration of the training process, a randomly chosen set of images from the training sample forms a batch. For AlexNet we send 50 images as one batch to the GPU to compute and update the network weights, while 22 are chosen for GoogLeNet. These batch sizes are chosen based on the network size and GPU memory limitations. After the network is loaded into the memory of the GPU, we choose a batch size that maximizes the usage of the remaining memory on the GPU. Figure \ref{fig:5PHiResAccuracy} shows both the loss and accuracy curves during the training process of the five particle classification task. The observed curves are consistent with what one would expect during an acceptable training course. 

In addition, for classification tasks, we train the networks with lower resolution images where the training and test images were downsized by a factor of two in both wires and time ticks to study how the network performance changes. As a final test, we train both AlexNet and GoogLeNet for a two-particle classification task using only a sample consisting of $\mu^-$ and $\pi^-$ images. This is to compare $\mu^-$ versus $\pi^-$ separation performance between networks trained on a sample containing only two particles versus the five-particle sample, where in the latter the network is exposed to a richer variety of features.  

\subsubsection{Particle Detection}

For single particle detection, we train the Faster-RCNN network~\cite{fasterrcnn} designed for object localization within images. The output of this network differs from the classification networks described above.  For a given image, the Faster-RCNN network returns $N$ classification predictions along with a rectangular shaped, minimum area bounding box for each image, which indicates where in the image the network estimates an object is located.  The number of boxes returned, $N$, is programmable. The inputs provided during training are different from the classification network as the task is different. To train the network, we provide for each training image a truth label and a {\it ground truth} bounding box.  (The designation of {\it ground} indicates the fact that the truth is defined by the user and might not necessarily correspond to an ideal box, if it even exists.) For MicroBooNE images, this is a defined rectangular sub-region of a plane view that contains all charge deposition based on simulation information.  A ground truth bounding box is used to train the network for a regression of the location and size of bounding box proposals.

\begin{figure}[tb]
\vspace{-0.1in}
  \centering
  \includegraphics[width=0.75\textwidth]{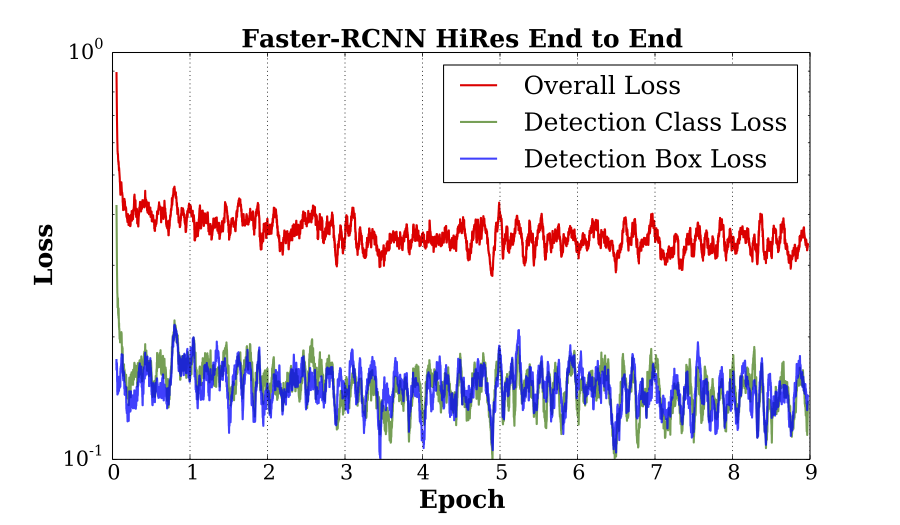}
\caption{Training loss for cost functions minimized by SGD in the Faster-RCNN end to end scheme. The classification loss (green curve) quickly falls at the beginning of network training, and slowly decreases until epoch 5. The bounding box regression loss (blue curve) decreases slowly over time. The overall detection loss (red curve) is a combination of the class and box loss and decreases until epoch 5, then levels off. }
  \label{fig:frcnnlog}
\end{figure}

One nice feature of the Faster-RCNN network is that its architecture is modular, meaning that it is composed of three parts: a base network that is meant to provide image features, a region proposal network that uses the features to identify generic objects and put a bounding box on it, and a classification network that looks at the features inside the bounding box, to classify what is in the bounding box. 
This modular design means that there is freedom to choose the base network.  This is because the portion of the network that finds objects in the image puts bounding boxes around them, and then classifies the object inside the box is intended to be appended to a base network which provides relevant image features.  In this study, we used AlexNet trained for five particle classification described above as the base network. We append the Faster-RCNN portion of the network after the fifth convolutional layer of AlexNet. 
We transfer the parameters from the classification portion of AlexNet and use them to initialize the classification portion of the Faster-RCNN network since it will be asked to identify the same classes of objects.
We train the whole network in the approximate joint training scheme \cite{fasterrcnn} with stochastic gradient descent. 
The loss minimized for the Faster-RCNN comes from both object detection and classification tasks.
This multi-task loss over the course of our training is shown in Figure~\ref{fig:frcnnlog}. We assess the performance of the detection network in a later subsection.

\subsection{Five Particle Classification Performance}
\label{sec:5PPerformance}

\begin{figure}[tb]
  \centering  
  \includegraphics[width=0.495\textwidth]{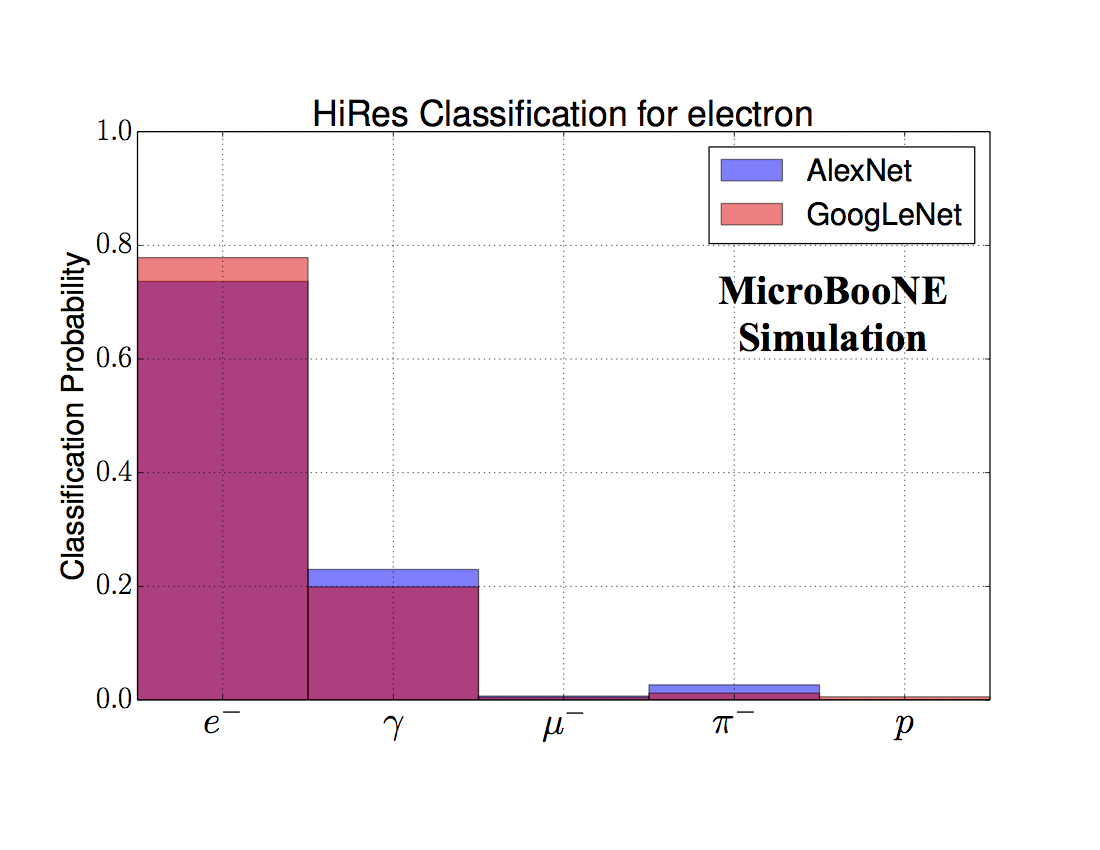}
  \includegraphics[width=0.495\textwidth]{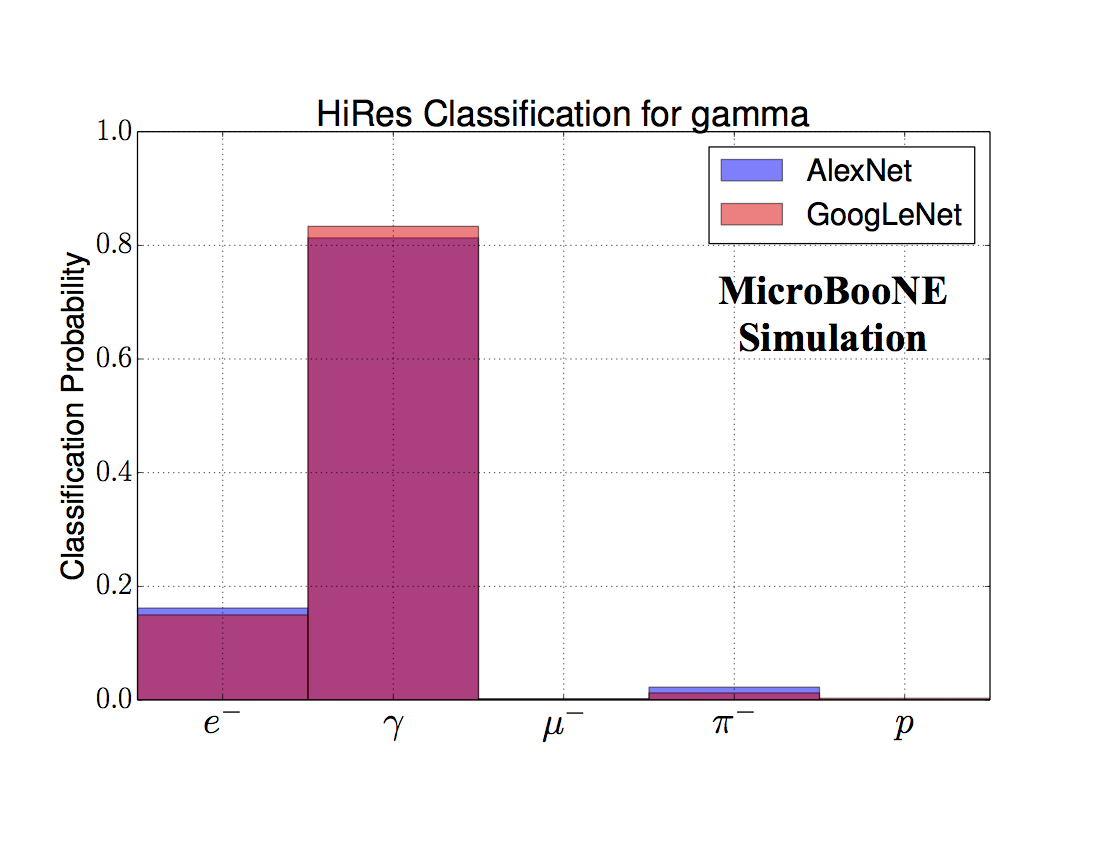}
  \includegraphics[width=0.495\textwidth]{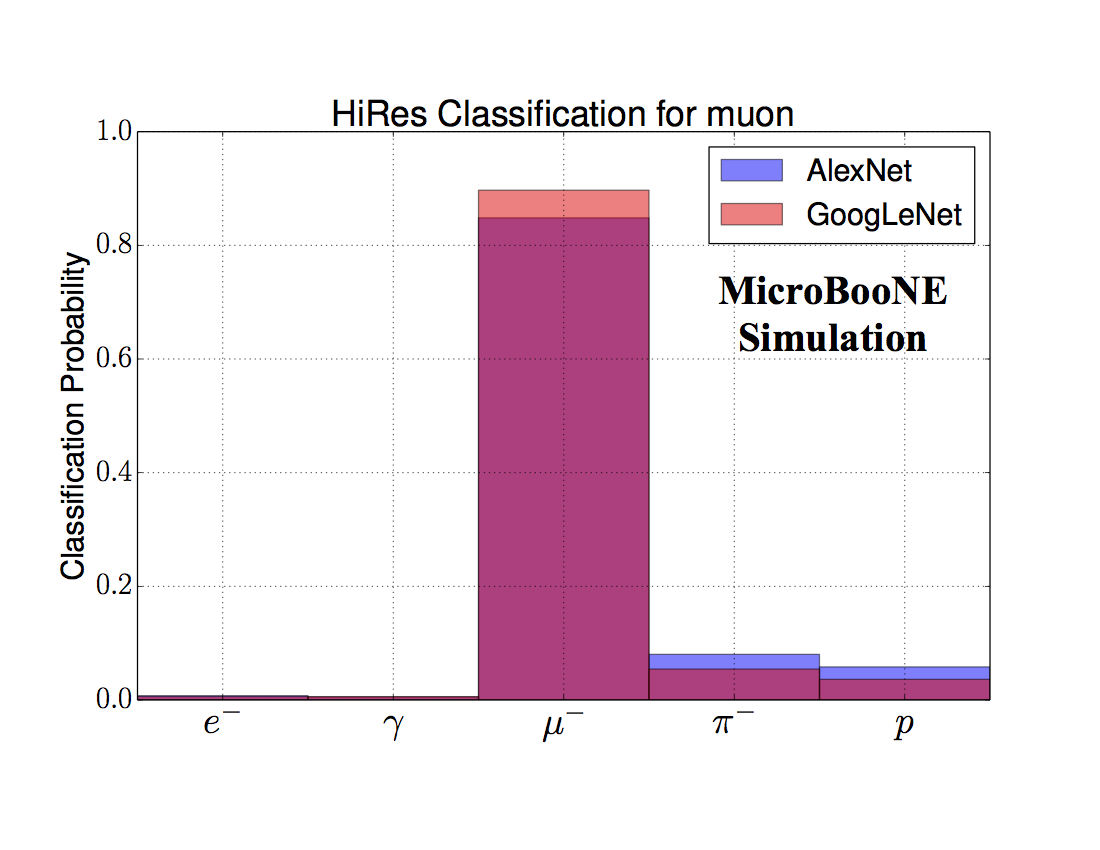}
  \includegraphics[width=0.495\textwidth]{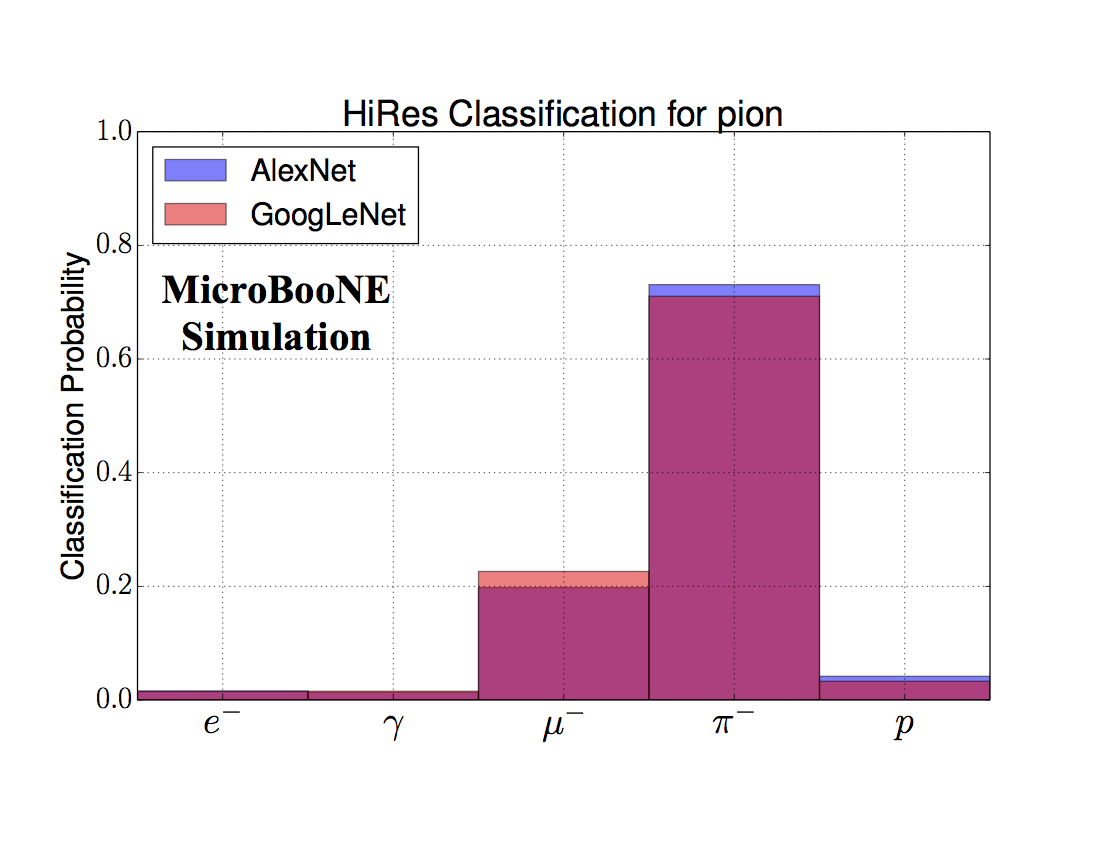}
  \includegraphics[width=0.495\textwidth]{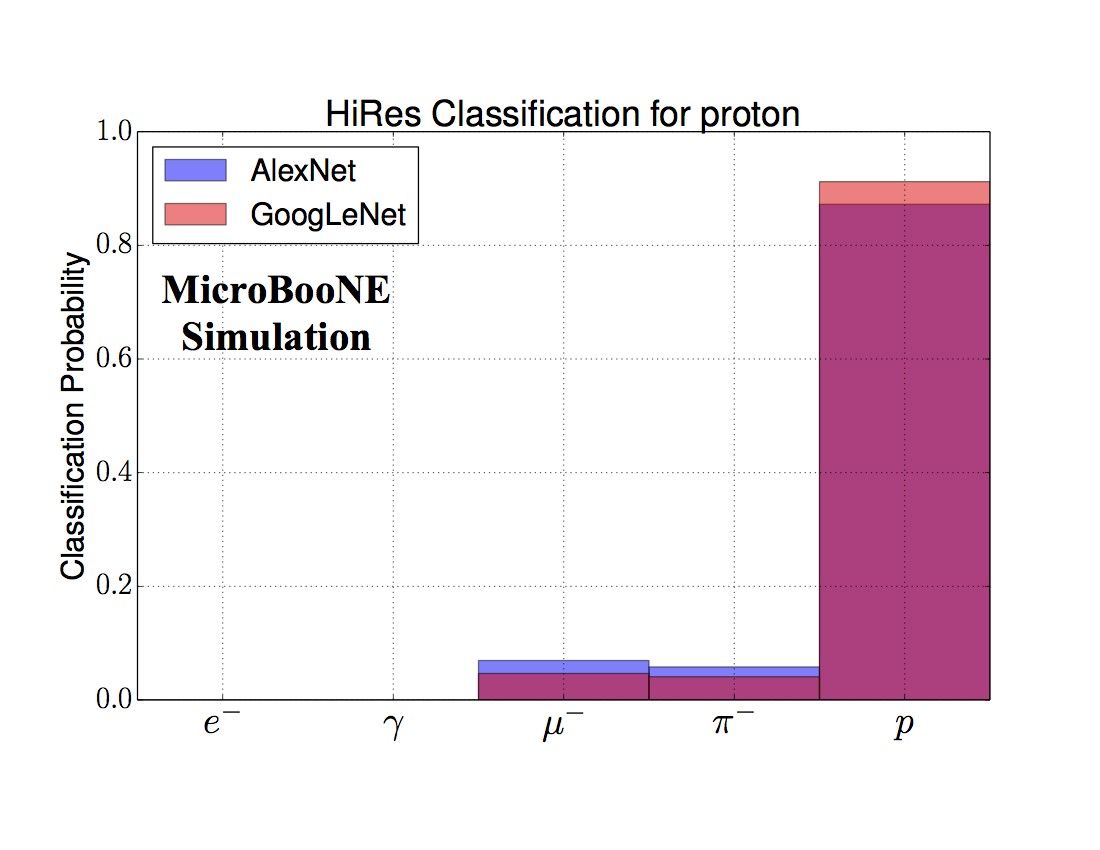}
\caption{Particle classification probabilities by true particle type from the high-resolution, five-class, single particle study. Both models struggle with electron and gamma separation as well as muon and charge pion separation.
}
  \label{fig:5PHiResPerformance}
\end{figure}

\begin{figure}[tb]
  \centering  
  \includegraphics[width=0.495\textwidth]{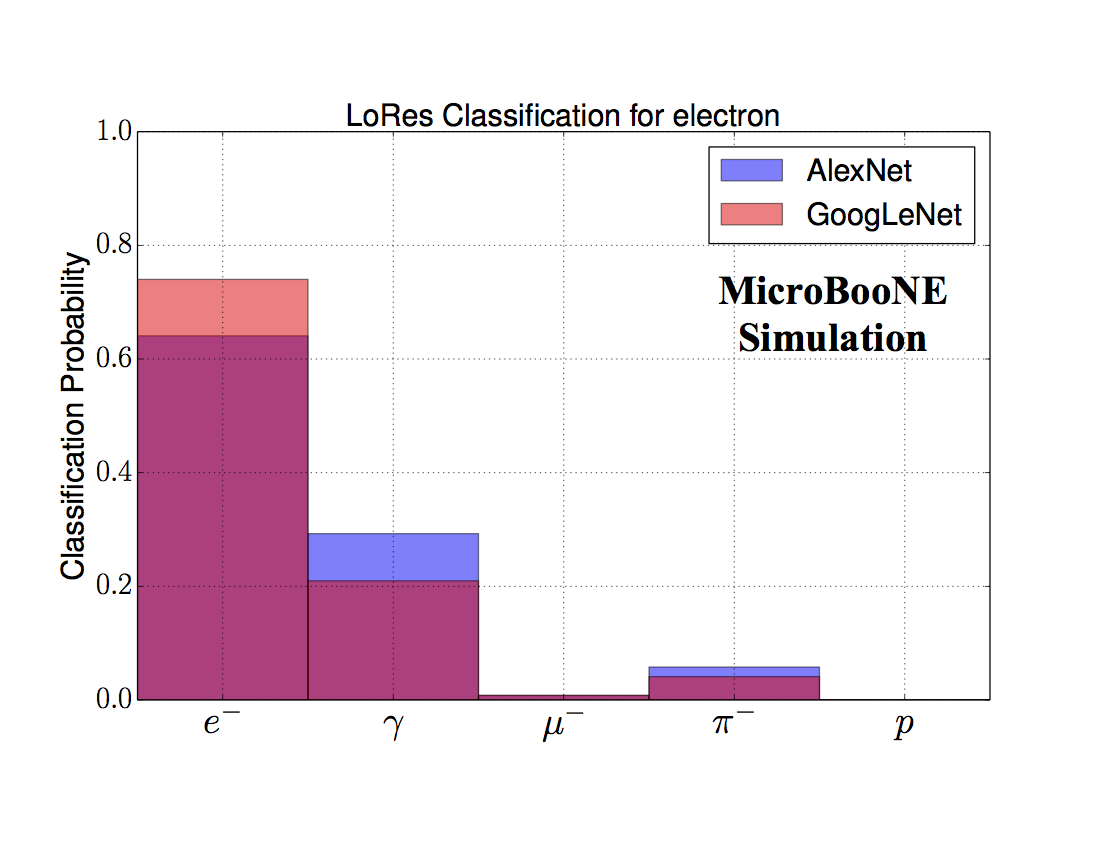}
  \includegraphics[width=0.495\textwidth]{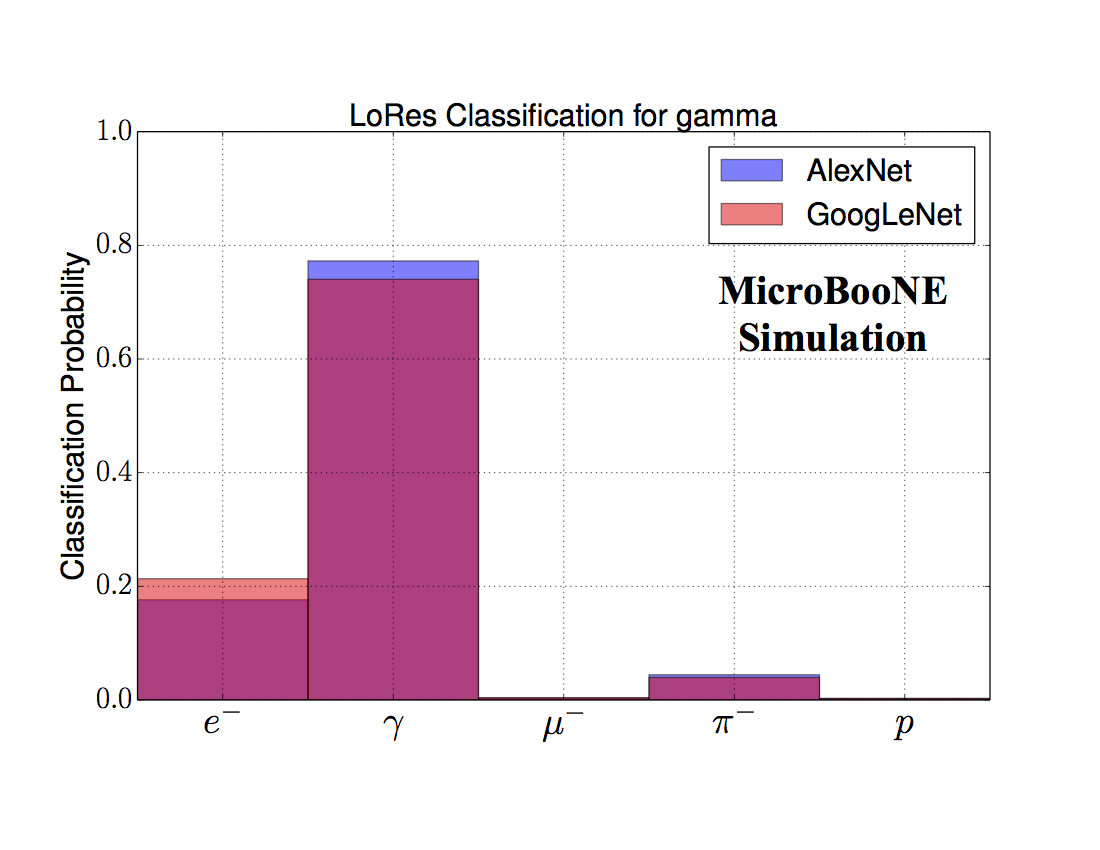}
  \includegraphics[width=0.495\textwidth]{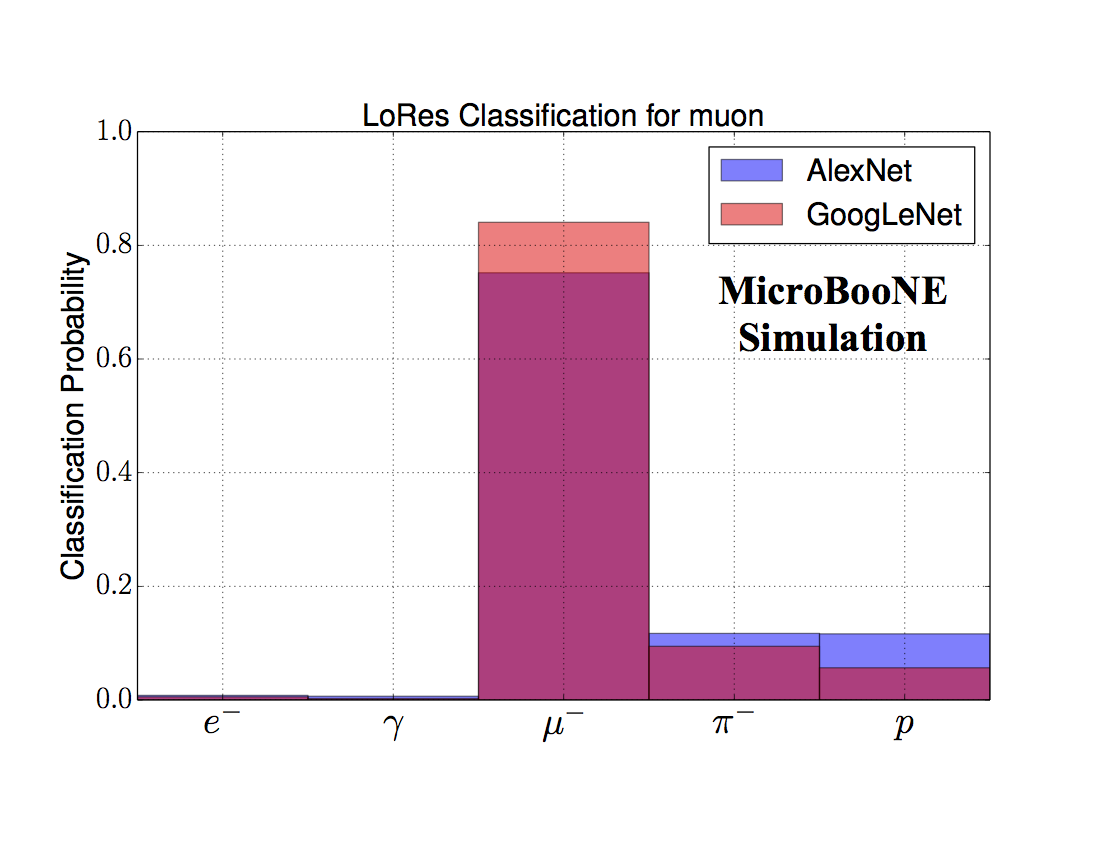}
  \includegraphics[width=0.495\textwidth]{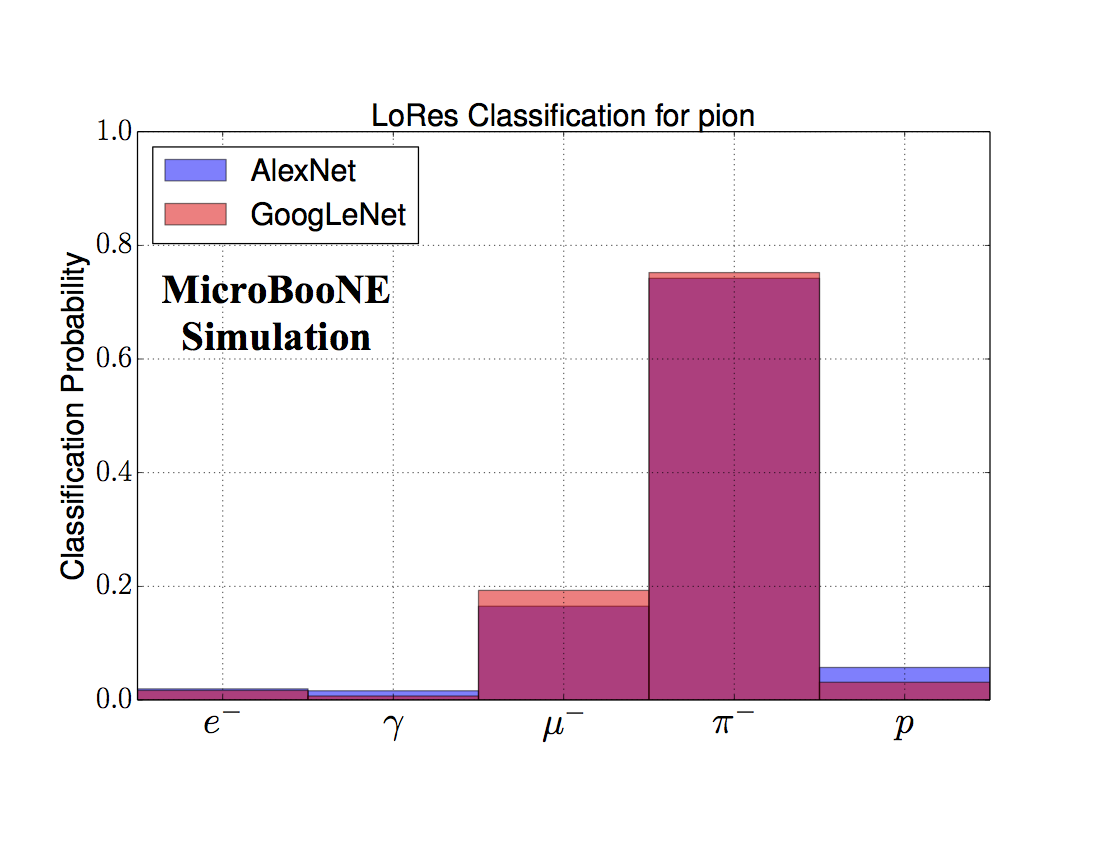}
  \includegraphics[width=0.495\textwidth]{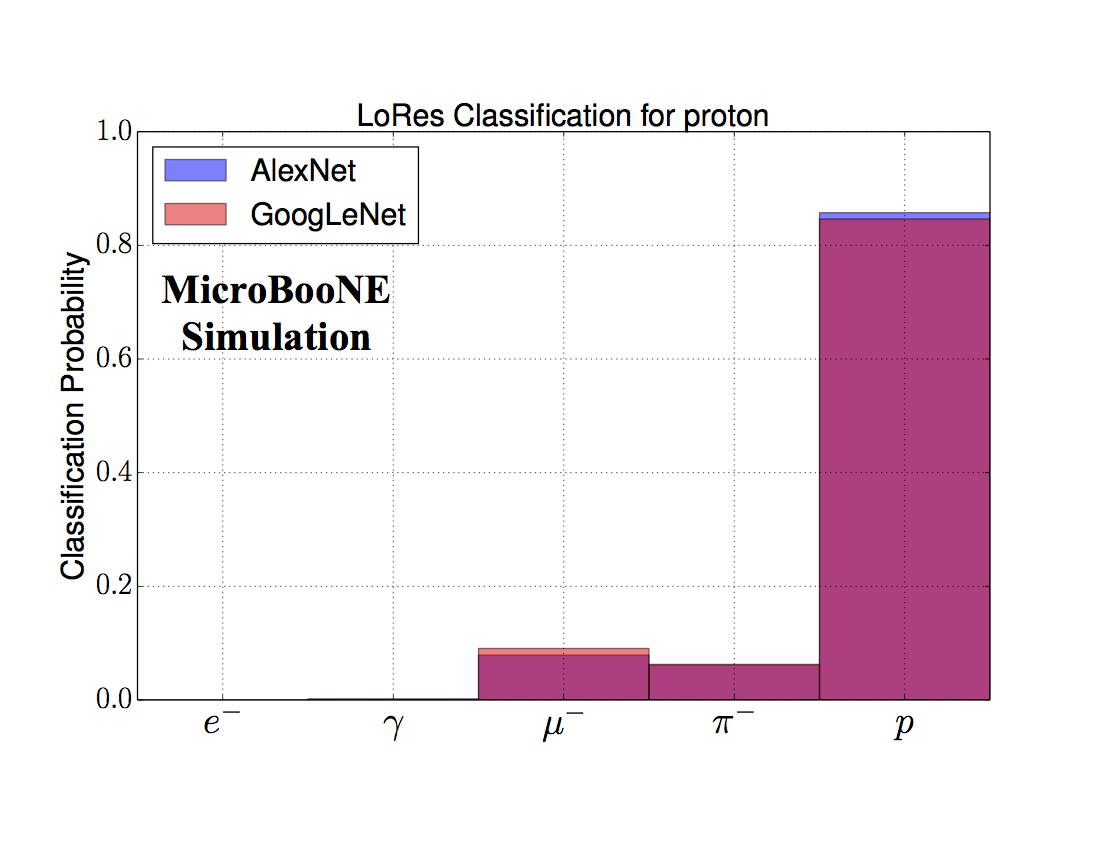}
\caption{Particle classification probabilities by true particle type from the Low-resolution, five-class, single particle study. Overall the classification performance is worse than the higher-resolution study whose results are shown above in Figure~\ref{fig:5PHiResPerformance}. The exception is for $\pi^-$.}
  \label{fig:5PLowResPerformance}
\end{figure}

Figure \ref{fig:5PHiResPerformance} shows the classification performance using the network trained on the 5 particle types described in section~\ref{sec:5PSamplePrep} using image sizes of 576 by 576 pixels. 
Each plot in the figure corresponds to a sample of events of a specific particle type, and the distribution shows the fraction of events classified as each of the 5 particle types, given by the highest score for each hypothesis.
A result using images that have been further downsized by a factor of 2 is shown in figure~\ref{fig:5PLowResPerformance}.  The classification performance for each particle type as well as the most mis-identified particle types, with mis-identifying score, are summarized in tables~\ref{tab:5PPerformance} and \ref{tab:5PMistake}. 

\paragraph{Learning Geometrical Shapes}
The networks have clearly learned about geometrical shapes of showers and tracks irrespective of image downsizing. Both figures~\ref{fig:5PHiResPerformance} and \ref{fig:5PLowResPerformance} show that the network is more likely to be confused among track-shaped particles ($\mu^-$, $\pi^-$, and proton) and also among shower-shaped particles ($e^-$ and $\gamma$) but less between those categories. 

\paragraph{Learning $\bm{dE/dx}$}
$e^-$ and $\gamma$ have similar geometrical shapes, i.e. they produce showers, but differ in the energy deposited per unit length, called $dE/dx$, near the starting point of the shower. 
The fact the network can separate these two particles fairly well may mean the network has learned about this difference, or some other difference in the gamma and electron shower development. It is also worth noting that downsizing an image worsens the result, likely because downsizing, which involves combining neighboring pixels, smears out the $dE/dx$ information.

\paragraph{Network Depth}
We expect GoogLeNet to be capable of learning more features than AlexNet overall, because of its more advanced, deeper model. Our results are consistent with this expectation as can be seen in the results table (Table~\ref{tab:5PPerformance}). However when comparing how AlexNet and GoogLeNet are affected by downsizing an image, it is interesting to note that AlexNet performs better for those particles with higher $dE/dx$ ($\gamma$ and proton), which suggests AlexNet is using the $dE/dx$ information more effectively than GoogLeNet after image is downsized by a factor of 2. 

\begin{table}[tb]
\begin{center}
\begin{tabular}{clllll}
 & \multicolumn{5}{c}{Classified Particle Type} \\
\cline{2-6}
\vspace{-0.1in}\\
\vspace{0.05in}
Image, Network & $e^-$ [\%] & $\gamma$ [\%] & $\mu^-$ [\%] & $\pi^-$ [\%] & proton [\%] \\
\hline 
\vspace{-0.1in}\\
\vspace{0.05in}
HiRes, AlexNet    & 73.6 $\pm$ 0.7 & 81.3 $\pm$ 0.6 & 84.8 $\pm$ 0.6 & 73.1 $\pm$ 0.7 & 87.2 $\pm$ 0.5 \\
\vspace{0.05in}
LoRes, AlexNet    & 64.1 $\pm$ 0.8 & 77.3 $\pm$ 0.7 & 75.2 $\pm$ 0.7 & 74.2 $\pm$ 0.7 & 85.8 $\pm$ 0.6 \\
\vspace{0.05in}
 HiRes, GoogLeNet & 77.8 $\pm$ 0.7 & 83.4 $\pm$ 0.6 & 89.7 $\pm$ 0.5 & 71.0 $\pm$ 0.7 & 91.2 $\pm$ 0.5 \\
\vspace{0.05in}
LoRes, GoogLeNet & 74.0 $\pm$ 0.7 & 74.0 $\pm$ 0.7 & 84.1 $\pm$ 0.6 & 75.2 $\pm$ 0.7 & 84.6 $\pm$ 0.6 \\
\bottomrule
\end{tabular}
\end{center}
\caption{Five particle classification performances. The very left column describes the image type and network where {\it HiRes} refers to a standard 576 by 576 pixel image while {\it LowRes} refers to a downsized image of 288 by 288 pixels. 
The five remaining columns denote the classification performance per particle type. Quoted uncertainties are purely statistical and assume a binomial distribution.}
\label{tab:5PPerformance}
\end{table}

\begin{table}[tb]
\begin{center}
\begin{tabular}{clllll}
 & \multicolumn{5}{c}{Classified Particle Type} \\
\cline{2-6}
\vspace{-0.1in}\\
\vspace{0.05in}
Image, Network & $e^-$ [\%] & $\gamma$ [\%] & $\mu^-$ [\%] & $\pi^-$ [\%] & proton [\%] \\
\hline 
\vspace{-0.1in}\\
\vspace{0.05in}
HiRes, AlexNet    & $\gamma$ 23.0 & $e^-$ 16.2 & $\pi^-$ 8.0 & $\mu^-$ 19.8 & $\mu^-$ 7.0 \\
\vspace{0.05in}
LoRes, AlexNet    & $\gamma$ 29.3 & $e^-$ 17.6 & $\pi^-$ 11.7 & $\mu^-$ 16.5 & $\mu^-$ 7.9 \\
\vspace{0.05in}
 HiRes, GoogLeNet & $\gamma$ 19.9 & $e^-$ 15.0 & $\pi^-$ 5.4 & $\mu^-$ 22.6 & $\mu^-$ 4.6 \\
\vspace{0.05in}
LoRes, GoogLeNet & $\gamma$ 21.0 & $e^-$ 21.3 & $\pi^-$ 9.4 & $\mu^-$ 19.3 & $\mu^-$ 9.1 \\
\bottomrule
\end{tabular}
\end{center}
\caption{The most frequently misidentified particle type for the five particle classification task. Following table~\ref{tab:5PPerformance}, the very left column describes the image type and network where {\it HiRes} refers to a standard 576 by 576 pixel image while {\it LowRes} refers to a downsized image of 288 by 288 pixels.  
The five remaining columns denote the classification performance per particle type. 
Each table element denotes the most frequently mistaken particle type and its mis-identification rate. }
\label{tab:5PMistake}
\end{table}

\newpage
\clearpage
\newpage
\clearpage
\newpage

\subsection{\boldmath$\pi^-/\mu^-$ Separation}
\label{sec:PiMuTraining}

\begin{figure}[tb]
  \centering  
  \vspace{-0.4in}
  \includegraphics[width=0.75\textwidth]{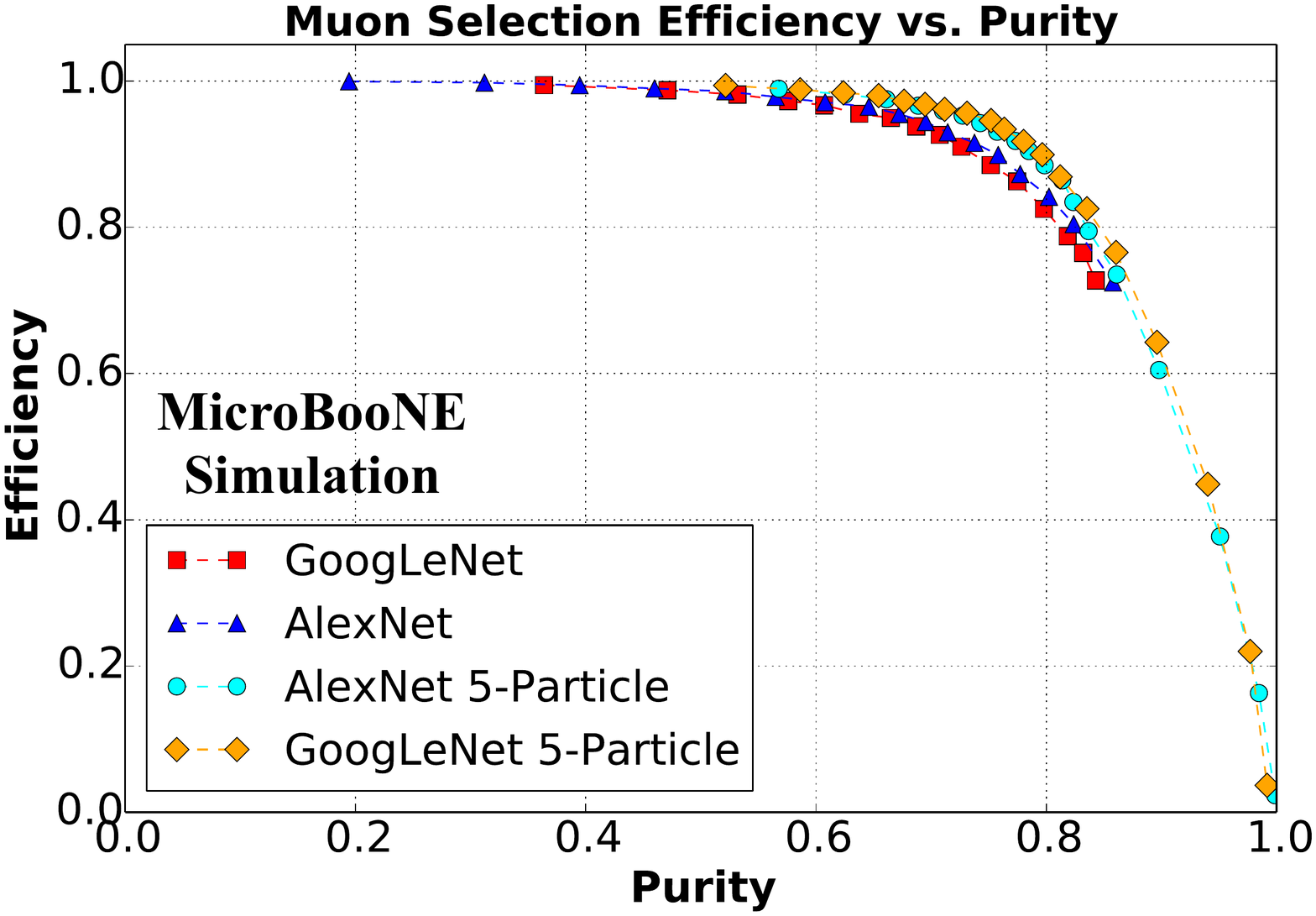}\\
  \vspace{-0.4in}
  \includegraphics[width=0.75\textwidth]{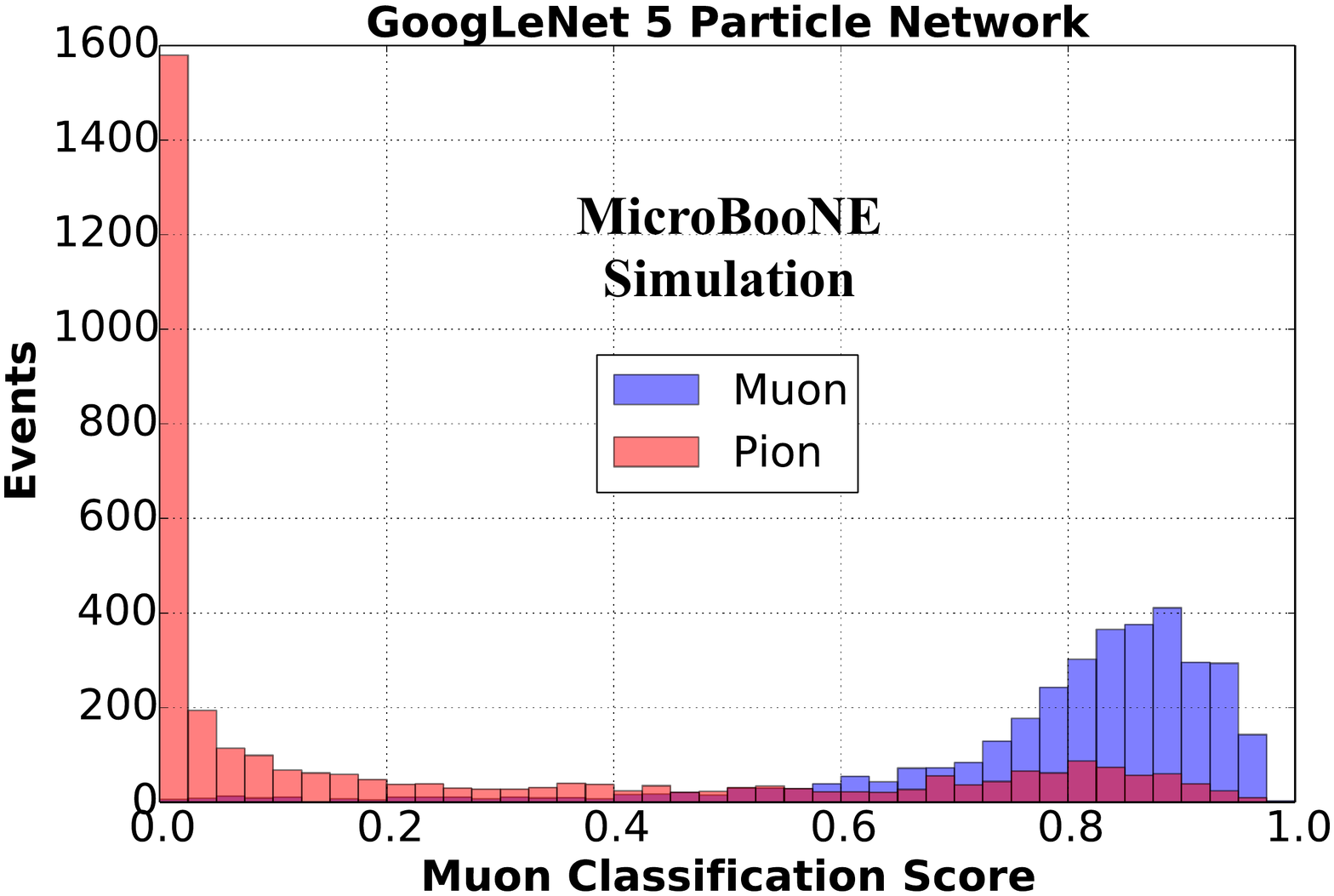}
  \vspace{-0.3in}
\caption{Top: $\mu^-$ selection EP curve for $\mu^-$ and $\pi^-$ images produced from the validation sample. Blue and red data points are AlexNet and GoogLeNet respectively trained with a sample set that only contain $\pi^-$ and $\mu^-$ as described in section~\ref{sec:PiMuTraining}. Orange and cyan data points are from GoogLeNet and AlexNet respectively, trained for five particle classification. Bottom: $\mu^-$ score distribution from GoogLeNet trained for five particle classification where the score is re-normalized for the purpose of $\mu^-/\pi^-$ separation.}
  \label{fig:PiMuSeparation}
\end{figure}

The top plot in figure~\ref{fig:PiMuSeparation} shows an efficiency vs. purity (EP) curve for $\mu^-$ selection for AlexNet and GoogLeNet trained for two-particle classification task from the validation sample where an efficiency is defined as
\begin{equation}
\text{Muon Selection Efficiency} = \frac{\text{Number of True Muons Selected As Muons}}
{\text{Total Number of Muons Scanned}}
\end{equation}
while a purity is defined as
\begin{equation}
\text{Muon Selection Purity} = 1 - \frac{\text{Number of True Pions Selected As Muons}}{\text{Total Number of Pions Scanned}}\text{.}
\end{equation}
Data points for the efficiency and purity result from using different cut values of the $\mu^-$ classification score to determine what was a muon, and plotted for the cuts ranging from 5\% to 95\% in 5\% steps. The lowest purity corresponds to the lowest selection score threshold cut. Blue and red data points are for AlexNet and GoogLeNet, respectively. The lack of data points at higher scores is the result of no events passing those score cuts. 
One can see both GoogLeNet and AlexNet have a similar performance as their curves are on top of each other.
 
\paragraph{Training with a More Feature-Rich Sample}
Figure~\ref{fig:PiMuSeparation} also contains orange and cyan data points that are obtained from GoogLeNet and AlexNet networks trained with the five-particle sample. However, because in this case we are interested in $\pi^-/\mu^-$ only, the $\pi^-/\mu^-$ classification score is computed by re-normalizing the sum of $\pi^-$ and $\mu^-$ classification score to 1 for these networks.
The fact that the network trained with five-particle classes performs better than the network trained only on two particle classes, i.e. solely on a $\pi^-/\mu^-$ sample, might appear counterintuitive, but this is one of the CNN's strengths: being exposed to other particle types, it can learn and generalize more features. It can then use this larger feature space to better separate the two particles.  Quoting a specific data point that lies in the outer most part of the orange curve, the network achieved 94.6$\pm$0.4\% muon selection efficiency with 75.2\% sample purity.

\paragraph{\boldmath$\mu^-/\pi^-$ Indistinguishability}
The bottom plot of figure~\ref{fig:PiMuSeparation} shows the score distribution of $\mu^-$ and $\pi^-$ from GoogLeNet trained for the five-particle classification task with a higher resolution image from the previous study. It is interesting to note that there is a small, but distinct set of $\pi^-$ events that follow the $\mu^-$ distribution. This makes sense since the $\pi^-$ has a similar mass to the $\mu^-$ and decays into $\mu^-$.  As a result, some $\pi^-$ can look just like a $\mu^-$. A typical way to distinguish a $\pi^-$ is to look for a nuclear scattering, which occurs more often for $\pi^-$ than for a $\mu^-$. There can also be a ``kink'' in the track at a point where the $\pi^-$ decays-in-flight into a $\mu^-$, although this is generally quite small. When neither is observable, the $\pi^-$ looks like a $\mu^-$, however when there is a kink or visible nuclear interaction involved, the $\pi^-$ is distinct. This can be seen by a very sharp peak for $\pi^-$'s in the bottom figure. 
The same reasoning explains why there are no $\pi^-$ above 97.5\% (with the statistics of this sample) because a $\mu^-$ can never be completely distinguished from the small fraction of $\pi^-$ that  neither decay nor participate in a nuclear scatter.

\subsection{\boldmath$e^-/\gamma$ Separation}

\begin{figure}[tb]
  \centering  
  \vspace{-0.4in}
  \includegraphics[width=0.75\textwidth]{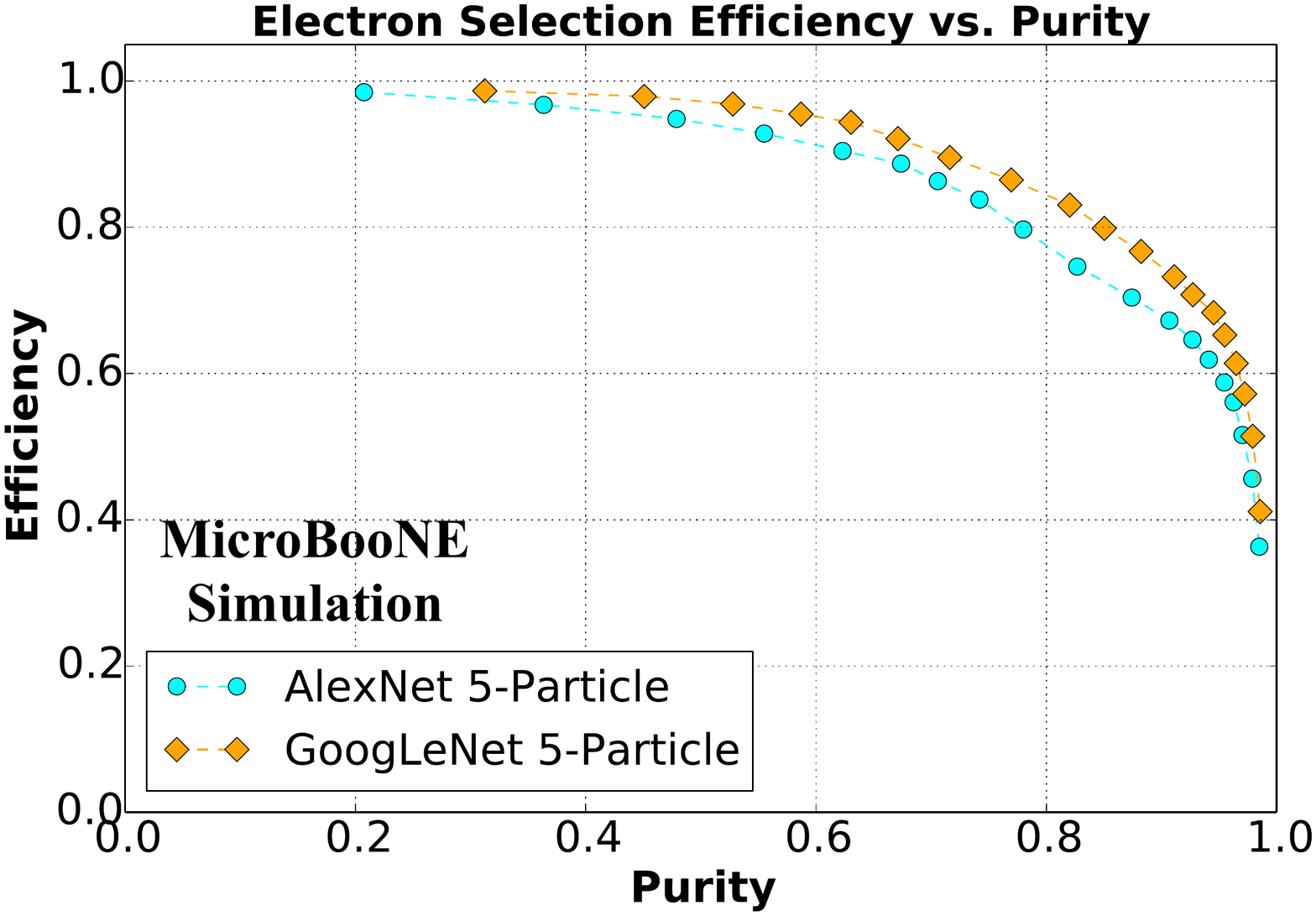}\\
  \vspace{-0.4in}
  \includegraphics[width=0.75\textwidth]{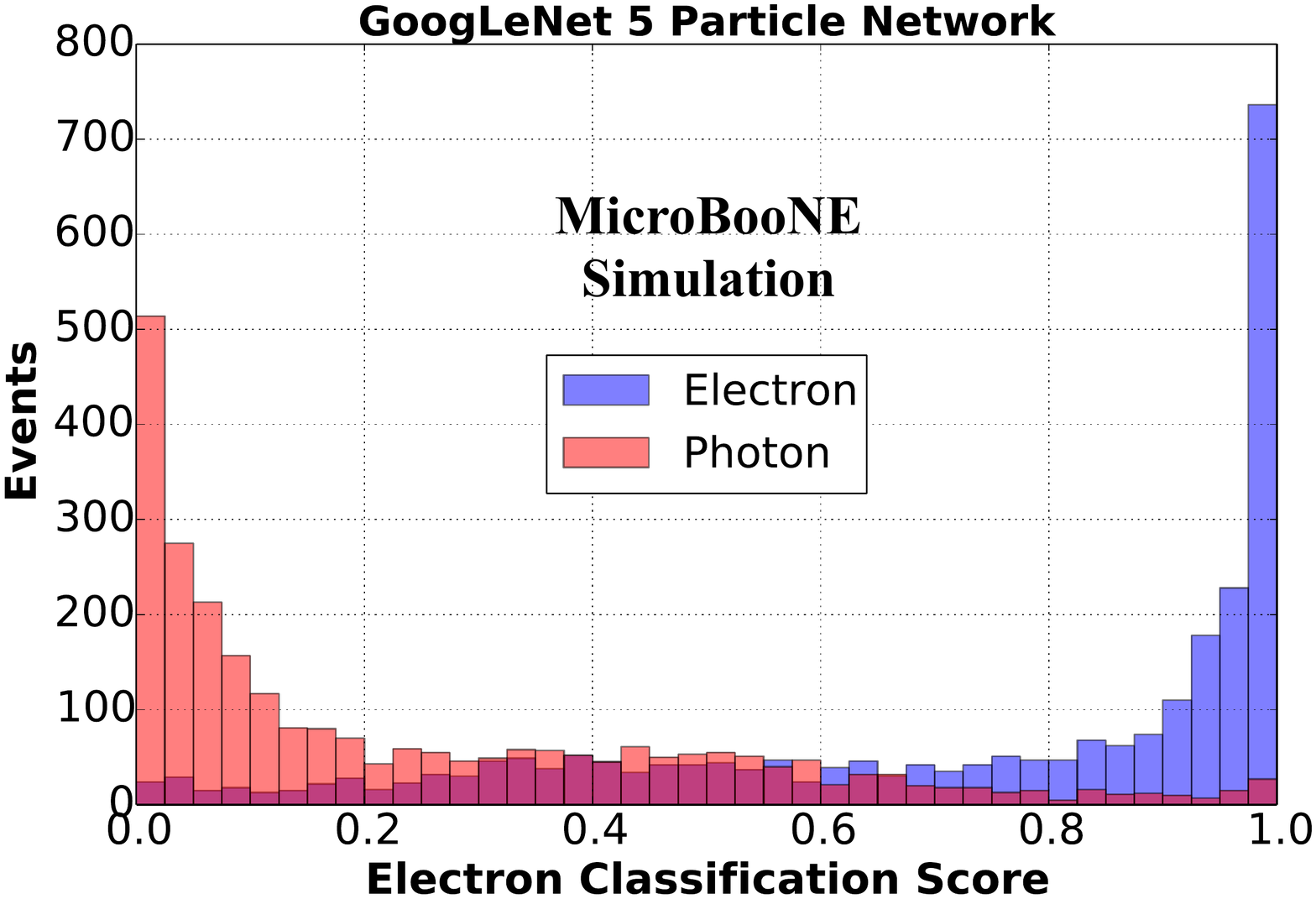}
  \vspace{-0.3in}
\caption{Top: $e^-$ selection EP curve for $e^-$ and $\gamma$ images made from the validation sample. Orange and cyan data points are from GoogLeNet and AlexNet respectively, trained for 5 particle classification. Bottom: $e^-$ score distribution from GoogLeNet trained for 5 particle classification where score is re-normalized for the purpose of $e^-/\gamma$ separation.}
  \label{fig:EGammaSeparation}
\end{figure}

We show a similar separation study for $e^-$ and $\gamma$ as we did for $\mu^-/\pi^-$. This time, however, we only show the results using the five-particle classification, since we saw those networks seem to perform better, presumably for similar reasons. The top plot in figure~\ref{fig:EGammaSeparation} shows the $e^-$ selection efficiency and purity from the validation set. The definition for an efficiency and purity is the same as how it was defined for the $\mu^-$ selection study. The outer-most point achieves an electron selection efficiency of 83.0$\pm$0.7\% with a purity of 82.0\%, although one might want to demand better separation at lower efficiency depending on the goals of an analysis.

\paragraph{\boldmath$e^-/\gamma$ Indistinguishability}
The bottom plot in figure~\ref{fig:EGammaSeparation} shows an electron classification score distribution for both $e^-$ and $\gamma$. The separation is not as strong as compared to $\pi^-/\mu^-$: the two types are essentially indistinguishable in the range of scores between roughly 0.3 to 0.6. We note that our high-resolution image has a factor of two in wire and six in time compression applied, and hence this might not be the highest separation achievable. It may be interesting to repeat more studies across different downsizing levels (including no downsizing) and study how this separation power changes. However, that is beyond the scope of this publication.

\subsection{Particle Detection Performance}

\begin{figure}[tb]
  \centering  
  \vspace{-0.4in}
  \includegraphics[width=1.0\textwidth]{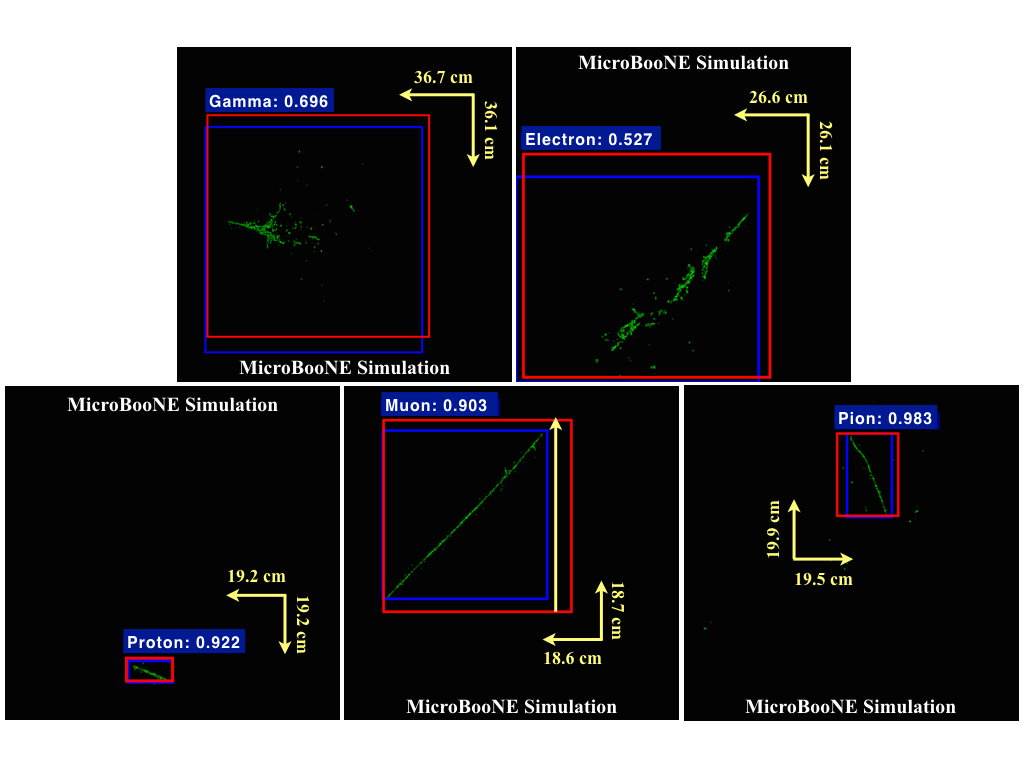}
  \vspace{-0.4in}
\caption{Example bounding boxes predicted by the CNN for each of the five particle classes. The blue box is the true bounding box. The red box is the network inferred bounding box. The detection score and inferred class sits atop the red box. It's interesting to note the network's ability to capture a shower-type particle's ionization charge within the detection box }
  \label{fig:detexamples}
\end{figure}

The goal of the Faster-RCNN detection network is to provide a bounding box around the energy deposition of the single particle.  Typical detection examples can be seen in figure \ref{fig:detexamples}. In the figure, the ground truth bounding boxes are also shown.  As done for all studies in this section, this analysis used the same training and validation sample described in section~\ref{sec:5PSamplePrep}. 

\begin{figure}[tb]
  \centering  
  \includegraphics[width=0.495\textwidth]{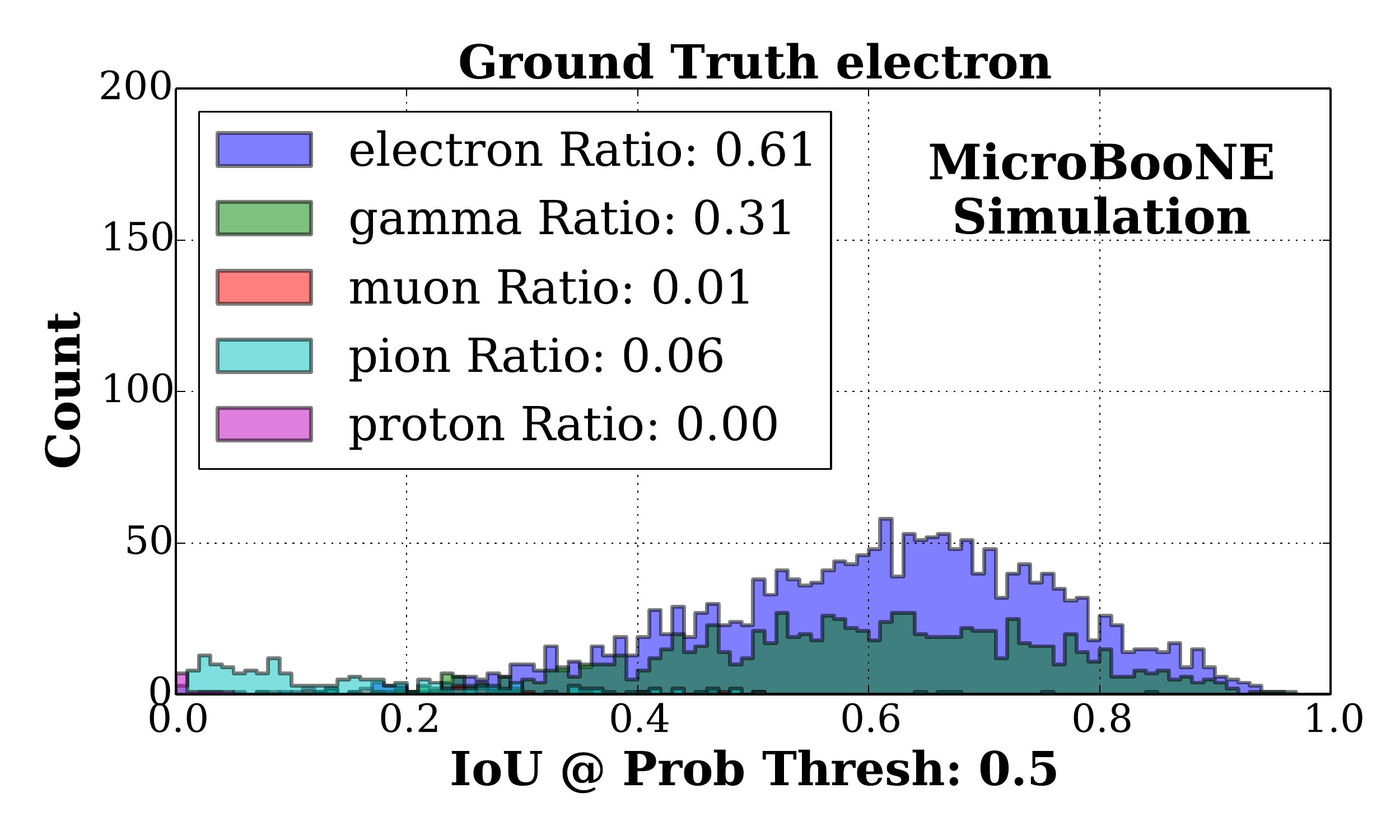}
  \includegraphics[width=0.495\textwidth]{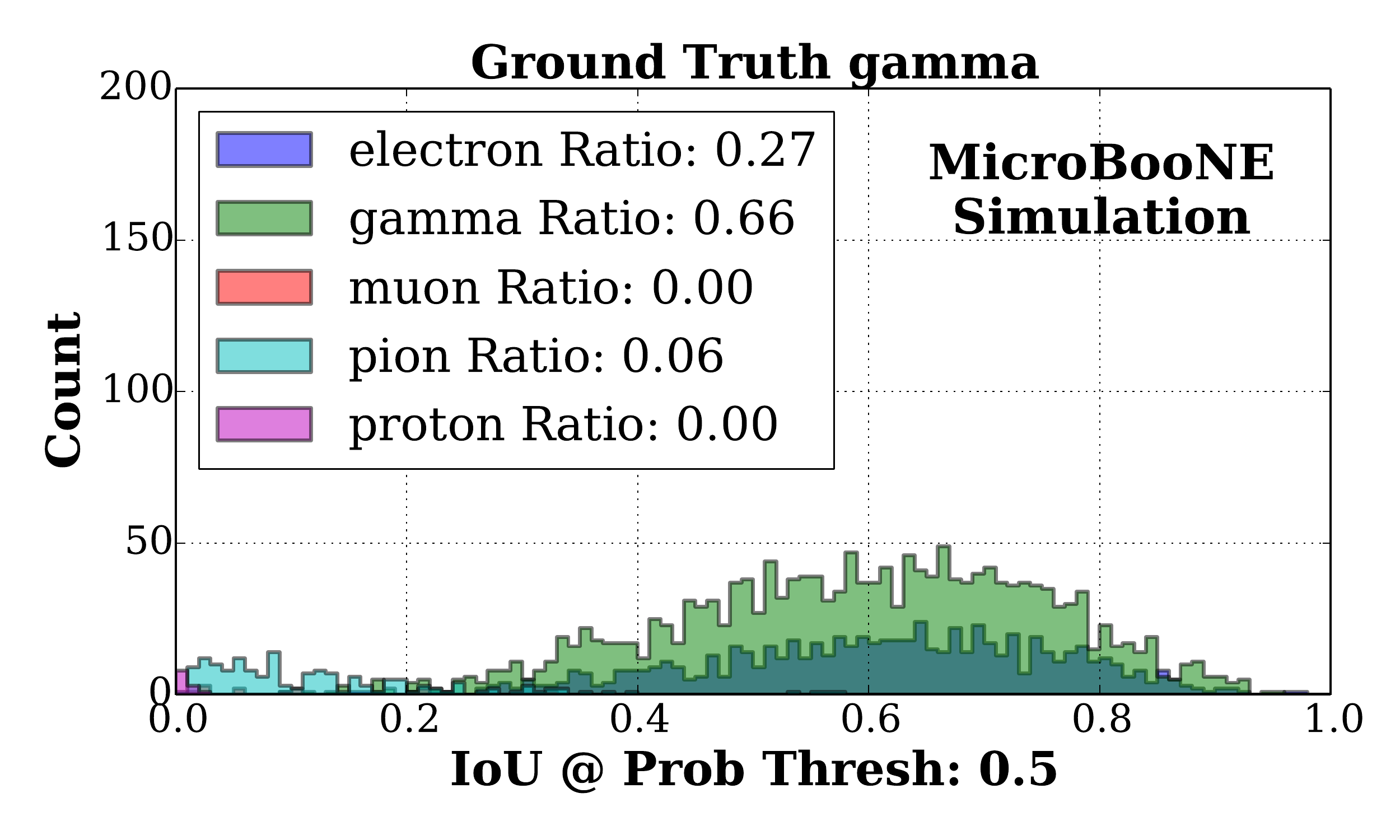}\\
  \includegraphics[width=0.495\textwidth]{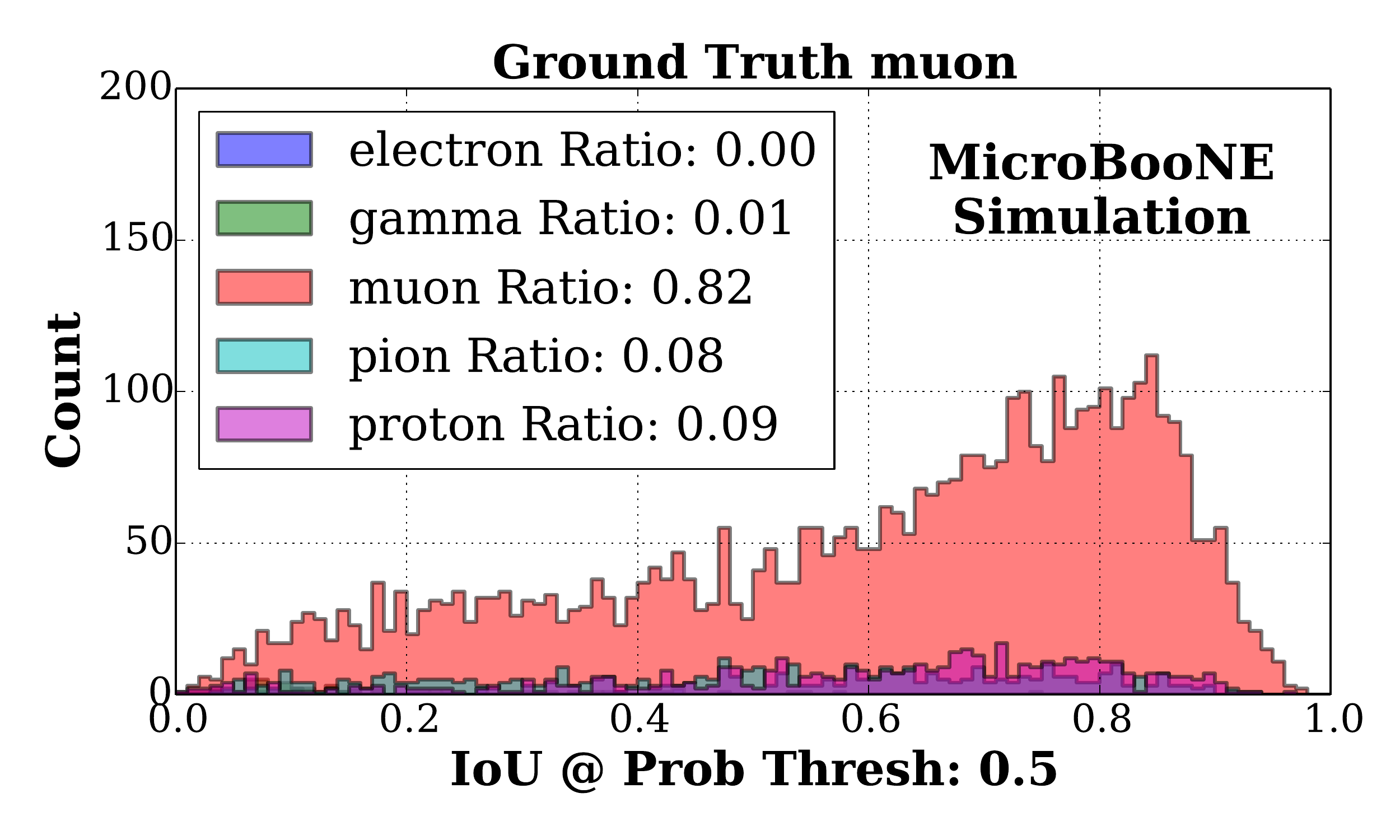}
  \includegraphics[width=0.495\textwidth]{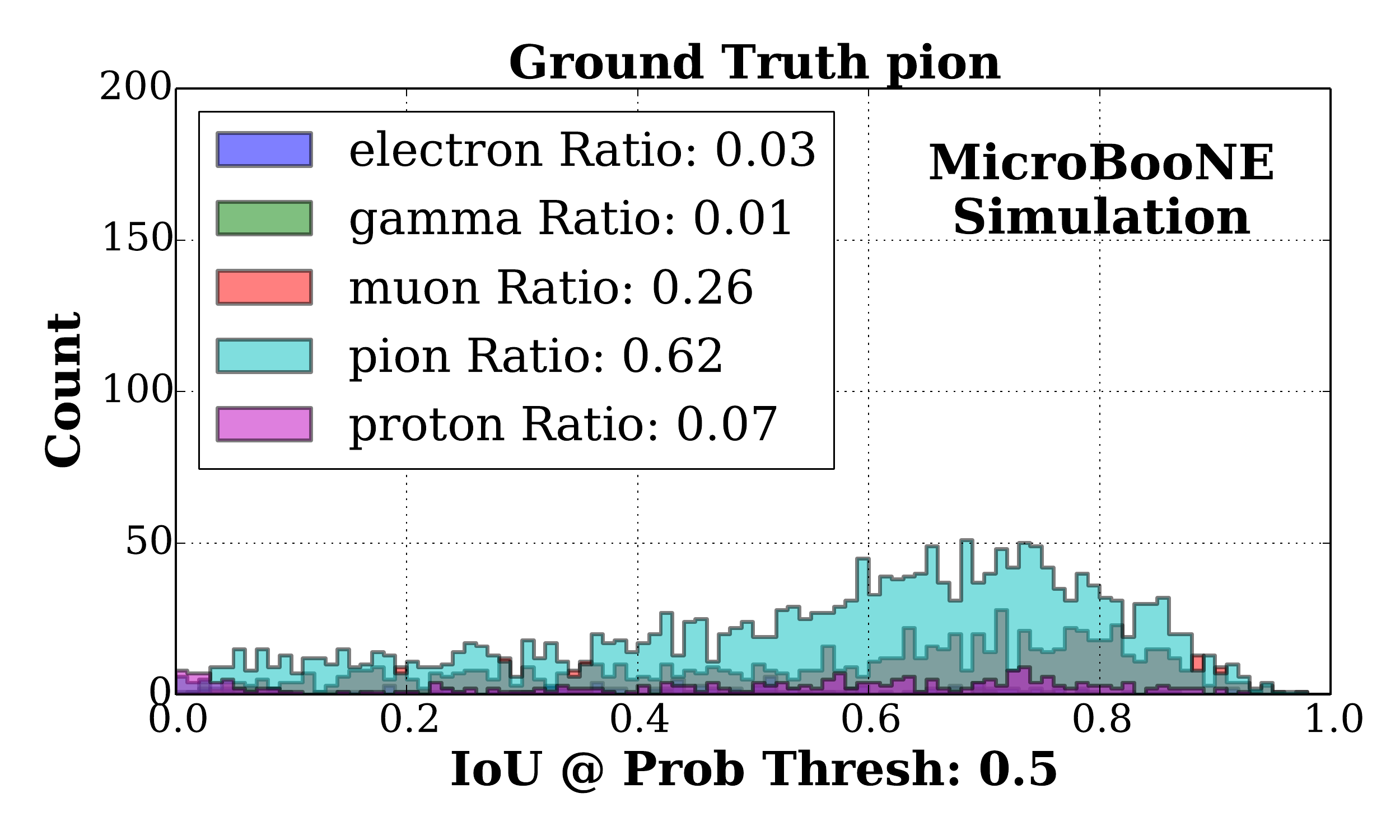}\\
  \includegraphics[width=0.495\textwidth]{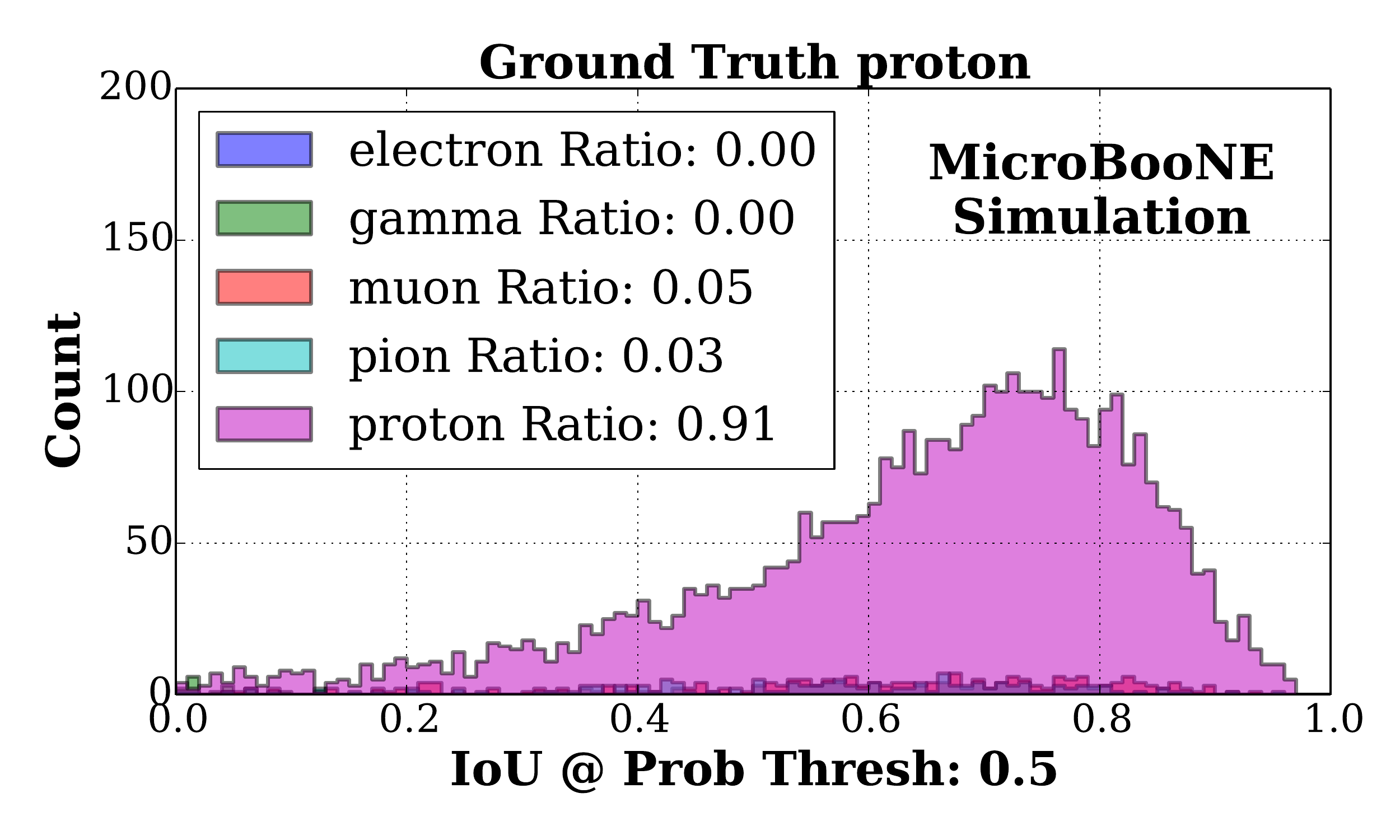}
\caption{
Intersection over union distributions for each of the 5 classes. The test sample is separated by true particle class and the detected fraction per particle type is reported in the legend. A valid set of detections is made with a score cut value of 0.5.
}
  \label{fig:iou}
\end{figure}

To quantify the Faster-RCNN detection performance on the single particle sample, we compute the intersection over the union of the ground truth bounding box and the predicted box with the highest network score. This is the standard performance metric used by object detection networks to compare with one another.  Intersection over union (IoU) is defined for a pair of boxes in the following way: the intersection area between two boxes is first computed by calculating the overlap area and then dividing by the difference between the total area of the two boxes and their intersection area. Specifically for two boxes with area $A_1$ and $A_2$,
\begin{align}
\text{IoU} = \frac{A_1\cap A_2}{A_1+A_2-A_1\cap A_2}.
\end{align}
This quantity is unity when the predicted box and the ground truth box overlap perfectly. In figure \ref{fig:iou}, we plot the IoU for the different five-particle classes. We separate the detected sample into the five different particle types and break down each sample by their top classification score. The true class label is in the title of the plot; the legend lists the five particle types that were detected for the sample and the class-wise fraction of all detections. For this plot, we make a cut on the network score of 0.5. We observe good detection accuracy and ground truth overlap on the $\mu^-$ and proton classes. If we consider classification only, $\mu^-$s and protons have the smallest contamination of other particle types. This could be a result of the strong classification performance of the AlexNet classifier model revealed in figure~\ref{fig:5PHiResPerformance}. The electron and $\gamma$ samples had expectedly similar contamination between the two as previously revealed by the pure AlexNet classifier. We also find a small contamination of $\pi^-$'s detection in the electron and $\gamma$ samples at the low IoU range indicating that some $\pi^-$ have features shared with electrons and $\gamma$'s. This is consistent with lower energy $\gamma$'s and electrons appearing track like in liquid argon. It is also interesting to note that both classes' IoU are similar, meaning the network is able to encapsulate the charge that spreads outwards as the shower develops. This means the model values the shower-like nature of the electron and gamma as essential and uses these learned features for classification. Lastly, the $\pi^-$ particle exhibits the least number of detections above a threshold of 0.5. We also find the largest contamination is from $\mu^-$. 

\subsection{Summary of Demonstration 1}

\subsubsection{Classification}
Our first demonstration shows that two CNN models, AlexNet and GoogLeNet, can learn particle features from LArTPC images to a level potentially useful for LArTPC particle-identification tasks. We also find that, as one might naively expect, downsizing an image has a negative effect upon particle classification performance for $e^-$, $\gamma$, $\mu^-$, and protons. The exception are $\pi^-$ images; this is not yet understood. GoogLeNet trained for five particle classification shows the best performance among tested cases for $e^-/\gamma$ (83.0$\pm$0.7\% efficiency and 82.0\% purity) and $\mu^-/\pi^-$ (94.6$\pm$0.4\% efficiency and 75.2\% purity) separation tasks.

\subsubsection{Particle Detection}

This demonstration shows that an AlexNet based Faster-RCNN object detector can be trained for a combined localization and particle classification tasks on LArTPC images. 
Similar to the image classification task described above, this network also distinguishes track-like particles ($\mu^-$/$\pi^-$/$p$) from shower-like particles ($e^-$/$\gamma$) fairly well. 
A difficulty in distinguishing $e^-$/$\gamma$ as well as $\mu^-$/$\pi^-$ particles are also observed in the detection network. A bounding box localization task, quantified by the IoU metric, gives similar results for each particle type, indicating reasonable localization of both track and shower like particles in high resolution images.



%
%
\section{Demonstration 2: Single Plane Neutrino Detection}
\label{sec:demo2}

In this section, we apply CNNs to the task of neutrino event classification and interaction localization using a single plane image. Neutrino event {\it classification} involves identifying on-beam event images that contain a neutrino interaction.  {\it Localization}, or {\it detection} as the task is referred to in the computer vision field, involves having the network draw a bounding box around the location of the interaction.  Detection, in particular, will be an important step for future analyses. If successful in identifying neutrino interactions, a high-resolution image then can be cropped around the interaction region, after which one can apply techniques, such as those in demonstration 1, at higher resolution. For the first iteration of this neutrino localization task, we ask the network to draw a bounding box in only a single plane image.
This is because the location of a neutrino interaction in all three planes requires the prediction of a bounding box in each plane that should be constrained to represent the same 3D box in the detector. Future work will ask the network to predict such a volume and project the resulting 2D bounding box into the plane image.

We take a two-step approach:
\begin{enumerate}
\item neutrino event selection and
\item neutrino interaction (ROI) detection (or localization) within an event image.
\end{enumerate}
Our strategy for the 1$^{\text{st}}$ step is to train the network with Monte Carlo neutrino interaction images overlaid with cosmic background images coming from off-beam data events. Accordingly, we define two classes for the classification task: 1) Monte Carlo neutrino data events overlaid onto cosmic data events, and 2) purely cosmic data events. For this to succeed, the Monte Carlo signal waveform, which is affected by particle generation, detector response, simulation, and calibration needs to match the real data. Otherwise a network may find a Monte-Carlo-specific feature to identify neutrino interactions, which may work for the training sample but may not work at all, or in a biased way, for real data. 

For a neutrino event classification task, we use a simplified version of Inception-ResNet-v2~\cite{Inceptionv4}. The network is composed of three different types of modules that are stacked on top of one another.  Because our images are larger (864$\times$756 pixels) than that used by the original network (299$\times$299 pixels), we must shrink the network so that we can train the network with the memory available on one of our GPUs. The three types of modules are labeled A, B, and C. 
Here, we only list our modifications from the original network in the reference: we reduced the number of Inception-A modules and Inception-C modules from 5 to 3, and the number of Inception-B modules from 10 to 5. 

For neutrino detection training, we use AlexNet as the base network for the Faster-RCNN object detection network, similar to what we did for demonstration 1. However, unlike our approach in demonstration 1, we train AlexNet+Faster-RCNN from scratch, instead of using the parameters found by training the network from a previous task since we found that we were not able to train AlexNet to a reasonable level of accuracy through such fine-tuning. Accordingly, the AlexNet+Faster-RCNN model trained in this study is specialized for detecting neutrino-vertex-like objects, instead of distinguishing neutrinos against cosmic rays. 

\subsection{Data Sample Preparation}
\label{subsec:sampleprep}

For this study, we generated simulated neutrino events without unresponsive wires first, and then we overlay an off-beam event image, which was recorded by the detector triggered out-of-time with the beam. Such off-beam images contain only cosmic-ray tracks.  The neutrino interactions are generated by passing the MicroBooNE neutrino flux simulation for neutrinos coming from Fermilab's Booster Neutrino Beam (BNB)~\cite{BNB} through the {\sc{genie}} event generator~\cite{genie}. 
While the simulation of the noise features observed in data is under development, we opted to utilize real detector data to characterize the impact of noise on this technique.
This is because the real data will have noise features and unresponsive wires throughout the image that a final application of the network must contend with. However this advantage does come at a cost.  
We note that there might be differences in the wire response modeling between the simulation and the real data that the network could use to identify neutrino interactions in the training sample. However, the topology of many neutrino interactions should be distinct enough to be used by the network. Further study to quantify this effect must be performed in order to apply the technique for high level physics analysis. However, the goal of this work is to demonstrate that this technique from computer vision can be applied to this task rather than to precisely quantify the expected performance.  

Upon overlaying the neutrino and off-beam cosmic-ray images, we remove the signals from the wires in the simulation image designated as unresponsive according to a bad wire record determined on a per-event basis for the cosmic data image. The determination of bad wires is performed both by referencing a list of known unresponsive wires and by various event-by-event checks of the signals from the wires.  
We create images with a downsizing factor of four for wires and eight for time ticks which creates an event image 864 by 756 pixels.

\begin{figure}[t]
  \centering  
  \includegraphics[width=1.0\textwidth]{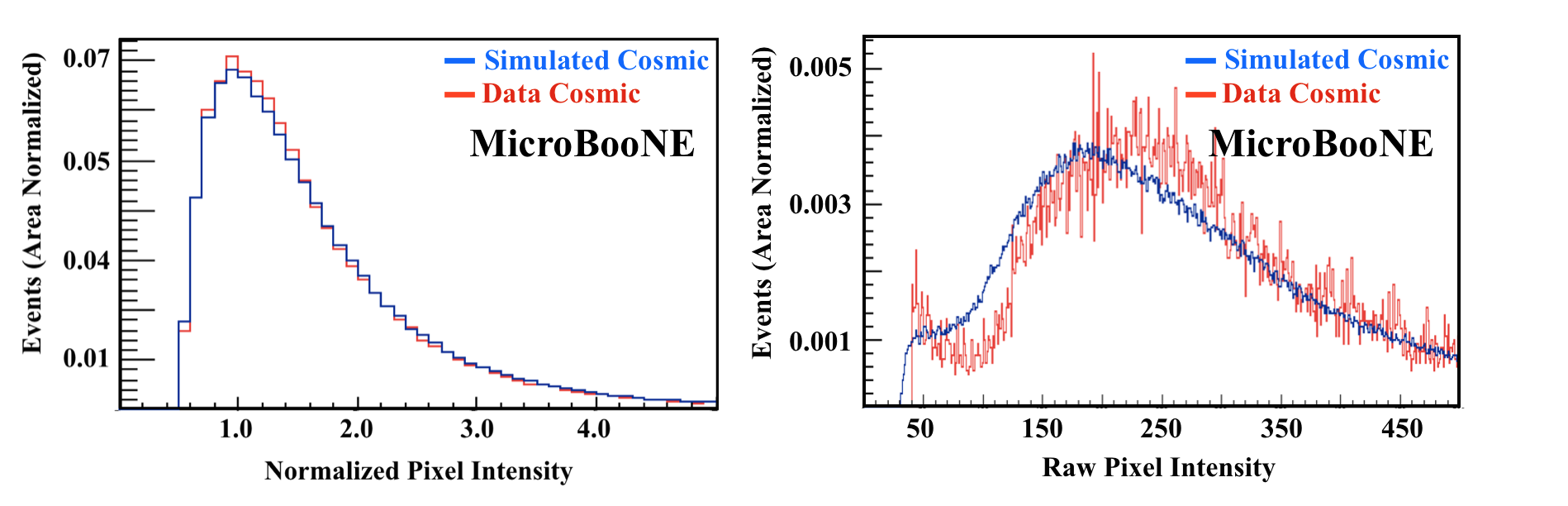}
  \vspace{-0.4in}
\caption{Left: area-normalized PI distributions for both cosmic-ray data (red) and simulated (blue) images.  The peak is mostly from the wire responses from muons, which are minimum ionizing particles (MIPs) in the detector.  Both PI values have scale factors applied to both the data and simulation such that the MIP peaks are normalized to one. A threshold is applied at 0.5 on the scaled PI value.  Right: the unscaled, area-normalized PI distributions.}
  \label{fig:CosmicADCScaling}
\end{figure}

\begin{figure}[t]
  \centering  
  \includegraphics[width=0.95\textwidth]{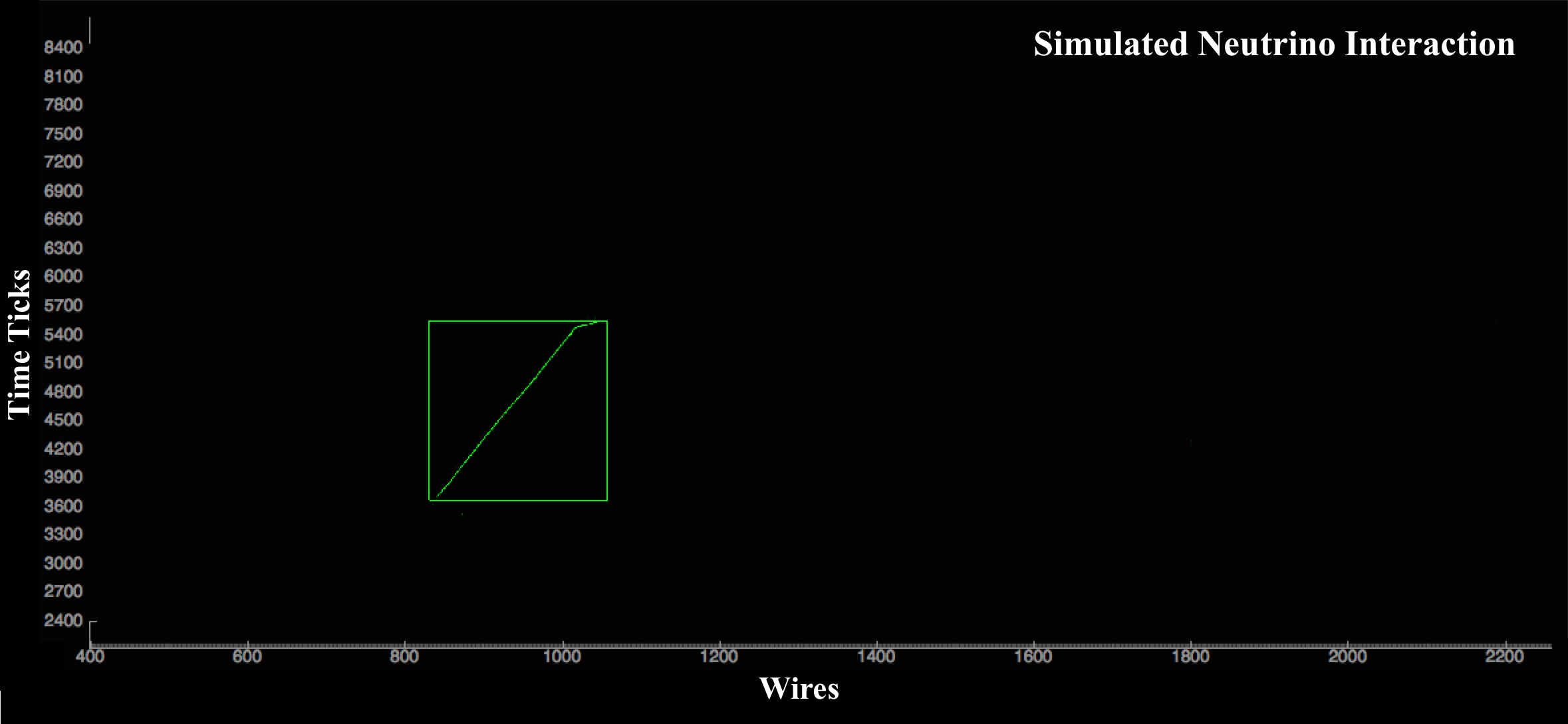} \\
  \vspace{0.2in}
  \includegraphics[width=0.95\textwidth]{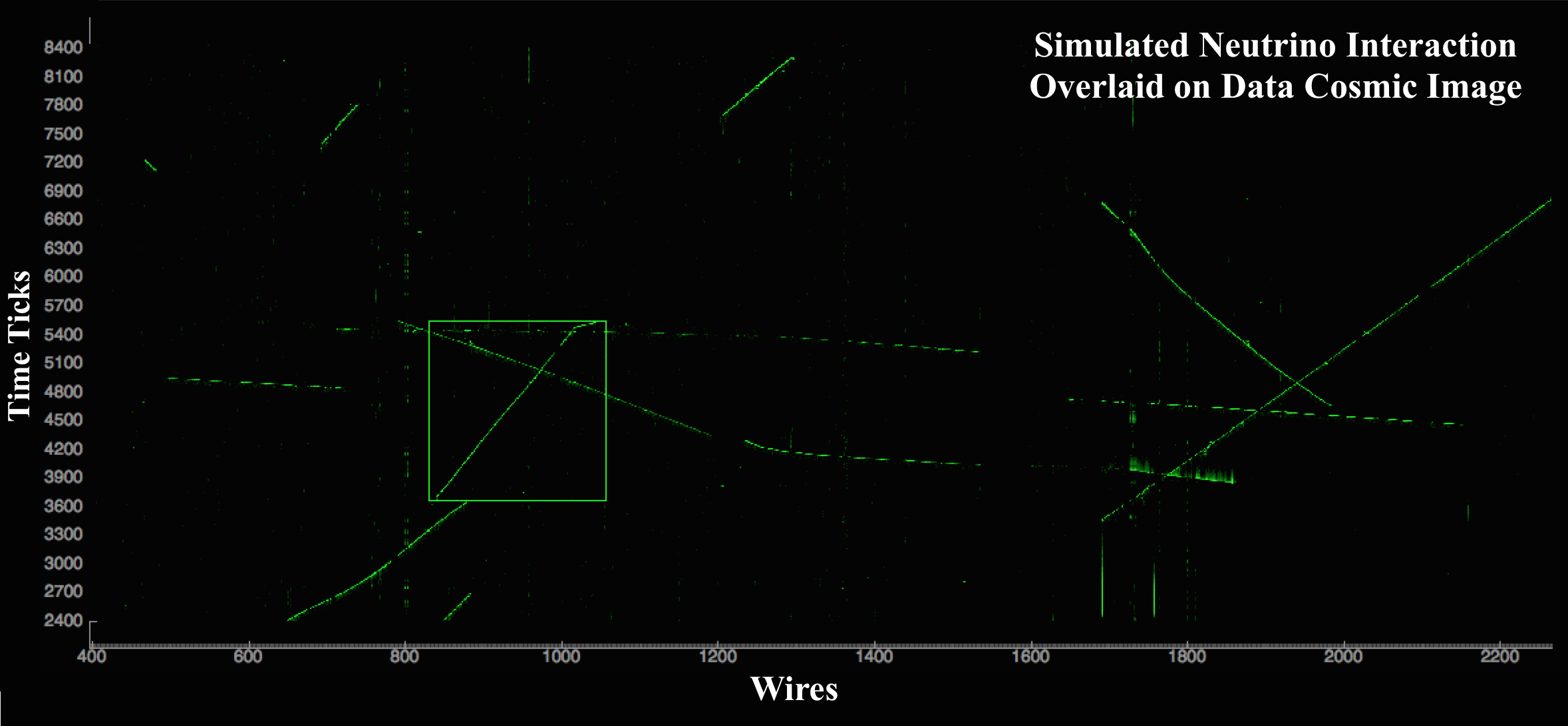}
\caption{Top: example event display of a neutrino interaction in the $Y$ plane view.  Square boxes indicate ROIs per plane created based on the charge deposition profile using simulation information. Bottom: the same event overlaid with a data cosmic image. The bounding box is used to train the detection network layer.}
  \label{fig:OverlayExample}
\end{figure}

In order to minimize the network's use of simulation-specific features, we calibrate the signal response between data and simulation. We have run a simple analysis algorithm to find the pixel intensity peak due to minimum ionizing particle (MIP) tracks in both data and simulated cosmic events.  The distribution of PI values is shown in figure~\ref{fig:CosmicADCScaling} (right) for images of data and simulated cosmic background events. We apply a scaling factor such that the peak amplitude becomes 1.0, and a threshold of 0.5 is used to reduce unmatched low PI noise components. The resulting PI distributions are shown in the left of the same figure. The $Y$ (collection) plane's scaling factor is 0.00475 for simulation and 0.005 for data. These scaling factors are applied to the simulated neutrino and the cosmic background data events, respectively, prior to an image overlay. An example event display image of a raw simulated neutrino as well as an overlaid image are shown in figure~\ref{fig:OverlayExample}.

We prepared an approximately 1:1 mixture of cosmic-only images and simulated-neutrino-overlaid images for both training (totaling 101,191 images) and validation (totaling 32,220 images).

\subsection{Training}
\begin{figure}[t]
  \centering  
 \includegraphics[width=0.495\textwidth]{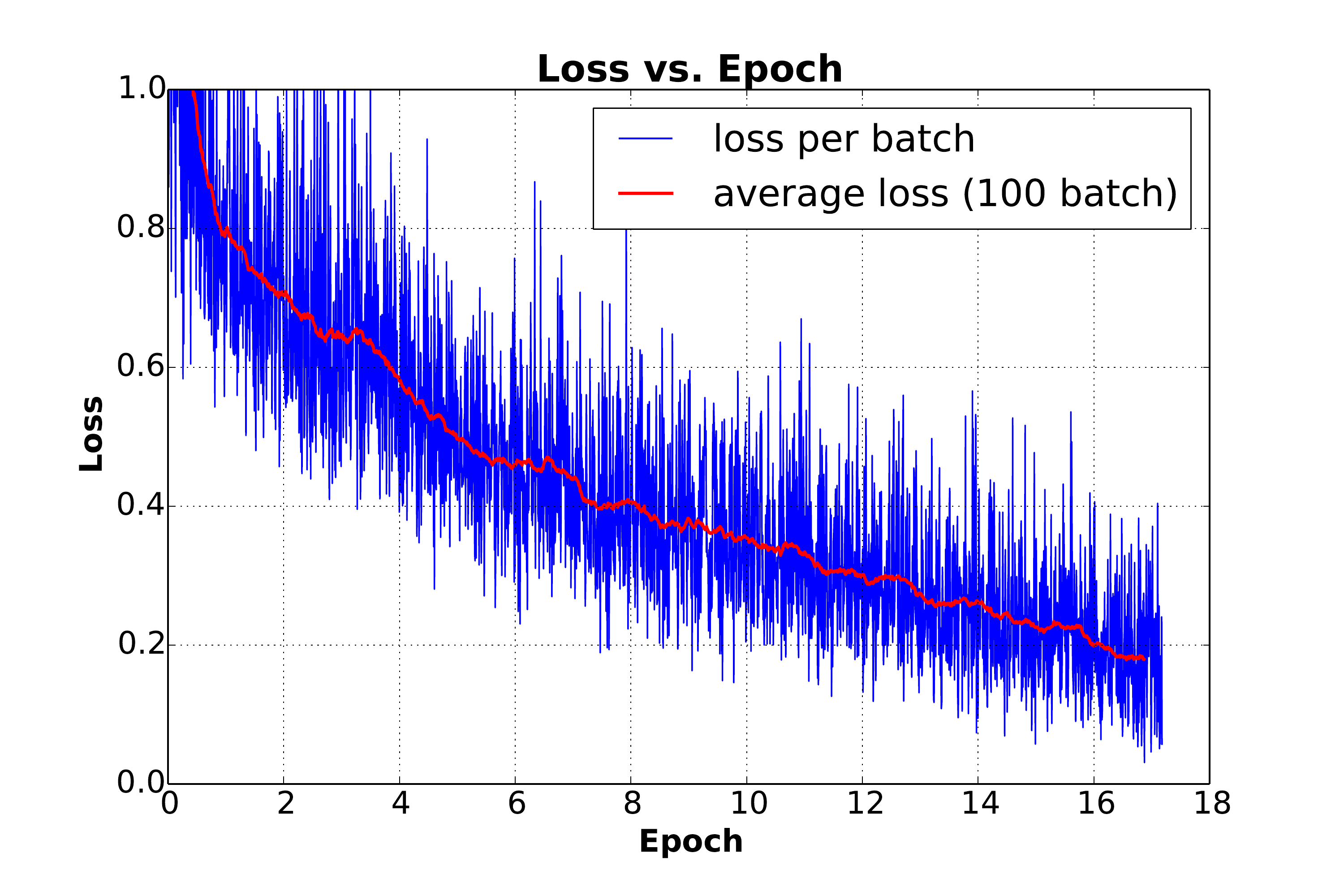}
 \includegraphics[width=0.495\textwidth]{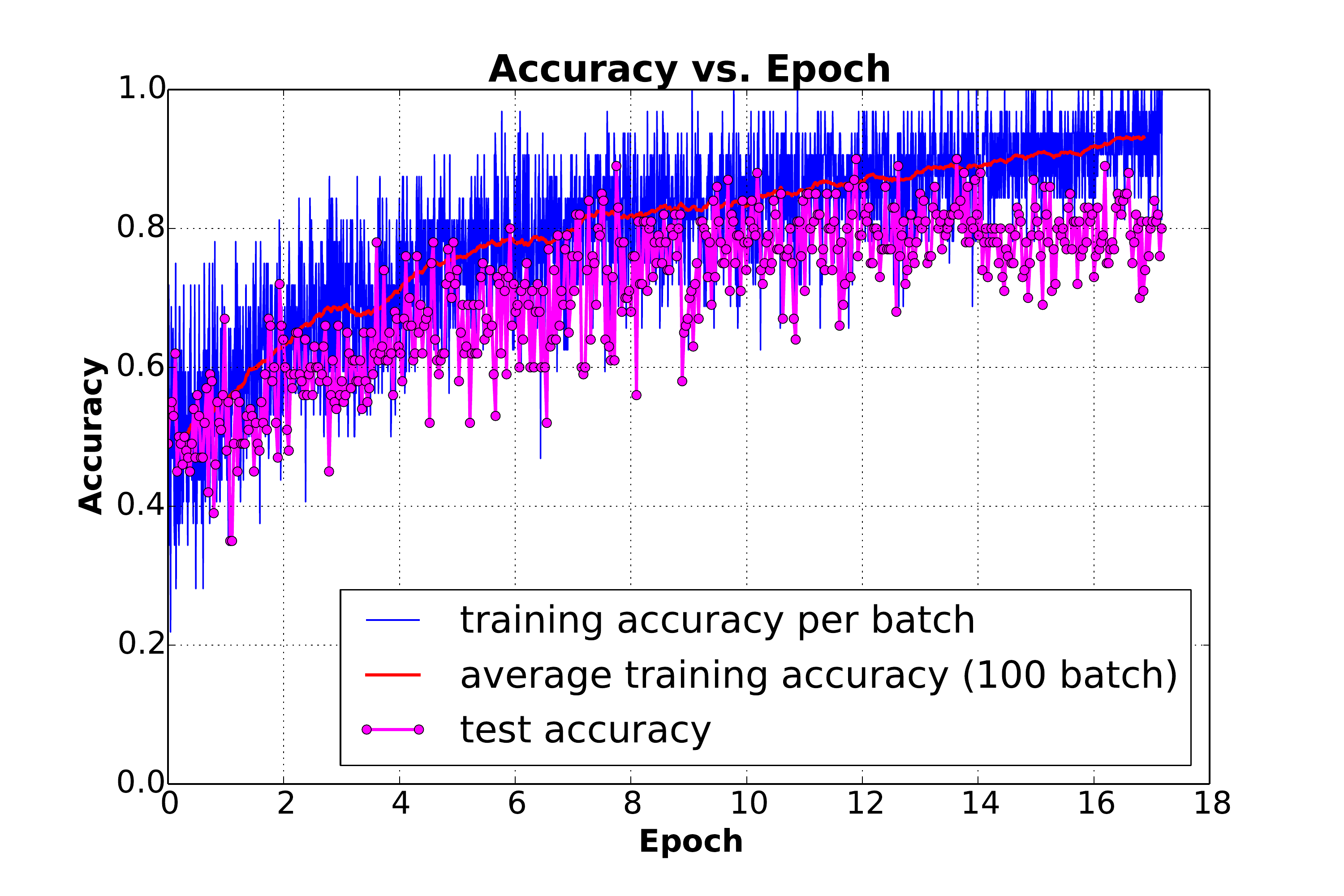}
 \vspace{-0.4in}
\caption{Left: 1-plane classification loss curve for our simplified version of Inception-ResNet-v2. Right: accuracy curve monitored during training. In both plots the blue data points are computed from the training process while red data points are our interpretation by taking an average of 100 consecutive blue data points. The magenta curve on the right plot is an accuracy computed using a test sample.}
  \label{fig:inception_resnet_accuracy}
\end{figure}

We next used our reduced Inception-ResNet-v2 for classification training of two samples: neutrino vs. cosmic events. At the data preparation stage during the training, we perform a random cropping of an image to a slightly smaller size along the time (vertical) axis in order to avoid over-training.  This is an example of data augmentation, which is one of the standard techniques employed in the field.  We found that if we did not present a randomly modified version of an image each time it is given to the network, the network would begin to over-train after a few epochs. Figure~\ref{fig:inception_resnet_accuracy} shows the classification accuracy reached the level of 80\%. The slightly lower accuracy of a test sample relative to the training sample points to a slight, but acceptable, over-training. Slight over-training is common practice as it is a sign that the number of parameters in a model, which typically scales with the model's capacity to learn, is just a bit larger than needed and, therefore, considered near optimal.

For detection training, the base network used in the Faster-RCNN architecture is the AlexNet model, where the region-finding and classification layers of Faster-RCNN are appended to the final fifth convolutional layer of AlexNet. This is similar to what was done for the single particle detection network. In this instance, we modify the allowed output classes to two, one for neutrino events and the other for an all-inclusive ``background'' class which includes cosmic rays. We train the AlexNet convolution layers from scratch, contrary to the single particle case, by re-initializing their weights and biases. We also re-initialize the two large fully connected layers at the end of the model, which sit downstream of the region-finding portion of the network, with random samples from a Gaussian distribution centered at zero with 0.001 standard deviation. We have empirically found that re-initializing the last fully connected layers with Gaussian weights, rather than copying those from the classification stage, especially in the case of neutrino detection, helps ensure that the detection-specific layers learn both bounding box regression and classification. 

Finally, when we train the detection model, we only provide the network with cosmic+neutrino images and ignore the cosmic-ray-only images. Each neutrino image contains a single truth bounding box around the neutrino interaction vertex, which encapsulates the bounding boxes associated with each daughter particle of the interaction. The truth information from the simulation is used to determine the bounding box.  We use the same strategy employed in the single particle demonstrations to find the bounding boxes for the daughter particles, i.e. we find the truth locations of energy deposited in the detector. We train with the standard batch size of 1 image~\cite{fasterrcnn}.

\subsection{Performance}
\begin{figure}[t]
  \centering  
  \includegraphics[width=0.495\textwidth]{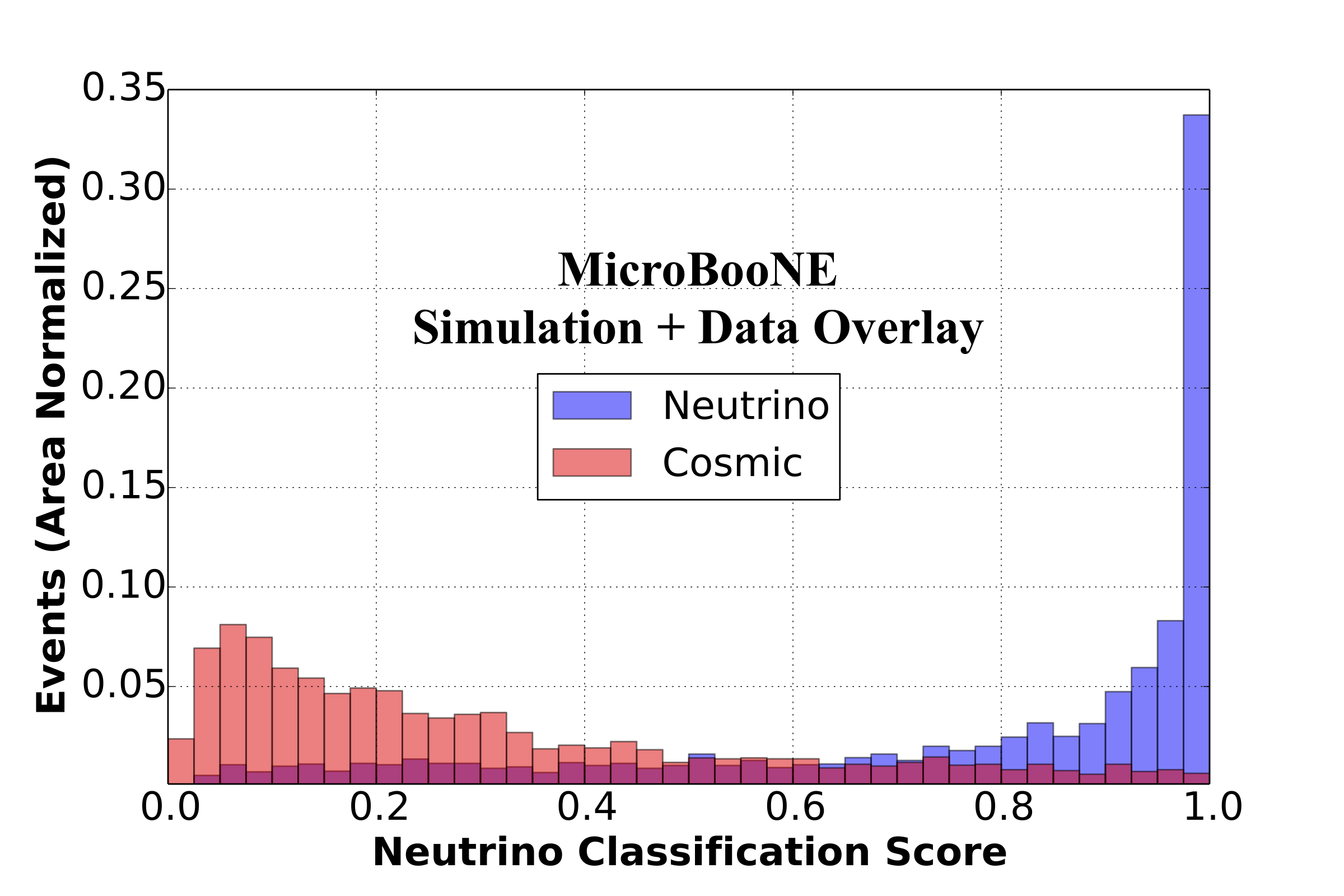}
  \includegraphics[width=0.495\textwidth]{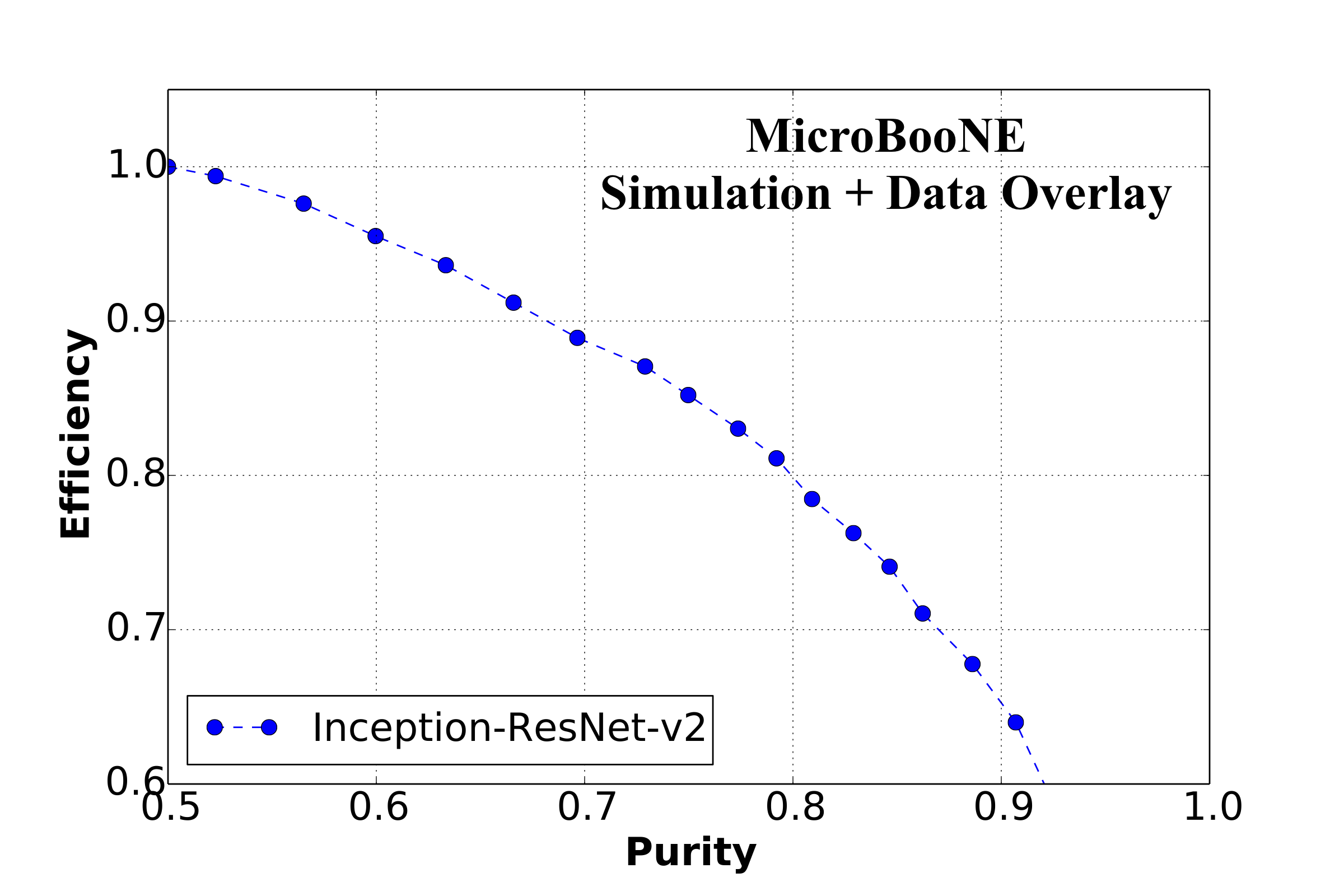}
  \vspace{-.4in}
\caption{Left: the neutrino score (horizontal axis) for the validation sample. The blue distribution is for neutrino events while red is for cosmic background events. Right: efficiency vs. purity curve for an equal number of cosmic background and neutrino signal events.}
  \label{fig:inception_resnet_performance}
\end{figure}

Figure~\ref{fig:inception_resnet_performance} shows the neutrino classification score distribution from the validation sample. The distribution matches our intuition, having a sharp peak where there are neutrino events. Cosmic background events are not expected to have a sharp peak since there is no specific feature to hone in within those images. The right plot in the same figure shows the overlaid neutrino event selection efficiency and purity for the case of having an equal number of neutrino events and cosmic-only events. A particular point on the curve achieved 87.1$\pm$0.5\% efficiency with 72.9\% purity for scores above 0.35. In MicroBooNE, however, before any selection, the expected cosmic to cosmic+neutrino event ratio is approximately 600 to 1. After applying a trigger that looks for scintillation light coincident with the expected beam window, this ratio is around 30 to 1. Given that this is single-plane performance using real data cosmic background events, we expect that the performance will improve once all three views are used, which is studied demonstration 3. 

\begin{figure}[H]
 \centering  
 \includegraphics[width=0.8\textwidth]{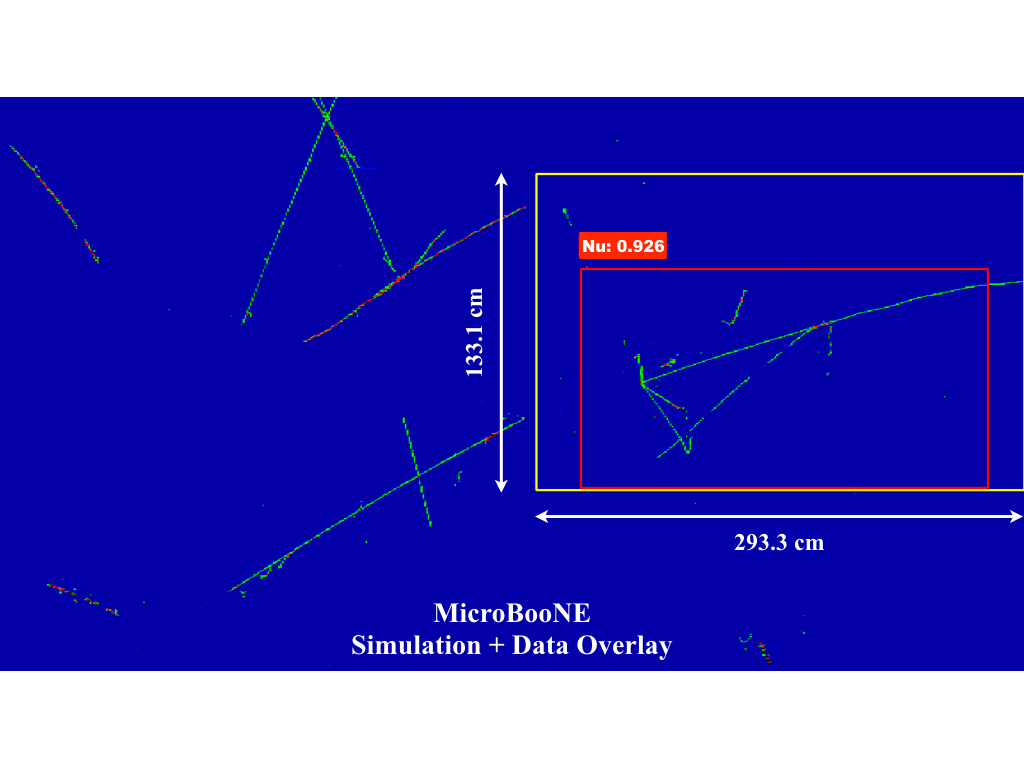}\\
 \vspace{0.5in}
 \includegraphics[width=0.8\textwidth]{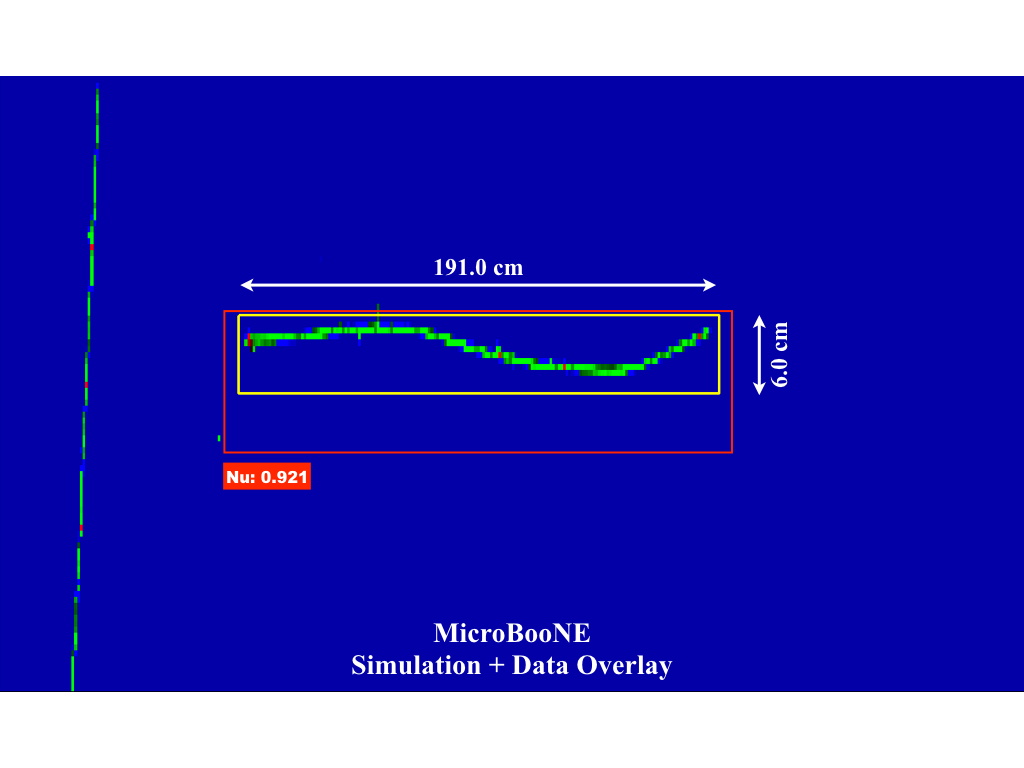}
\caption{Detected neutrino bounding box within an event image. Top: A $\sim 2$ GeV $\nu_\mu$ charged current interaction with a single $\mu$, three protons, and two charged $\pi$'s produced. Bottom: $\nu_\mu$ charged current interaction with a single $\mu$ and proton produced. The red predicted box extends in the correct dimension to encapsulate the full interaction. The yellow box shows the truth bounding box.}
  \label{fig:nudetect1}
\end{figure}

\begin{figure}[H]
 \centering  
 \includegraphics[width=0.8\textwidth]{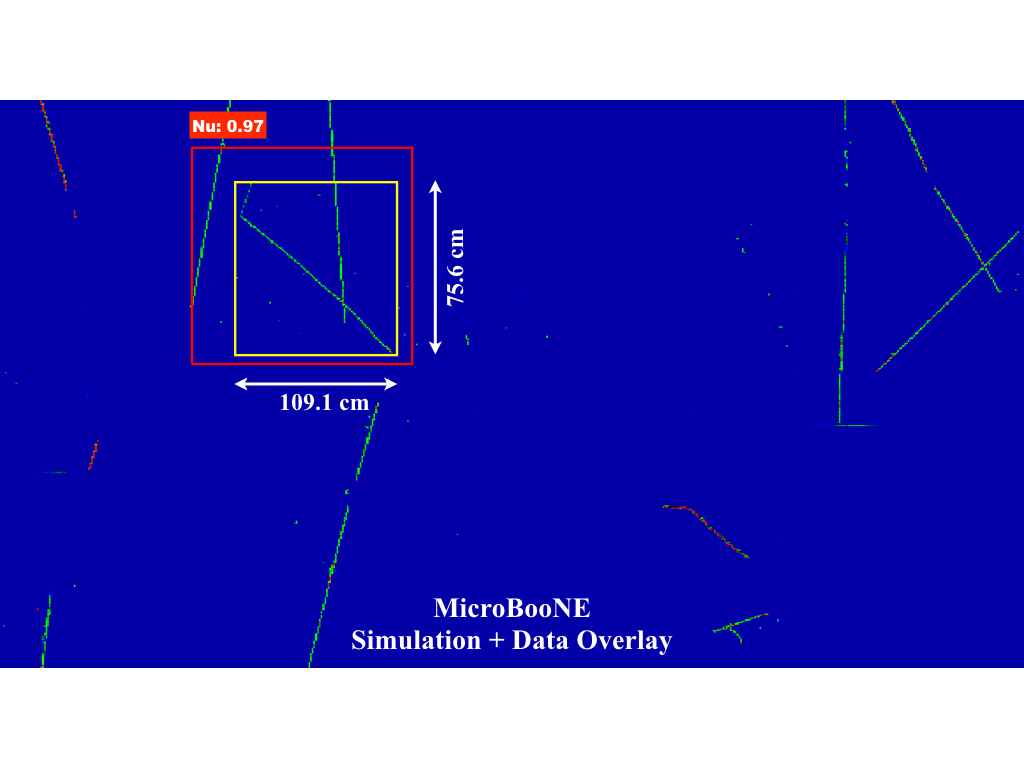}\\
 \vspace{0.5in}
 \includegraphics[width=0.8\textwidth]{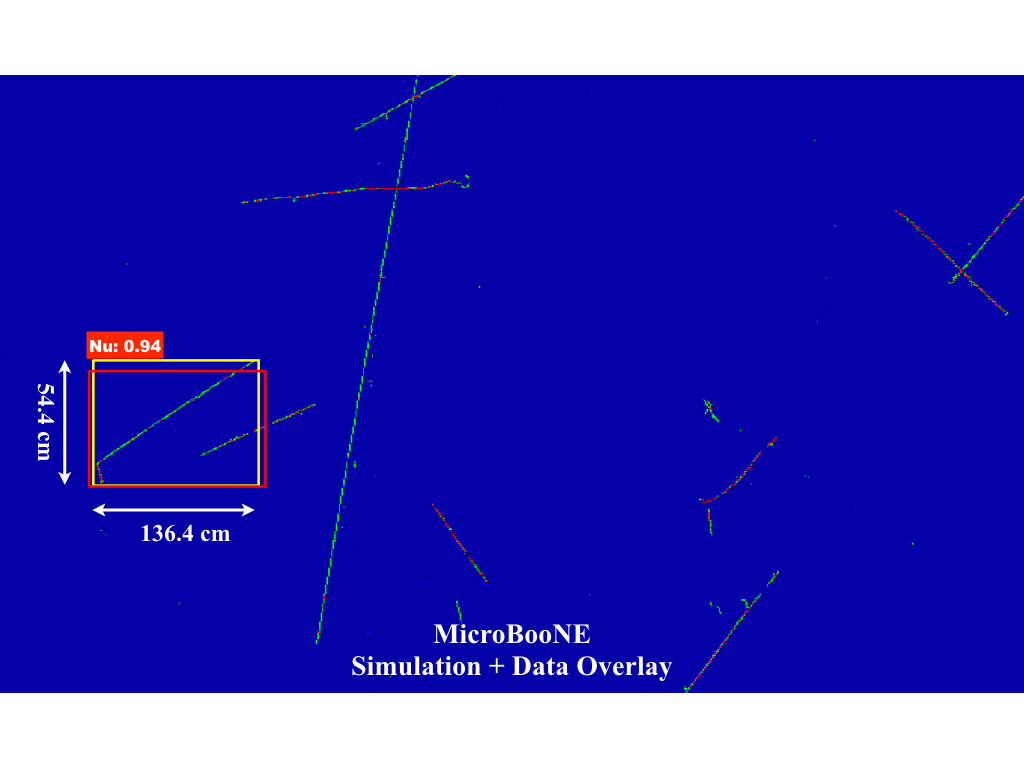}
\caption{Detected neutrino bounding box within an event image. Top: $\mu$ and charged $\pi$ produced from a $\sim 1$ GeV neutrino. The detection box (in red) appears to capture a neighboring cosmic ray, but maintains the overall shape of the ground truth box (in yellow). Bottom: $\nu_\mu$ charged current interaction event with a $\mu$ and proton produced.}
  \label{fig:nudetect2}
\end{figure}

\begin{figure}[H]
 \centering  
 \includegraphics[width=0.8\textwidth]{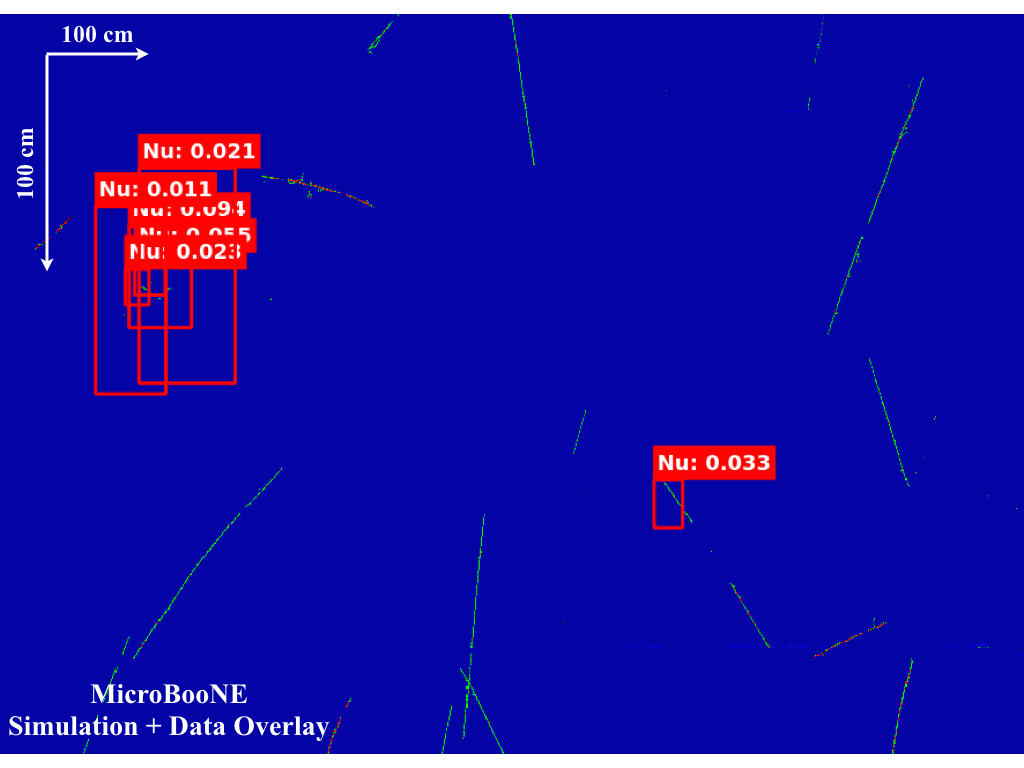}\\
 \vspace{0.5in}
  \includegraphics[width=0.8\textwidth]{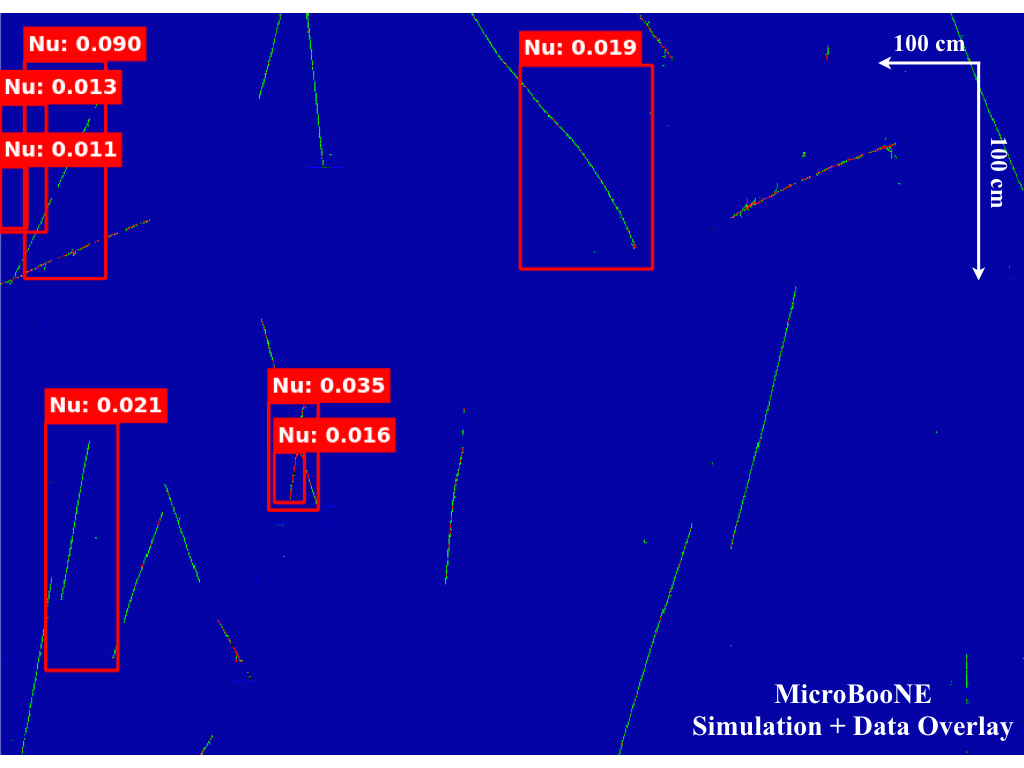}\\
\caption{Example of cosmic background events (both top and bottom) with detected neutrino bounding boxes with scores less than 0.1. One can see many low score candidate bounding boxes were made.}
  \label{fig:nudetect4}
\end{figure}

\begin{figure}[H]
 \centering  
 \includegraphics[width=0.8\textwidth]{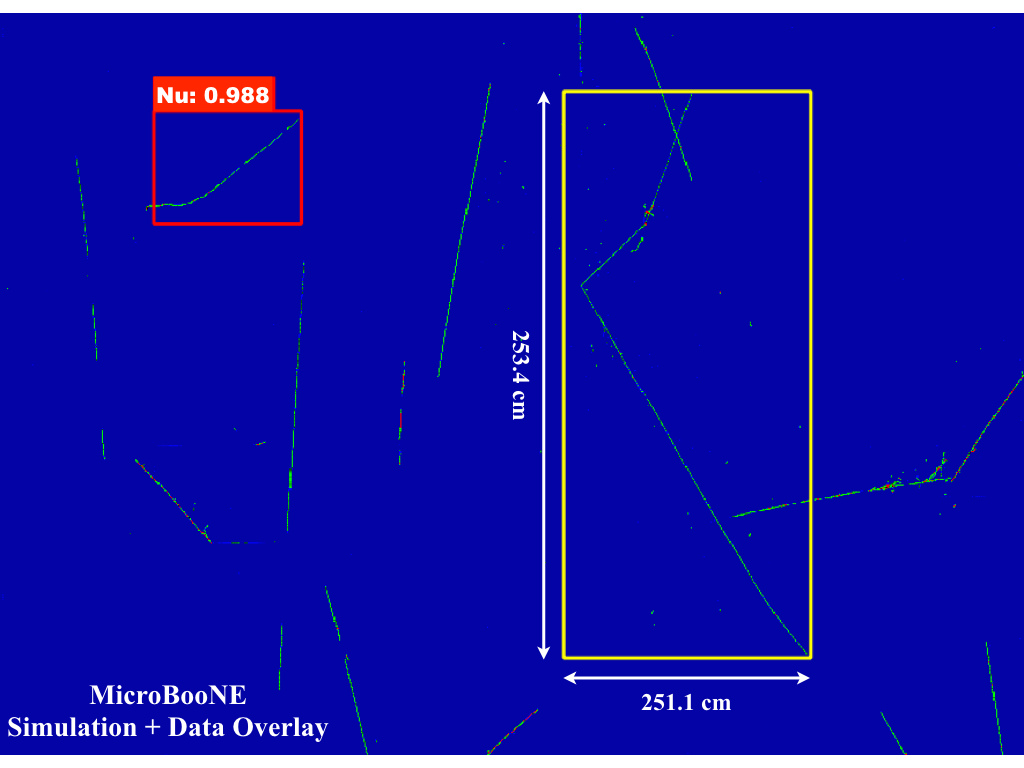}\\
 \vspace{0.5in}
 \includegraphics[width=0.8\textwidth]{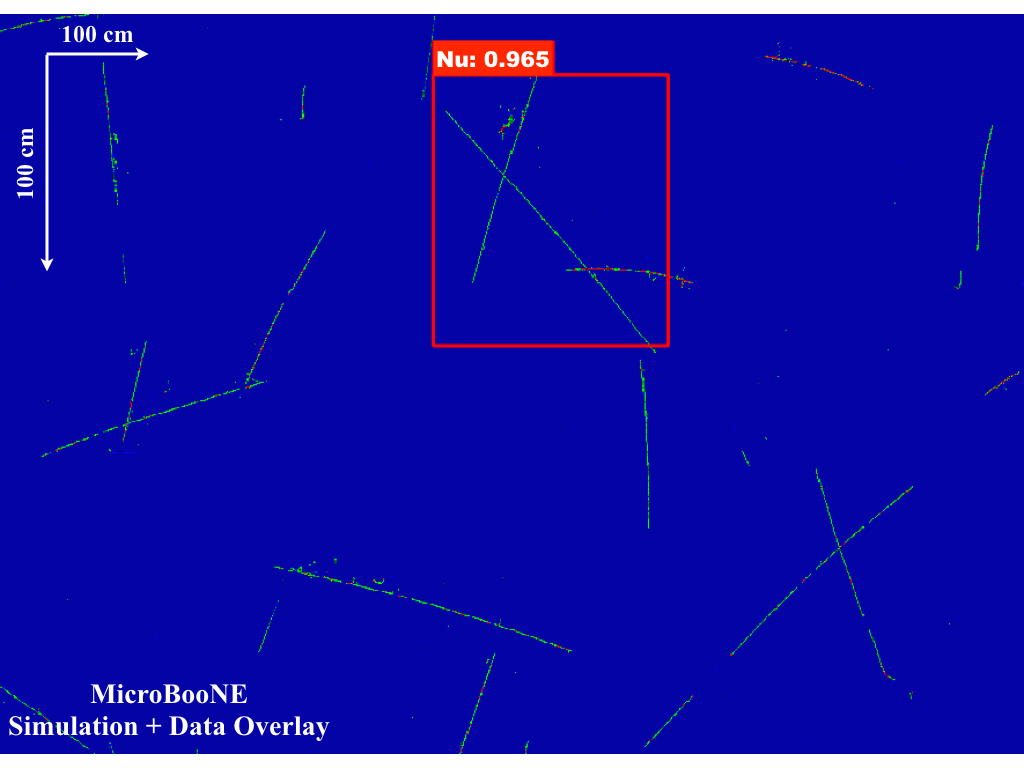}
\caption{Top: an example of a true neutrino (yellow box) event where the network found the highest score bounding box in the wrong location (red box). Bottom: an example of a cosmic-only event where the network found a bounding box (red) with a high neutrino score. There is no neutrino interaction in this event.}
  \label{fig:nudetect5}
\end{figure}

Example images showing regions that the network has identified as being neutrino-like are shown in figures~\ref{fig:nudetect1}-\ref{fig:nudetect5}.  The different figures show examples of where the network correctly identifies neutrinos with strong confidence (figures~\ref{fig:nudetect1} and figures~\ref{fig:nudetect2}), where the network identifies cosmic-ray tracks as neutrinos but with low confidence (figure~\ref{fig:nudetect4}), and where the network labels cosmic-ray tracks as neutrinos with high confidence (figure~\ref{fig:nudetect5}).  Each image in those figures is a high-resolution sample of the event readout. The images are single channel and are represented with false color scales, with blue representing lower PI values, green representing roughly the PI values produced by minimum ionizing particles, and red representing larger charge depositions. The yellow box in the figures are the ground truth bounding boxes around the neutrino determined from Monte Carlo information. The red box is the network prediction for the region containing the neutrino and the network score is labeled in white text above. Figures~\ref{fig:nudetect1} and \ref{fig:nudetect2} show example images where the CNN successfully located a neutrino with high score ($>$0.9). Figure~\ref{fig:nudetect5} shows example images with two types of mistakes: 1) finding a high score ($>$0.9) bounding box in a wrong location in a neutrino event, and 2) also in a cosmic background event where there is no neutrino. In either case, the bounding box contains an interaction topology that could be mistaken as a neutrino event. Thus, these are not boxes drawn randomly. Finally, figure~\ref{fig:nudetect4} shows examples of how the CNN finds many boxes with low neutrino score in a cosmic background event. Boxes shown in this figure are those with neutrino scores less than 0.1.

\begin{figure}[t]
  \centering  
  \includegraphics[width=0.8\textwidth]{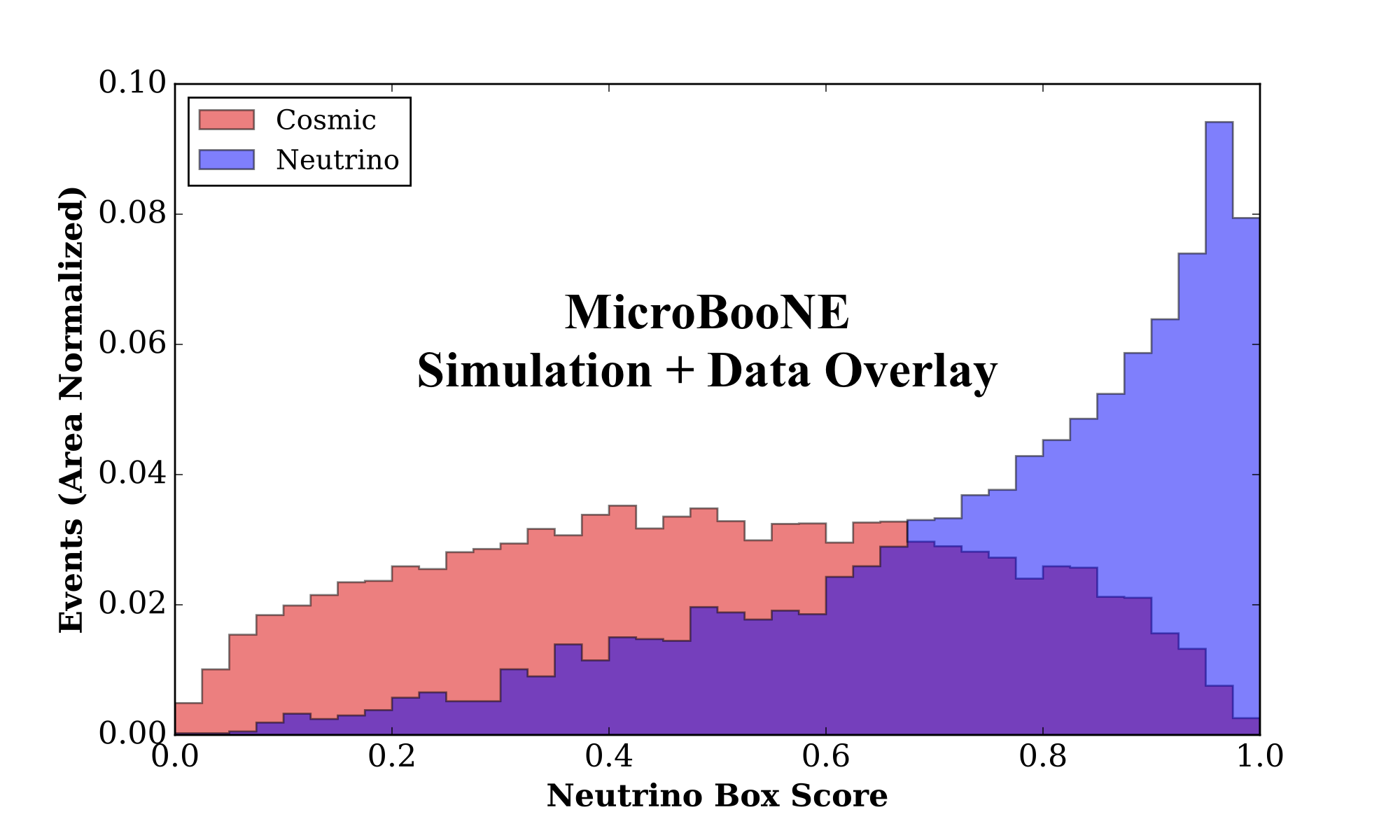}
\caption{Neutrino detection box score distribution for neutrinos (blue) and cosmics (red), normalized by area. The peaked neutrino distribution shows that the detection network has some neutrino versus cosmic event discrimination power in addition to its ability to locate neutrino interaction-like regions.}
  \label{fig:nu_detscore}
\end{figure}

Figure~\ref{fig:nu_detscore} shows the distribution of neutrino bounding box scores predicted by Faster-RCNN per event for both neutrino+cosmic and cosmic-only images. We can see that the network is successfully finding a more neutrino-like bounding box in neutrino events than cosmic background events. Moreover, because the network is not trained to specifically discriminate cosmic events, it finds a bounding box with a moderate score value among cosmics. This is a good sign, as it indicates that the network is not simply exploiting a minor difference of data and simulation.

\newpage

\newpage

\newpage

\newpage

\newpage

\subsection{Summary of Demonstration 2}

In this study, we demonstrated how a CNN can be used to perform event classification using only one plane view from MicroBooNE simulated neutrino events overlaid on cosmic ray data events.
In particular, using a simplified Inception-ResNet-v2 architecture and starting with an equal number of signal and background events, we quoted a selection efficiency of $87.1$\%$\pm$0.5\% with 72.9\% purity for neutrino events by cutting on a neutrino score of 0.35 (figure~\ref{fig:inception_resnet_performance}). We also demonstrated that CNNs, in particular using the AlexNet+Faster-RCNN model in this study, can successfully learn neutrino interaction features and localize them in an event view. These are important steps for an analysis chain using CNN techniques. Neutrino event classification and detection networks could be at the beginning of an analysis chain, providing a method for event selection and locating interactions in a large LArTPC detector, which can be cropped at a higher resolution and passed to downstream algorithms.

\clearpage
%
%
\section{Demonstration 3: Neutrino Event Identification with 3 Planes and Optical Detector}
In this section, we report on the study of neutrino event classification using the full MicroBooNE detector information including TPC three-plane data in combination with optical detector data. We employ a novel network architecture based on ResNet~\cite{ResNet} with a data augmentation technique for combining TPC and optical detector data in an event image. We first describe sample preparation and network training methods. The general strategy of using a simulated neutrino image overlaid on a cosmic background image taken from data stays the same as in the previous demonstration.

\subsection{Network Design}
\label{sec:3plane_nuid_design}

The network we designed for the three-plane neutrino classification task is based on the ResNet network~\cite{ResNet}.  This network features the repeated use of what are referred to as residual convolutional modules.  These modules have been demonstrated to help networks train more quickly~\cite{Inceptionv4}. As described in more detail below, the input images are chosen to be 768$\times$768 pixels with 12 channels as a third dimension. This is a relatively large amount of input data for a network compared to existing models. This compelled us to employ a network model based on a truncated ResNet network. This constraint arises because of the memory limitation of the GPUs, which is 12 GB.  The more slowly one reduces the size of the output feature maps at each layer, the more memory that is required at each given layer. This means that there can be fewer layers and that each layer can learn fewer filters.  More layers and filters mean that the network can learn more and, in principle, attain a higher performance.  However, by preserving the resolution of the feature maps, the network is exposed to more detailed features in the image. Fully exploring the space of the network configurations is reserved for future studies.  

\begin{figure}[t]
  \centering
  \includegraphics[width=1.0\linewidth]
  {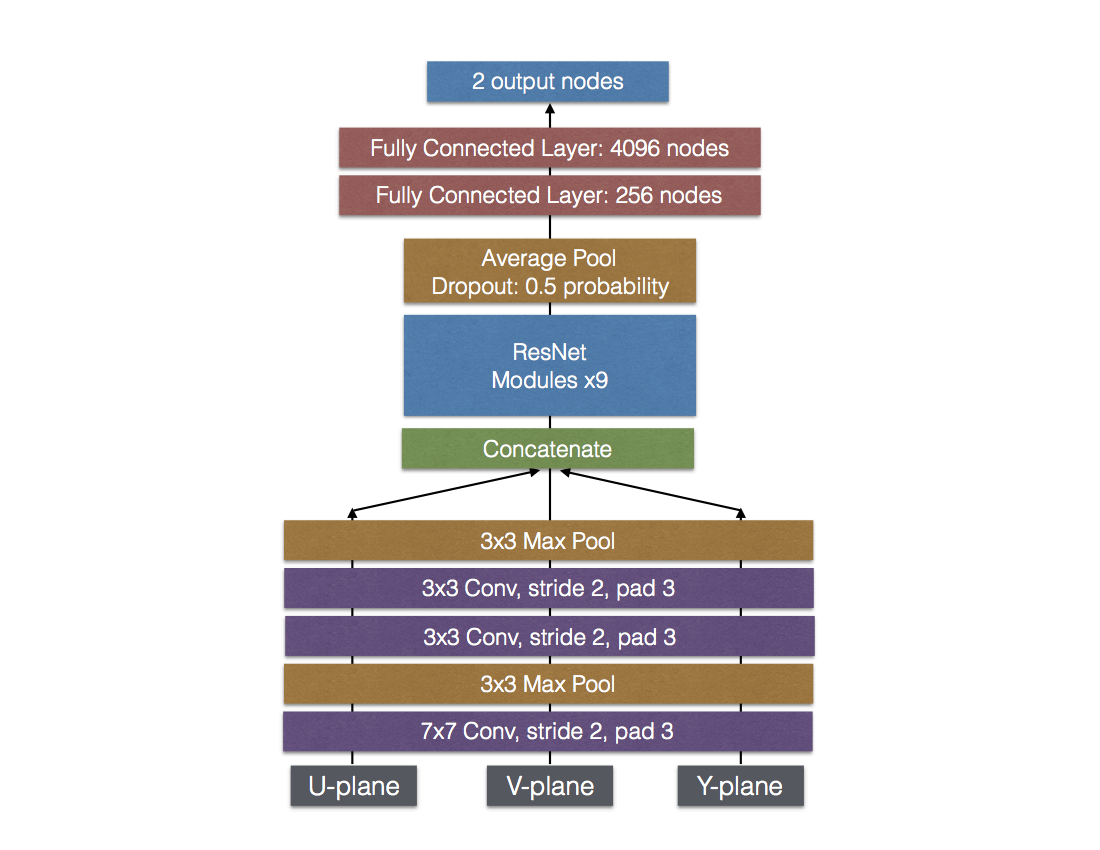}
  \vspace{-0.3in}
  \caption{Neutrino ID Network. The data input layer is at the bottom and takes three wire plane views separately. The process flows from the bottom to the top where the final decision of a neutrino or cosmic event is made.}
  \label{fig:ubresnet}
\end{figure}

The full neutrino ID network shown in figure~\ref{fig:ubresnet} uses all three TPC planes. The three planes are passed separately through the {\it stem} portion of the network which consists of three convolutional layers, along with a couple of pooling layers to reduce the size of the output feature map.  All convolutional layers (in both the stem and elsewhere) use a technique known as {\it batch normalization}~\cite{BatchNorm} and are followed by a ReLU.  Note that the convolutional layers in the stem are the same for all three views. In other words, the parameters of those layers are shared.   The output feature maps of the stem are then concatenated and passed through nine residual modules. This is followed by an average pooling layer that incorporates a technique known as {\it dropout}~\cite{dropout}. After the dropout layer, two fully-connected layers are used to classify the output features and determine if the event has a neutrino or not. For more details on the network and different component layers included please see appendix~\ref{app:3planenuid}.

\subsection{Sample Preparation and Training}

The network is asked to classify images into two classes: a cosmic-only image and a cosmic+neutrino interaction image.  For the cosmic-only images, we use off-beam, PMT-triggered events. When producing cosmic+neutrino interaction images, we overlay simulated neutrino interactions with off-beam events.  Just as in demonstration 2, the neutrino interaction is generated using the BNB flux~\cite{BNB} and {\sc{genie}} interaction model.  The neutrino interaction time is made to fall uniformly within the expected beam time window.  



When selecting simulated BNB neutrino interactions to overlay, we apply quality cuts to (1) select events with an interaction vertex inside inside the TPC, (2) select events with neutrino energy above 400 MeV, and (3) select $\nu_\mu$ charged-current  (CC) events using generator-level information. The second and third cuts are meant to ensure that the amount of charge deposited by the neutrino interaction is sizable.

As discussed in the previous section, we are forced to down-sample the MicroBooNE event images to a lower resolution in order to fit our network model within the limits of the GPU memory.  Unfortunately, as shown within demonstration 1, down-sizing has a negative effect on the performance. However, for this first study, we choose an image size that balances resolution by still being fairly large (768$\times$768 pixels) while still allowing us to construct a fairly deep network capable of viewing the images from all three planes and fitting onto a single GPU.

In addition to the three images from the TPC planes, we also provide the network with additional images that provide information from the PMTs along with the minimum ionizing particle (MIP) and heavily ionizing particle (HIP) charge scales.  This technique is inspired by the algorithm, AlphaGo, which was used to play the board game Go~\cite{alphago}.  In this example, the CNN responsible for estimating the best move was provided an image of the board along with supporting information such as the number of pieces surrounding a given location on the board.  Analogously, we provide three additional supporting images for each wire plane's image.  

The first is an image marking pixels with PI values consistent with a MIP.  The second is a similar image, marked with PI values consistent with a HIP.  The idea here is to help the network determine the proximity of a muon (a MIP) and a proton (a HIP), which usually occur together in a neutrino interaction. In principle this is information that the network can learn by itself. However, since we have reduced the image-size early in the network, the calorimetric information of the image, with which the network would have learned to identify the vertex, is somewhat lost. We therefore provide these binary (MIP/HIP) images to solve this issue.  The third is an image weighted by the location and amplitude of PMT pulses that occur in time with the beam window.  The purpose of this PMT image is to help the network only look at regions of the detector that are consistent with the portion of the detector where scintillation light, coincident with the arrival of the neutrino bean, is detected by the PMTs.  This is to help narrow down the region in the detector to which the network should pay attention. 

\begin{figure}[t]
  \centering
  \includegraphics[width=1.0\linewidth]{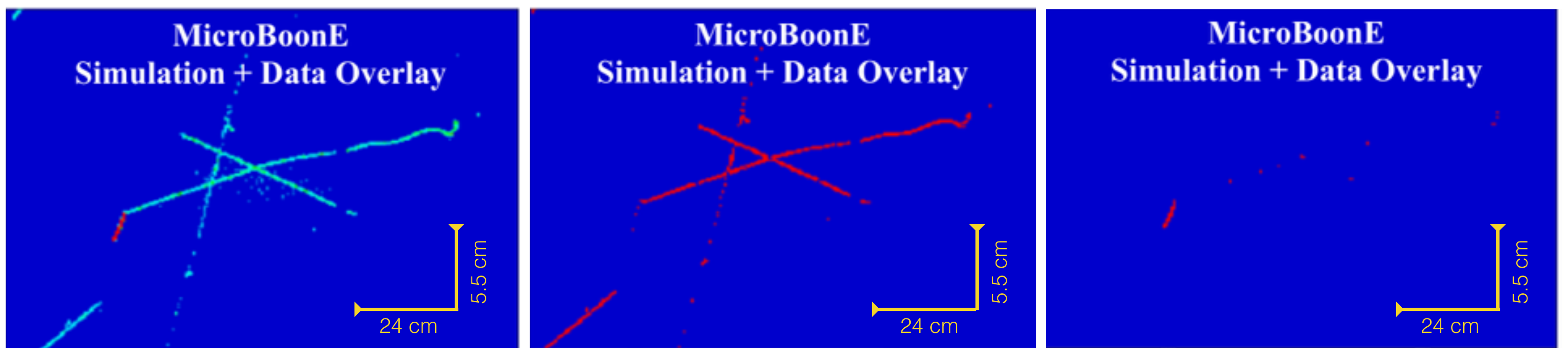}
  \caption{Comparison of a TPC image (left), MIP-created image (center), and HIP-created image (right). This example is for the collection plane.  The MIP and HIP images are made by assigning pixel values of 1.0 to any pixel that has an PI value consistent with a minimum ionizing particle and heavily ionizing particle, respectively. 
For the neutrino interaction, one can see that the proton track has more HIP pixels, while the muon track consists mostly of MIP pixels. 
  A constant drift velocity of 0.11 cm/$\mu$s was assumed to determine the Y-scale.  The constant 0.3 cm wire pitch was used to determine the X-scale.}
  \label{fig:hipmip}
\end{figure}

The images providing the supporting MIP and HIP information are made by assigning a value of 1.0 to each pixel whose PI value falls within a certain range.  Figure~\ref{fig:hipmip} shows an example of both type of images.  For the MIP image, the PI value must fall between 10 and 45. For the HIP image, the PI value must be greater than 45.

\begin{figure}[tb]
  \centering
  \vspace{-0.4in}
  \includegraphics[width=0.85\linewidth]{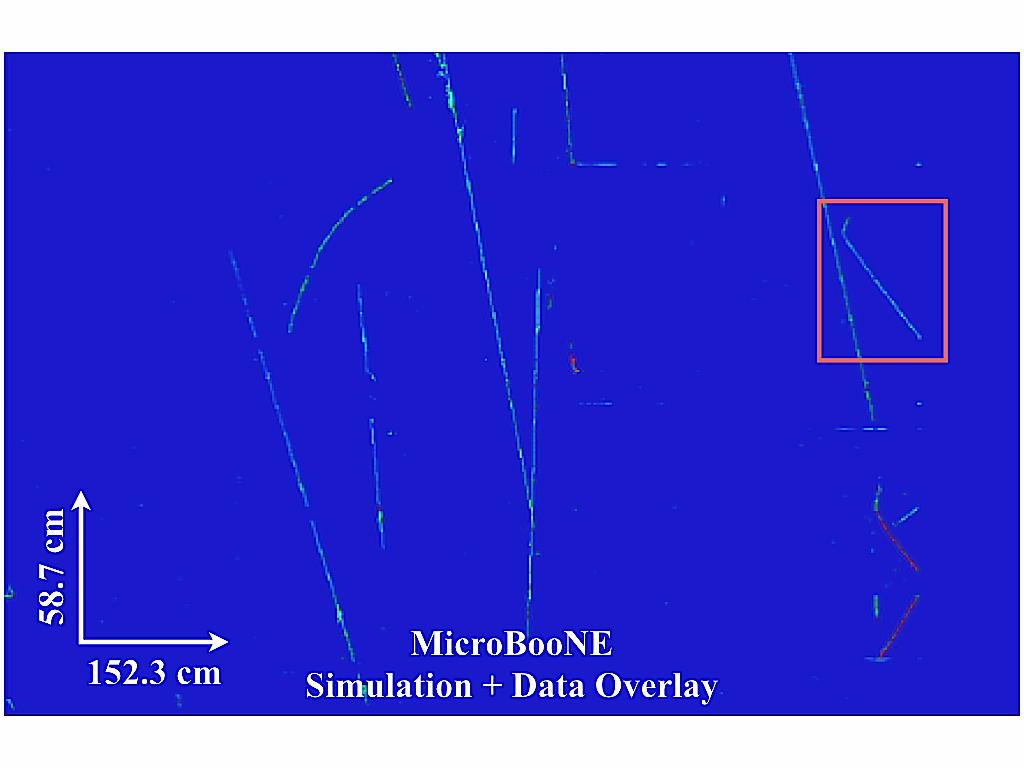}\\
  \vspace{-0.2in}
  \includegraphics[width=0.85\linewidth]{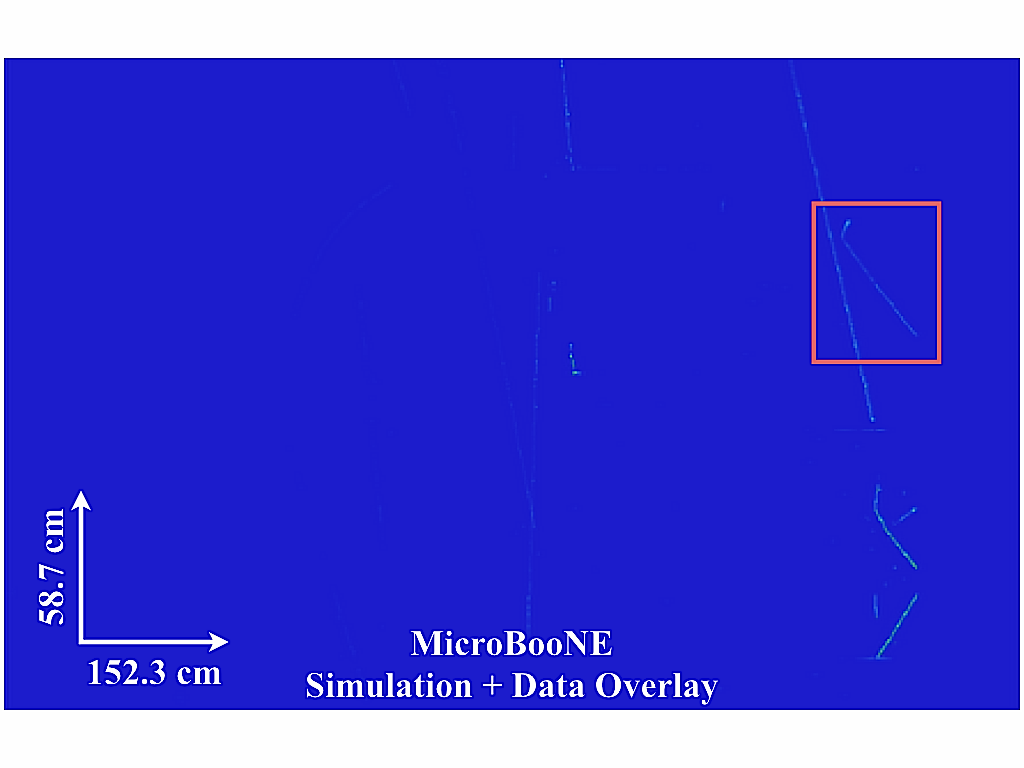}
  \vspace{-0.2in}
  \caption{Comparison of a TPC image (top) and PMT-weighted image (bottom). This example is for the collection plane.  The PMT-weighted image is made by weighting the TPC image using information about the amount and location of charge seen by the PMTs inside the beam window. As this example demonstrates, the PMT-weighted image tells the network where it should focus its attention. The box (drawn by hand) indicates the neutrino interaction.}
  \label{fig:pmtweighted}
\end{figure}

The PMT-weighted image is made by weighting the charge on each wire by (1) the amount of charge seen by each PMT and (2) the distance each PMT is to the wire in question.  Figure~\ref{fig:pmtweighted} compares an image of the TPC charge with its corresponding PMT-weighted image.  The weight for each wire, $i$, is given by
\begin{equation}
W_{i} = \sum_{p}^{N_{PMTS}} w_{ip}q_{p}.
\end{equation}

In the above, $w_{ip}$ is the PMT distance weight between wire, $i$, and PMT, $p$, and is defined by
\begin{equation}
w_{ip} = \frac{1}{D_{ip}^\alpha}/\hat{w_{i}},
\end{equation}
where $D$ is the shortest distance between wire, $i$, and PMT, $p$, and $\alpha$ is 2 for the $Y$-plane and 1 for the $U$ and $V$ planes.  The factor, $\hat{w_{i}}$ is the largest-valued weight for wire, $i$, and is used to normalize the set of of distance weights for a given wire.  For the second factor in $W_{i}$, $q_{p}$ is the PMT charge weight.  It is defined as
\begin{equation}
q_{p} = Q_{p}/\hat{Q},
\end{equation}
where $Q_{p}$ is the amount of charge in PMT, $p$, inside the beam window, and $\hat{Q}$ is the maximum $Q_{p}$ used to normalize the set of weights. The PMT-weighted image acts like what is known in the deep learning field as a {\it soft-attention model}, helping to indicate on which parts of the image to focus.

In total, we prepare a training data set with 35,146  images and a validation data set with 14,854 images. For each image set, half were cosmic-event-only from real data, and the other half were simulated neutrino overlaid on a separate set of cosmic-only images.

The network is trained using the optimization algorithm RMSprop~\cite{rmsprop} with an initial learning rate of $1.0\times10^{-4}$.  A batch of size seven is used, which is the maximum number that could fit on a single GPU.  The training is run for about 75 epochs. Note that each time the network sees a particular image, it is modified slightly. This technique is known as data augmentation and is standard practice for preventing the network from over-training. The images are originally 768$\times$768 pixels.  During image preparation, all images are given 10 pixels worth of padding to both ends of the image in the time dimension, making the output image 768$\times$788 pixels.  The values for these padding pixels are set to zero in all channels.  During training, the images are randomly cropped back down to 768$\times$768 pixels before being passed to the network.  This shifts the image in time, while preserving the wire-dimension.  We found that without doing this random cropping, the network over-trains within several epochs.

\begin{figure}[t]
  \centering
  \includegraphics[width=0.495\linewidth]{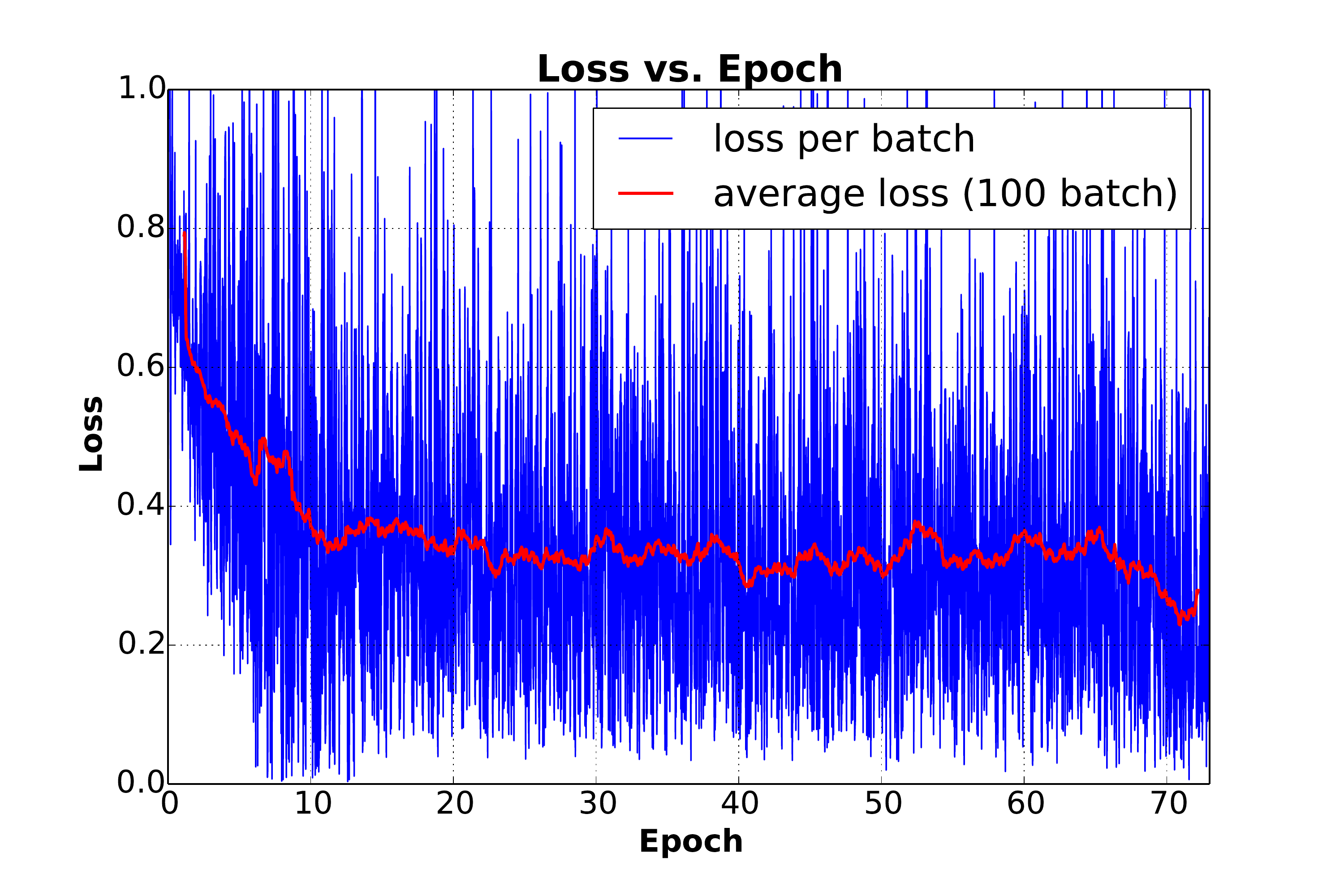}
  \includegraphics[width=0.495\linewidth]{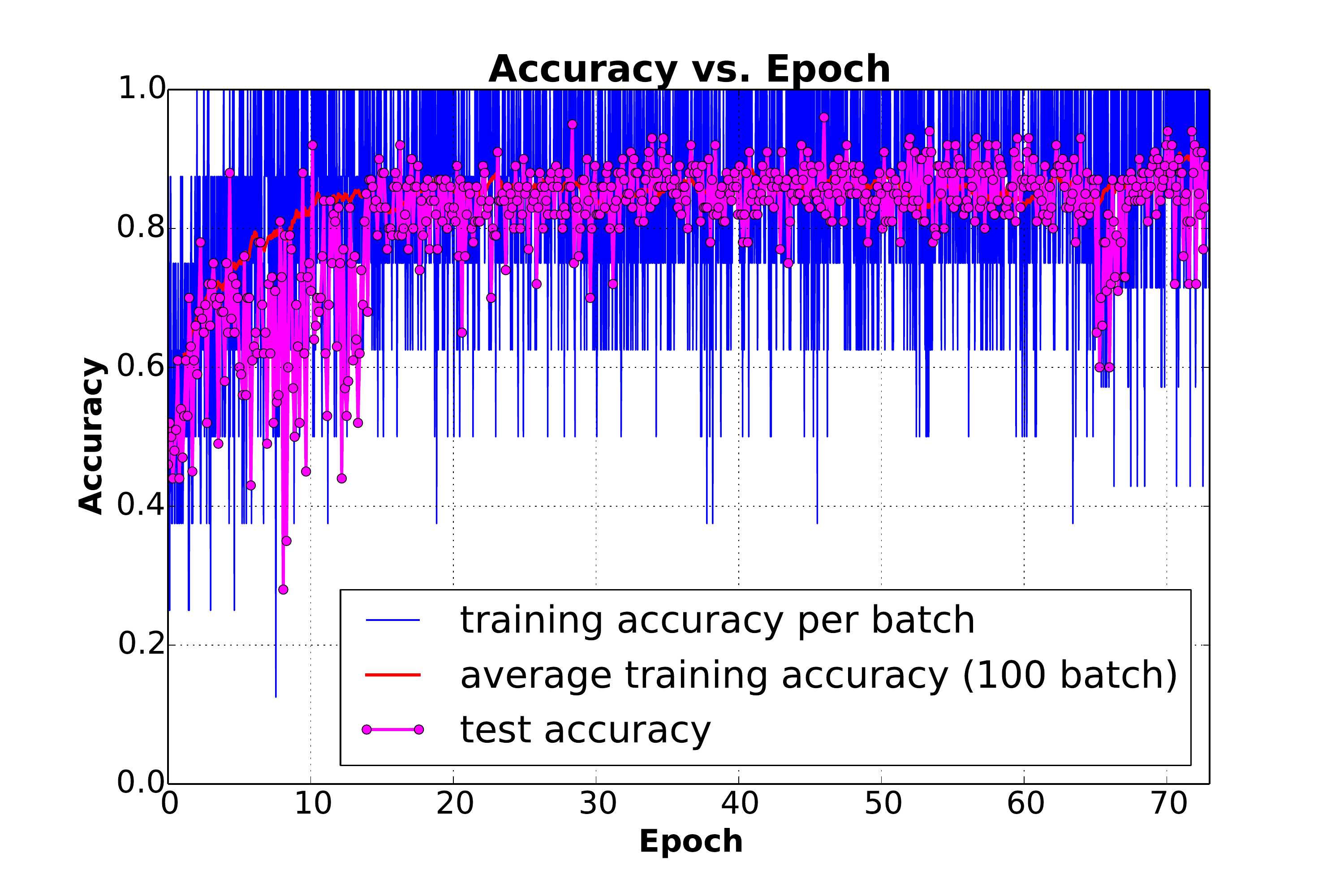}
  \vspace{-0.4in}
  \caption{Loss vs. epoch (left) and accuracy of training and validation data sets vs. epoch (right).  The dip around epoch 70 is due to a change in the training parameters, made to try and help training.}
  \label{fig:training_curves}
\end{figure}

Figure~\ref{fig:training_curves} shows the loss curve along with the training and validation set accuracy which reached a little over 85\% and improved on the performance of the single plane training.  The accuracy of the training and validation sets are close in value throughout and at the end of the training, indicating that the network had not yet over-trained when the training was stopped.  The training took about two days on a single Titan X GPU. Note that the dip in the validation set accuracy around epochs 65 and 70 occurred because we stopped the training and changed the learning rate in order to see if the training was in a local minimum of the loss function.  This did not seem to make a noticeable difference in the training for validation accuracy in the end.

\subsection{Results}

\subsubsection{Performance on Validation Data Set}

To analyze the performance of the network, we use the validation data set. This set is not used to optimize the parameters of the network, but rather to monitor the performance over time. To do this, we pass every image in the validation data set through the network and record its neutrino class score.  Unlike the training stage, which positions the crop randomly at test time, we crop out only the padding at the ends of the images, described in the previous section.

\begin{figure}[t]
  \centering
  \includegraphics[width=0.8\linewidth]{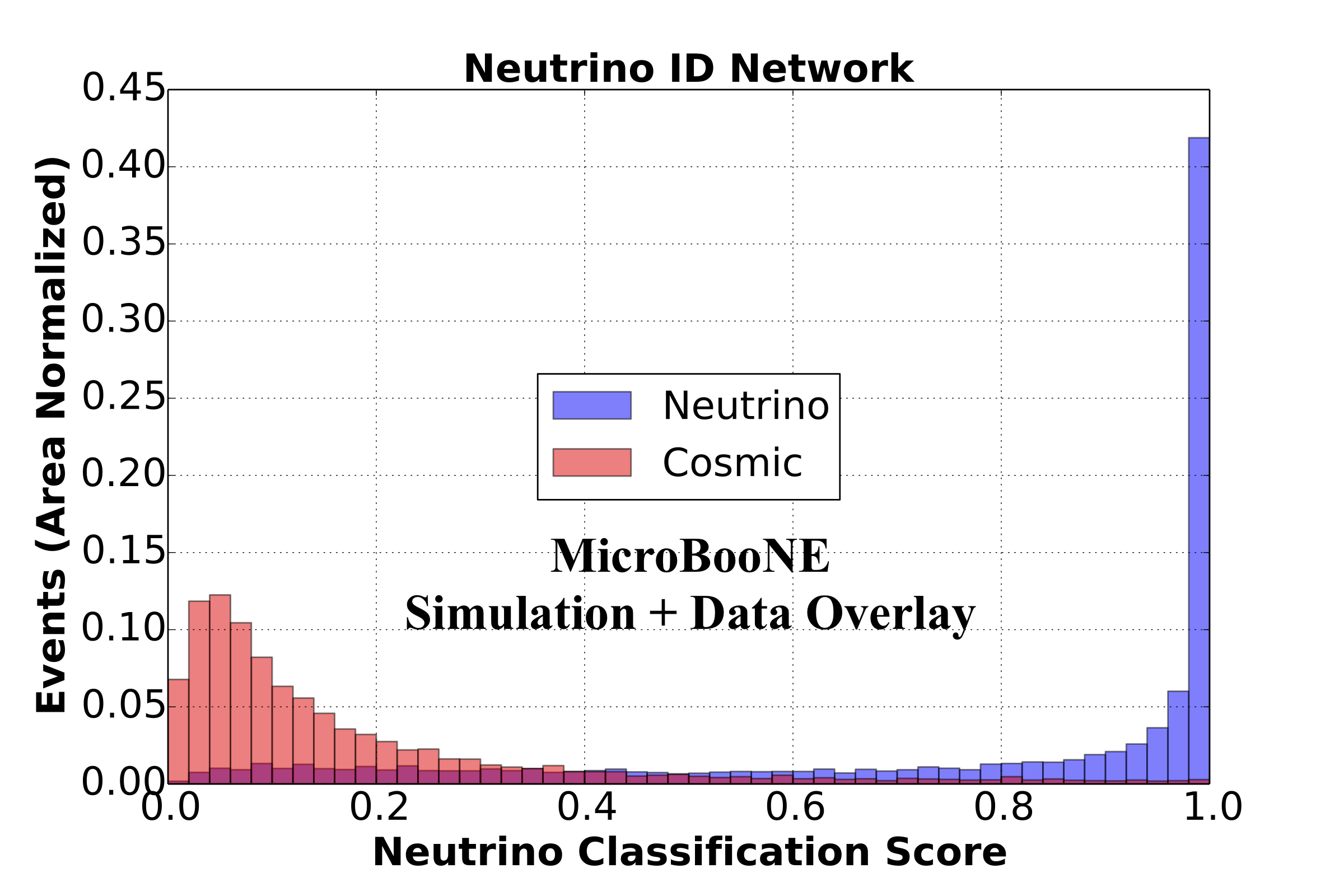}
  \caption{Distribution of neutrino class score for cosmic-only (red) and simulated neutrino overlaid on data cosmic (blue) images.  The cosmic-only and neutrino+cosmic distributions are area normalized.  Events come from the validation data set (which is not used to train the network).}
  \label{fig:nuid_dist}
\end{figure}

\begin{figure}[t]
  \centering
  \includegraphics[width=0.8\linewidth]{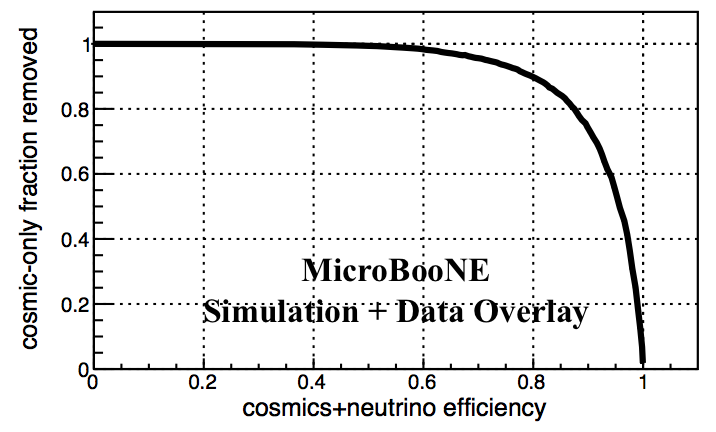}
  \caption{Fraction of cosmic-only events that are removed vs. neutrino event efficiency.  Each point on the curve is for some cut value on the neutrino class score.
  }
  \label{fig:networkperformance}
\end{figure}

Figure~\ref{fig:nuid_dist} shows the distribution of neutrino scores for true cosmic-only images (red histogram) and true cosmic+neutrino images (blue histogram), where both histograms are area-normalized separately.  One can see that there is good separation between the two types of event. 
This is quantified in figure~\ref{fig:networkperformance} by plotting the fraction of cosmic-only events removed versus the neutrino+cosmic event efficiency.
A cut on the neutrino score is varied to generate the curve. 
Requiring 90\% neutrino+cosmic event efficiency, 75\% of cosmic-only events are rejected.
A selection that uses this network will likely require approximately 99\% or higher cosmic-only event rejection which results in 60\% neutrino+cosmic selection efficiency because the number of cosmic-only events is expected to be much higher than events that contain a neutrino in the data.
The  point in the purity vs. efficiency curve closest to 100\% in both categories is about 85\% efficiency with 85\% cosmic-only rejection. Note that the efficiency is relative to the selected neutrino events for the training sample: $\nu_\mu$ CC interactions with neutrino energy greater than 400 MeV.  


As discussed earlier in section~\ref{subsec:sampleprep}, we perform our training and tests using images that combine simulated neutrino interactions overlaid onto an image coming from an off-beam data event.  This is to ensure that the network contends with realistic detector noise and unresponsive channels.  However, we also know that the simulation of the detector response is not perfect and there might be features that the network can learn to identify simulated neutrino images.  Therefore, we have studied the response of the networks on a set of images coming from real beam-on neutrino candidate events.   These events are selected by an independent analysis using ``classic'' reconstruction techniques~\cite{ccnote} aimed at selecting $\nu_\mu$ charged current inclusive events. This selection is not completely pure, but most of the events in this set are expected to be real neutrinos, and hence we use it to check the network responses on real data images.  In addition, this selection does not completely overlap with the sample we used to train the network. This study was performed in order to address an important question of whether the network trained with a mixture of simulated neutrino and data background can be generalized, at least in part, to work on data-only images. In this preliminary study, we apply a cut on neutrino score of greater than 0.95 and find that the network selects events with neutrino candidates.  However, we find that the efficiency of the selection is lower than expected from Figure~\ref{fig:networkperformance}, where we expect it to be 50\% at the neutrino score cut value of 0.95.  While a fraction of the interactions selected will not correspond to the topology for which the network was trained, we postulate that there is a difference to be resolved in the simulated and actual detector response. Improvements in the simulation of the detector response and noise will be required in order to achieve the expected performance. This is a focus of on-going efforts and will be reported in future work. Despite this, the study confirms that there are topological features that the network can learn to find neutrino interactions in our data.

\section{Conclusion}
In this work, we have successfully demonstrated that CNNs can be trained to perform particle classification, particle and neutrino detection, and neutrino event identification.  In particular, a single particle classification has shown a performance of identifying $e^-$ with an efficiency of 83\% and 82\% purity in a mixture of single $e^-$ and single $\gamma$ events. A $\mu^-$ with efficiency of 95\% and purity of 75\% was achieved in an equal mixture of single $\mu^-$ and single $\pi^-$ events. Using full detector information, demonstration 3 showed a capability of neutrino event selection with 85\% neutrino+cosmics event efficiency for 85\% cosmic-only rejection. 

While these results show that CNNs can be applied to analyze LArTPC images and have the potential for good performance, there is much more to be explored.  For one, our studies overlaid simulated neutrino interactions onto off-beam data events. This was to train the networks with a realistic depiction of detector noise, something a successful application must contend with.  However, before being able to quantify the network performance with high enough accuracy for physics-level analyses, we will need to perform careful, separate analyses to ensure that simulated signal waveforms carry the same features in real data. 
This is one example of the many systematic uncertainty studies that have not been explored.
Another type of potential systematic uncertainty may arise from using simulated neutrino events which might cause a model-dependent bias to what a network learns. Careful study is required to quantify the bias, and to minimize it wherever possible. Alternatively, one might instead develop a model-independent approach for detecting neutrinos in data. Such studies are a requirement for physics analyses using this technique and developing new methods to quantify this will be explored in future work. 

Overall, this work takes the first steps in exploring CNNs and the broader discipline of the field of deep learning, a collective term for algorithms and architectures that employ deep neural networks, to LArTPCs. We have shown the following:
\begin{itemize}
\item It is necessary to consider a strategy to either crop or downsize images for a large LArTPC detector, and study their effect. We have shown one possible method (demonstration 1).
\item As long as GPU memory constraints are not an issue, it is best to limit downsizing as the networks use $dE/dx$ information for classification. A downsizing of the images by a factor of 2 showed a clear negative effect (demonstration 1).
\item For a particle classification task, using weights for a network trained for a greater variety of features might perform better. This was seen in the $\mu^-/\pi^-$ study, where the network trained for five-particle classification outperformed the same network trained for only the $\mu^-/\pi^-$ sample (demonstration 1).
\item The Faster-RCNN architecture can be used for neutrino interaction detection in a large event image (demonstration 2).
\item It is possible to combine different types of detector information to enhance the features in the image data. We have shown one method to combine PMT and TPC information as an illustration (demonstration 3).
\item A multi-view architecture can be employed to process multiple TPC plane-views for neutrino interaction selection, and to enhance the performance achieved compared to using single plane information only (demonstration 3).
\item Modeling of the wire response and noise characteristics of the detector need improvement before a network trained on simulated images can be fully generalized to real data images. Likewise, techniques  to mitigate the impact of this disagreement should be developed.
\end{itemize}
There are certainly many other avenues, besides the ones listed here, to study and possibly improve performance.  However, the proof-of-principle tests conducted in this work show that CNNs can deliver results worthy of further exploration and provide a useful starting-point for those interested in developing their own CNNs for the reconstruction of neutrino events in liquid argon TPCs.


\section*{Acknowledgments}
This material is based upon work supported by the following: the U.S. Department of Energy, Office of Science, Offices of High Energy Physics and Nuclear Physics; the U.S. National Science Foundation; the Swiss National Science Foundation; the Science and Technology Facilities Council of the United Kingdom; and The Royal Society (United Kingdom). Additional support for the laser calibration system and cosmic ray tagger was provided by the Albert Einstein Center for Fundamental Physics. Fermilab is operated by Fermi Research Alliance, LLC under Contract No. DE-AC02-07CH11359 with the United States Department of Energy.


\newpage

\appendix


\section{Details of the Components in the Three-plane Neutrino ID Network}
\label{app:3planenuid}

This appendix provides more details of some of the techniques incorporated into the network described in section~\ref{sec:3plane_nuid_design}.

\begin{figure}[t]
  \begin{center}
  \includegraphics[width=0.75\textwidth]{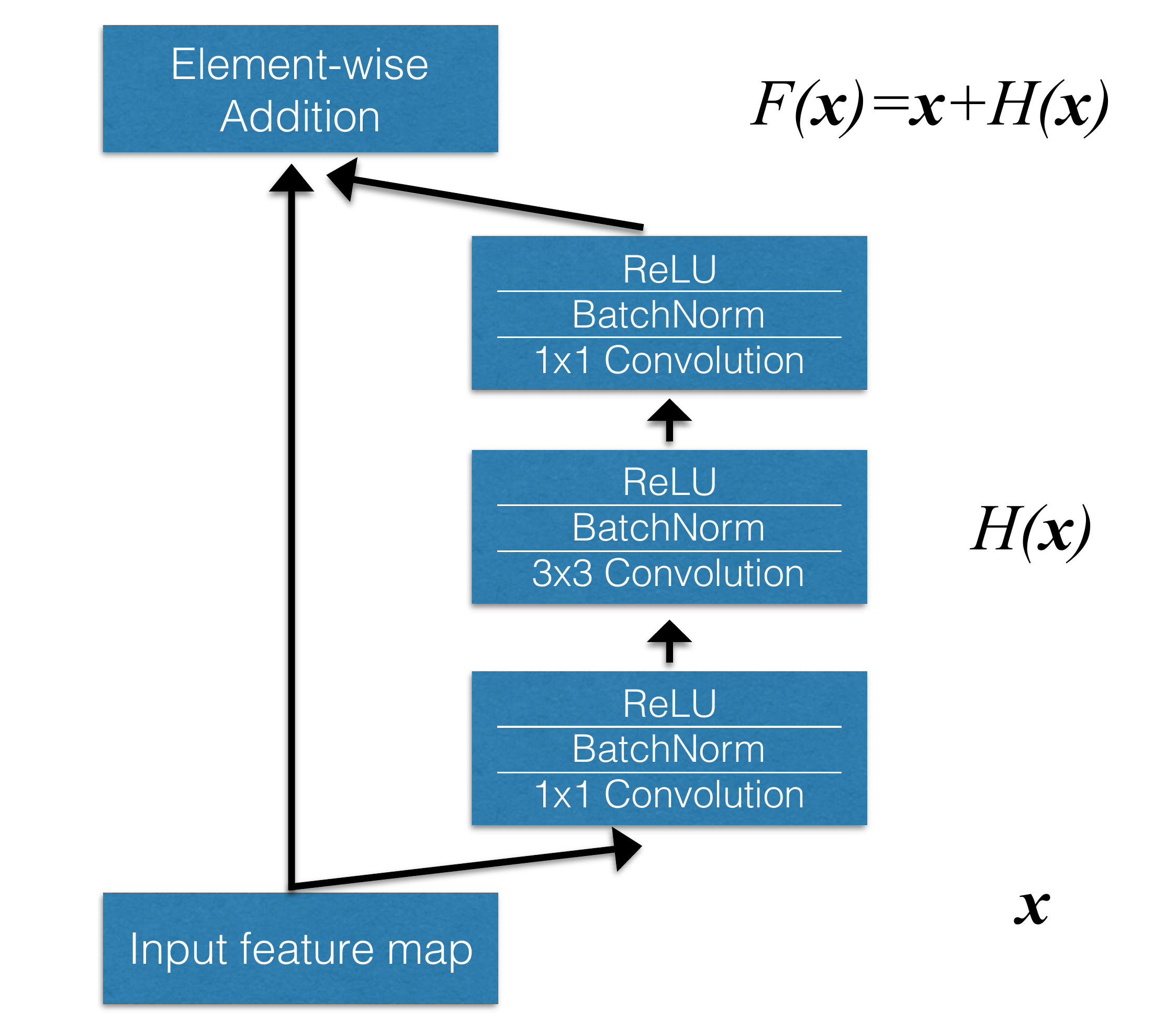}
  \end{center}
  \caption{A ResNet module. This is the main unit used for the neutrino ID network in demonstration 3.
  \label{fig:resnetmod}}
\end{figure}

The primary type of network layer employed is known as the residual convolutional module~\cite{ResNet}. The basic idea is that instead of training the network to optimize a function, $F(x)$, over the feature space, $x$, it is easier for a network to optimize residuals, $H(x)$, such that $F(x) = x + H(x)$.  A diagram of the residual module employed in our network is given in figure~\ref{fig:resnetmod}. The insight of the authors of ref~\cite{ResNet} arose from their experience in training very deep networks where the number of layers surpasses 100 or more. What they found is that these very deep networks perform worse than shallower ones.  This seemed counter-intuitive, because if additional layers were not useful, a network in principle should just learn to use the identity convolution to pass existing feature maps forward in the network and retain the same performance, all things being equal.  This insight led them to hypothesize that with deep networks it was difficult (or just required too much time) for convolutional layers to optimize to the identity if it needed.  Therefore, the residual module, which can be initialized near $H(x)\approx 0$, allows a layer to start at the identity convolution and optimize from there.

Another proposed reason this network structure is useful is that during training the amount of error made by the network can back-propagate more easily through the network.  As can be seen in figure~\ref{fig:resnetmod}, information passing backwards can flow both through the convolutional layers and through a so-called {\it bypass} connection directly back to the input feature map.  The bypass protects the  information from potentially small derivative values coming from the convolutional layers. Other studies showed that non-residual, very deep networks, while they do not necessarily perform worse than residual networks, train more slowly than residual networks~\cite{Inceptionv4}.   That residual networks train faster corroborates both hypotheses as to why residual network structures are effective and, thereby, more than warrant their use. Based on these findings, we use these types of modules in the network of demonstration 3 above.

Our version of the residual module follows that used in ref.~\cite{ResNet}. We use a 3$\times$3 convolution that is both preceded and followed by 1$\times$1 convolutions. The output of each convolutional layer is passed through a ReLU. It is believed that this $(1\times 1)-(3\times 3)-(1\times1)$ structure has about the same complexity as two 3$\times$3 convolutions but contains more non-linearities and fewer free parameters.  This is desirable as it is hypothesized to increase the separation power while reducing over-fitting.  The convolutional layers also use a technique known as batch normalization, or BatchNorm~\cite{BatchNorm}.  

	BatchNorm is a technique where, for a batch during training, the activation of a layer is normalized to have a mean of zero and a variance of one.  This technique is believed to reduce learning time by preventing the distribution of activations of each layer to change, something the authors in ref.~\cite{BatchNorm} refer to as internal covariance shift.  BatchNorm also seems to possibly provide a degree of regularization and thereby helps to reduce overfitting~\cite{BatchNorm}.	
    
    The final technique incorporated into the network is known as {\it dropout}, where parts of the feature map are set to zero randomly during each backward pass.  This technique is thought to be very effective in limiting over-training~\cite{dropout}. It is also considered to force the network to develop features that can stand on their own, instead of relying on two features that must be activated in conjunction with others in order to lead to a certain classification outcome. 
Reducing {\it co-adapted features} promotes robustness of the network and has been observed to improve accuracy (but at the cost of longer training).
The hypothesized reason for increased accuracy is thought to be due to the network behaving like an ensemble of networks.  This is possibly due to the stochastic removal of features that can promote multiple paths for classification.
    

\section{Study of Simulated Neutrino Pixel Intensity Variations on Network Output}
\label{app:adcstudy}

At the time of this work, the MicroBooNE experiment is still in its early stages, having less than a year of beam data. Therefore, we expect that there will be a handful of detector effects that our simulation does not accurately model.  This is why, as explained in the body of the text, we chose to overlay simulated neutrino interactions onto images coming from off-beam data events.  The off-beam data events will contain the correct noise features and wire responses by construction. However, the neutrino interactions we overlay are made using simulated wire responses, and these simulated wire responses can differ from those in the data. We expect this to be a systematic uncertainty a final analysis must minimize.  If we are to apply our networks to data, we would expect the accuracy of the network to be affected by errors due to modeling the wire response.  However, at this proof-of-principle stage we are not primarily concerned about the precision on the neutrino data but instead are interested in establishing that there are features in real neutrino data that the networks can use to detect the presence of a neutrino interaction in the images. To this end, we performed the study described in this section. One way in which the simulated and observed wire responses might differ is in the amplitude of the wire pulses.  This might be due to wire-by-wire differences in data that are not modeled in the simulation.  

This study investigates whether the CNNs used in the paper are keying on PI differences to find the neutrino events. The study is performed by varying the amplitude of the simulated wire signals coming from the outgoing particles from the neutrino interactions that get overlaid into the off-beam cosmic ray images. We quantify the change in network performance as a function of the change in amplitude.   In particular, we are asking if the neutrino score substantially improves for small variations of the amplitudes within the level to which the simulation is tuned, and for large variations where the neutrino interaction looks substantially different from the cosmic overlay.  

\subsection{Sample Preparation}

This study varies the images passed to the CNNs trained in demonstration 2 and demonstration 3.  These are the single plane neutrino detection network and the three plane neutrino vs. cosmic event classifier, respectively.  Note that for this study we will evaluate only images with a simulated neutrino overlaid by cosmic data; there are no cosmic-only images. Also, we do not retrain any of the networks. The samples prepared in this study are only evaluated by the networks.

We produce several copies of each event, varying the PI of the simulated neutrino interaction in each copy. The PI values of the off-beam cosmic images, coming from data, are scaled the same in each copy.  The nominal simulated neutrino images are those that have been scaled to match the data as described in section~\ref{subsec:sampleprep}.  By scaling the same event several times, we can look at the effect of scaling on both the whole distribution and on an event-by-event basis.  

Although we think that the scaling function used to correct the simulation to match data is correct to better than 5\%, we look at an extreme range of scale factors.    This allows us to understand whether our results are stable within the level of our understanding and also to see a breakdown of the results for extreme changes.   Specifically,
for demonstration 2, we prepared approximately 3,000 events with shifts in scalings of  $0\%$, $\pm 11\%$, $\pm 21\%$, $\pm 32\%$ and $\pm 42\%$.
For demonstration 3, we prepared 8,000 events with scaling shifts of -50\%, 
-30\%, -20\%, -5\%, 0\%, +5\%, +15\%, and +50\%.

\subsection{Results of Demonstration 2: Single View Neutrino Detection}
For each PI scaling factor, we ran the detection network on a selection of $\sim3,000$ simulated neutrino events overlaid with data cosmic images. We use the same network parameters found from the initial training with the nominal scaling factor and the neutrino+cosmic data set reported in section~\ref{sec:demo2}. For each image, the detection network reports a variable number of detection boxes, each with a neutrino score. For each image we choose the detection box with the highest neutrino score and call this the {\it Nu score}, or neutrino score, in the plots for this section. 

\begin{figure}[H]
 \centering
 \hspace{-0.4in}
 \includegraphics[width=0.55\textwidth]{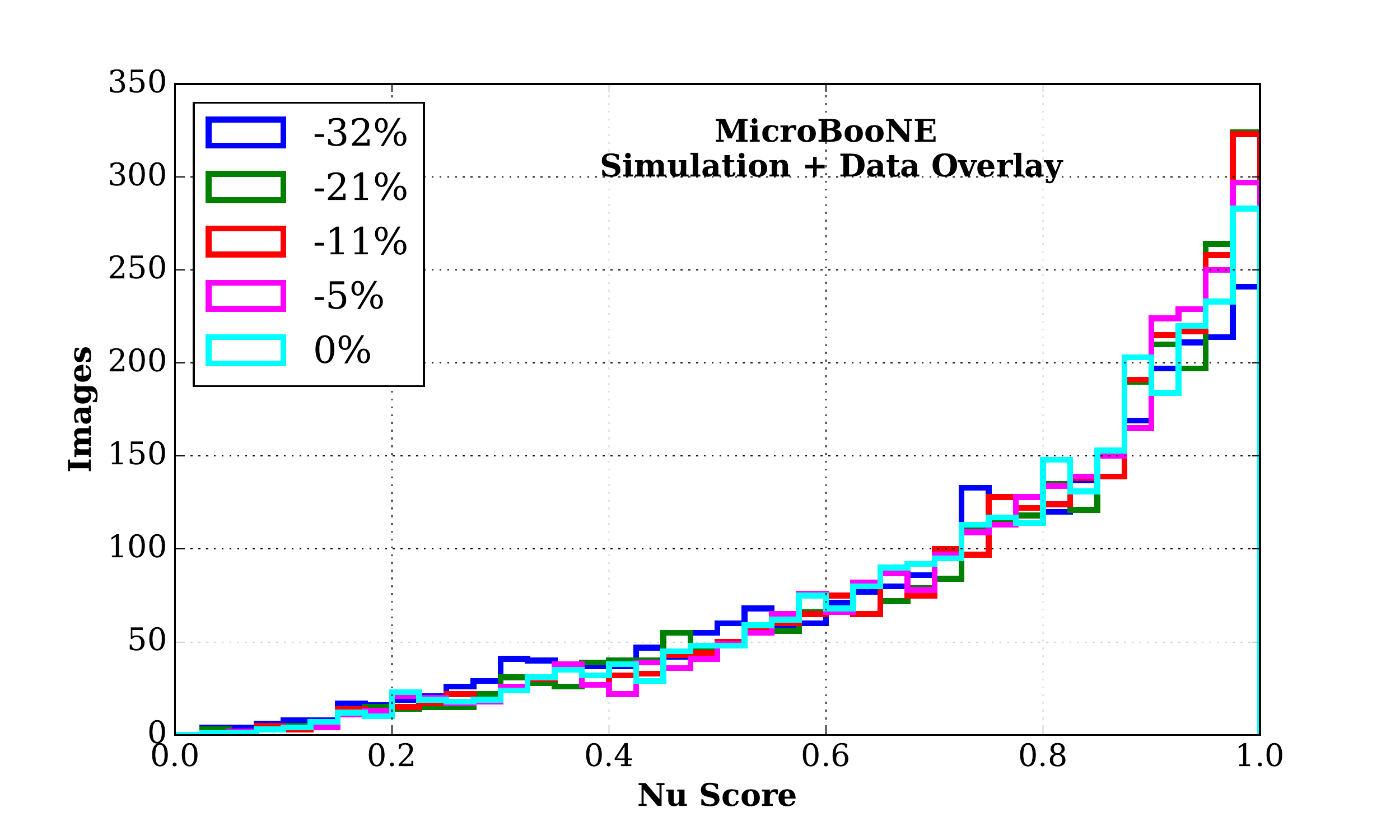}
 \hspace{-0.4in}
 \includegraphics[width=0.55\textwidth]{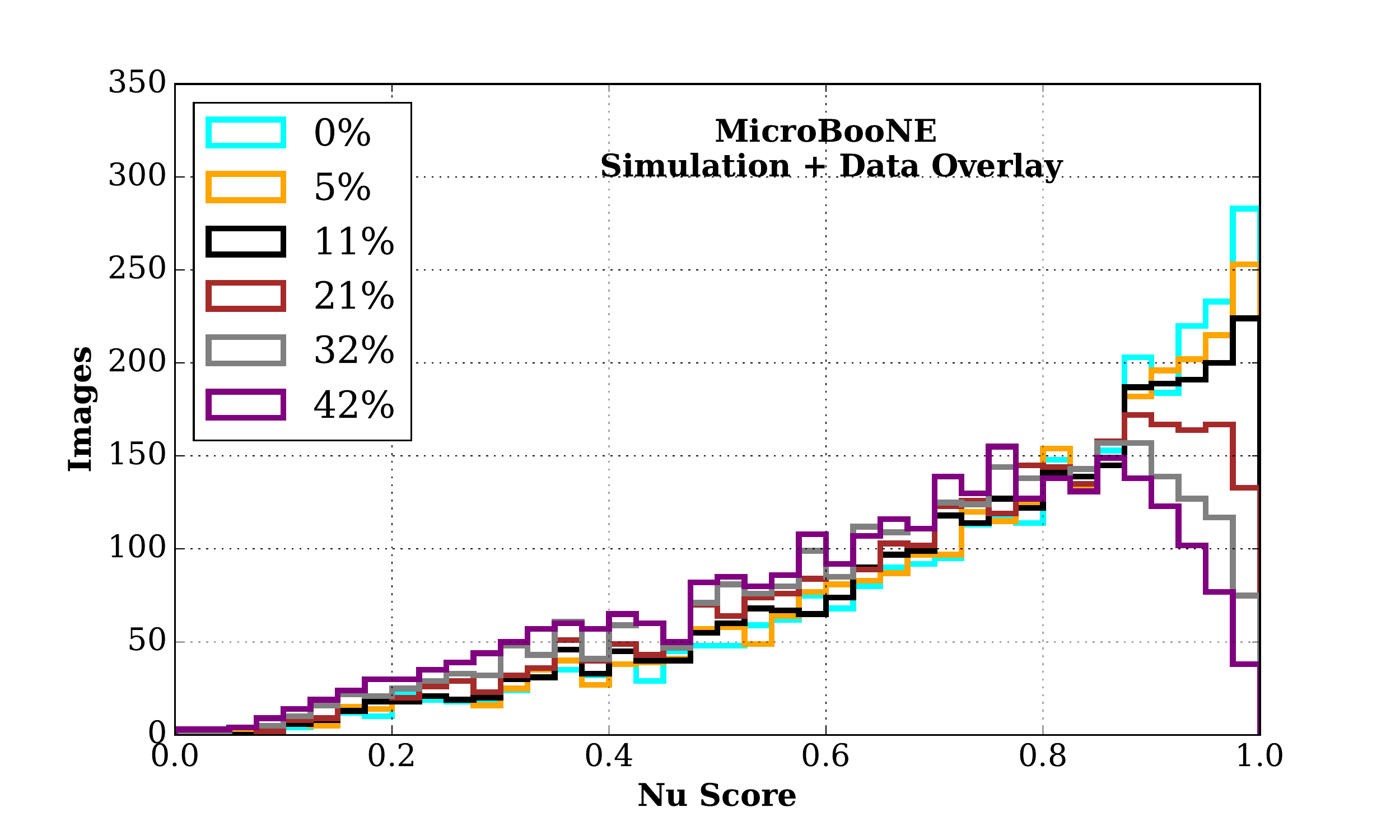}
  \caption{Demonstration 2 neutrino classification distribution for varying PI shift. \emph{Left}: Distribution of neutrino scores for negative shifts in PI related to the nominal (0.0\%) PI scale, for decreasing the PI scales. 
\emph{Right}: Distribution of neutrino scores for positive shift in PI relative to the nominal (0.0\%) PI scale.}
  \label{fig:shift_score}
\end{figure}

The distribution of the neutrino score is shown in figure~\ref{fig:shift_score} for the various perturbations in the PI scale. Qualitatively, we observe slight variations in the neutrino score for 5\% shifts in the PI scale. The distribution stays relatively stable against these small shifts. We note that positive shifts in the PI scale tend to shift high probability scores to lower values. We also find that moderately decreasing the PI scale provides some enhancement. However, for large shifts either in the positive or negative direction, the neutrino scores diminish.

\begin{figure}[H]
 \centering
 \includegraphics[width=0.45\textwidth]{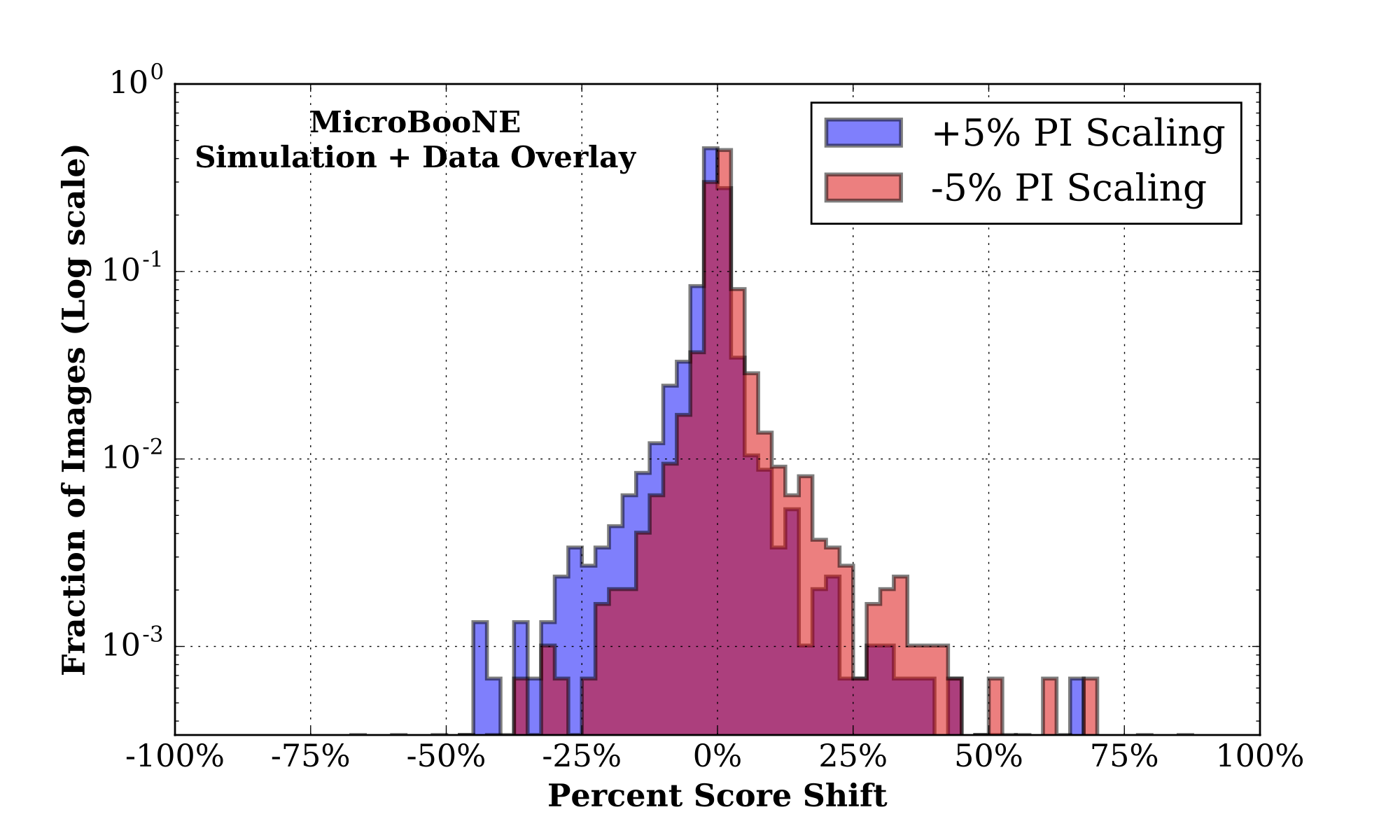}
 \includegraphics[width=0.45\textwidth]{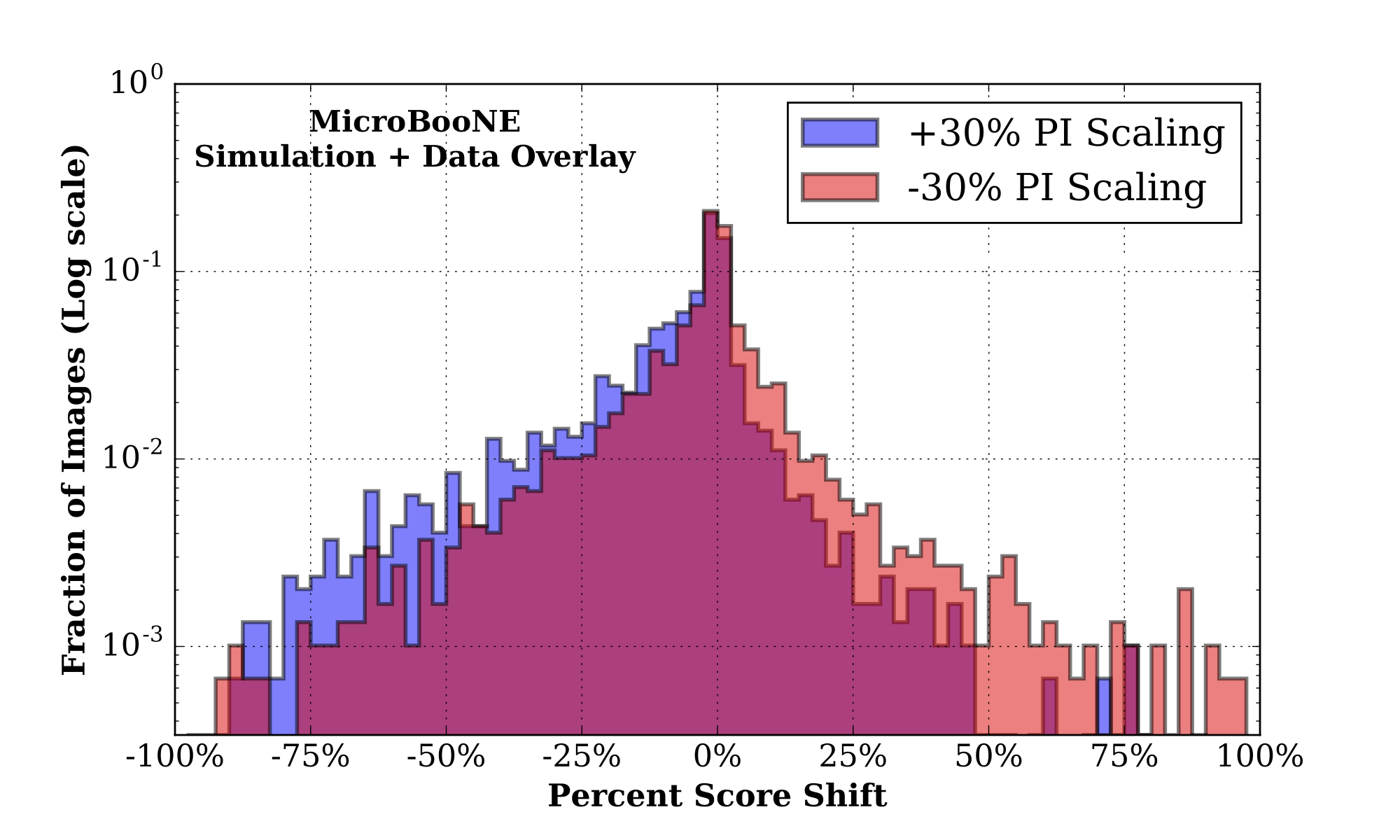}
  \caption{Demonstration 2 event-by-event shift in neutrino score. The blue, and red distributions are positive and negative PI scalings respectively.  \emph{Left}: Distribution of percent shift in neutrino score for smallest perturbation in PI scale. At least 50\% of events show less than 1\% deviation in neutrino score. \emph{Right}: Distribution of percent shift in neutrino score for the largest perturbation in PI scale.}
  \label{fig:shift_5_30}
\end{figure}

To understand the effect quantitatively, we show the event-by-event percent-shift in the neutrino score for small and large perturbations in PI in figure~\ref{fig:shift_5_30}. For the $\pm$~5\% shift in the PI scale the distribution remains within $\pm$~10\% of the nominal score. Large changes in PI scale degrade the event-wise neutrino score, pulling the distribution to lower neutrino scores as the PI scaling deviates from the training sample.   As an example of this effect, we show the $\pm$~30\% shifts.

\begin{figure}[H]
 \centering
 \includegraphics[width=0.75\textwidth]{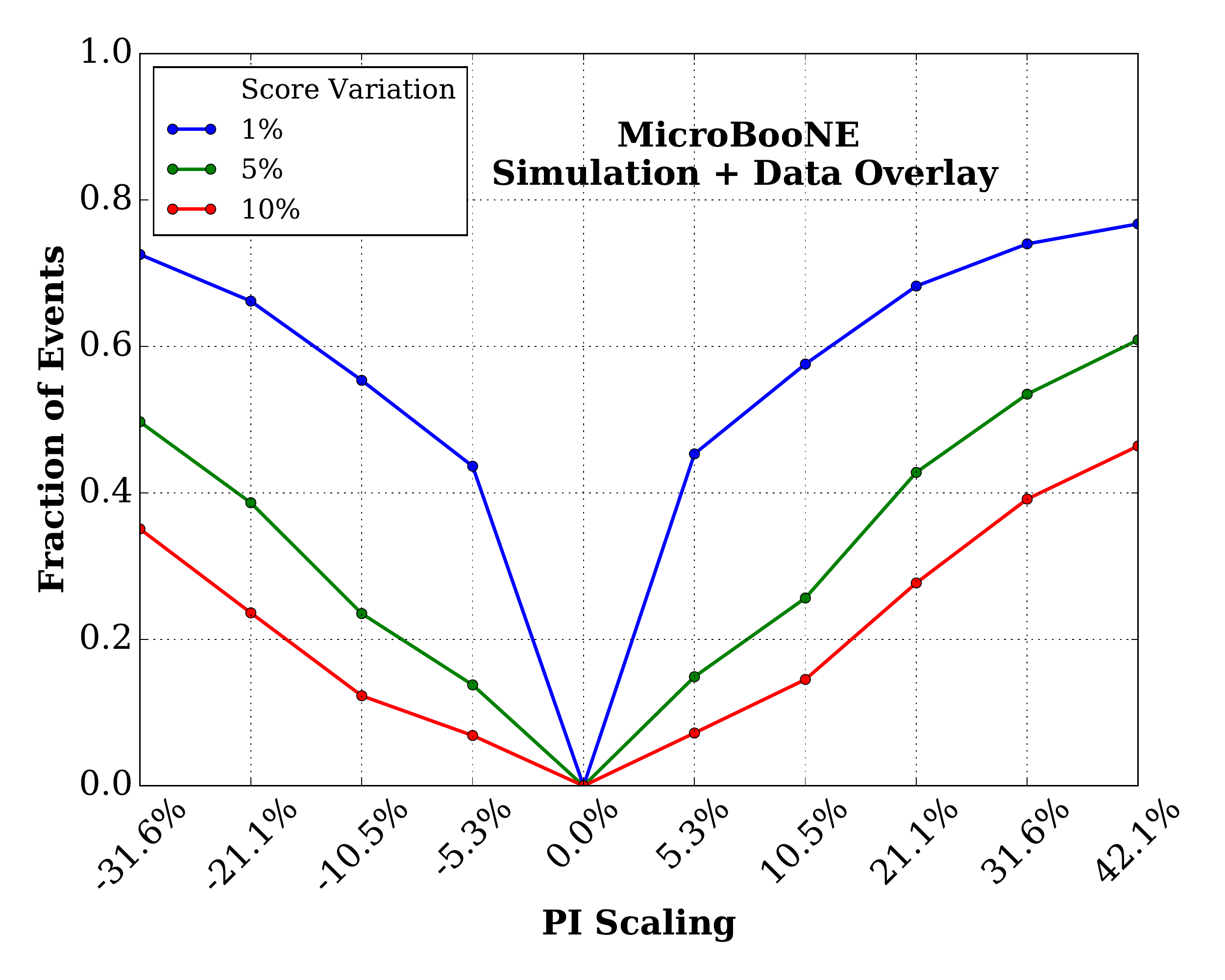}
  \caption{Demonstration 2 fraction of total events which shift in network score as a function of the PI scaling percentage. Three curves are shown for neutrino score variations by 1\%, 5\% and 10\%. The fraction of events which change in network score decreases as the threshold is increased. With a 5\% PI scaling we observe  approximately 10\% of the events shift in network score by 5\% or more (green curve).}
  \label{fig:shift_scale}
\end{figure}

The total fraction of events which shift in network probability by a fixed percentage or more is shown in figure~\ref{fig:shift_scale}. For small $(\pm 5\%)$ PI re-scalings, fewer than 10\% of the events or the events change the network score by an amount of $\ge 10\%$ (red curve). If we make a more stringent test, examining score-changes of $\ge 5\%$, more than 80\% of the event scores remain stable at or below this level.

If the network were relying on PI scale differences to identify neutrino events within the overlaid image, we would have expected much larger instabilities for small perturbations.   This gives us confidence that the network is relying on other features, such as the presence of a vertex, to identify neutrino events, and not substantially relying on PI scale differences between data and simulation to make its network prediction.  

\begin{figure}[H]
 \centering
 \includegraphics[width=0.45\textwidth]{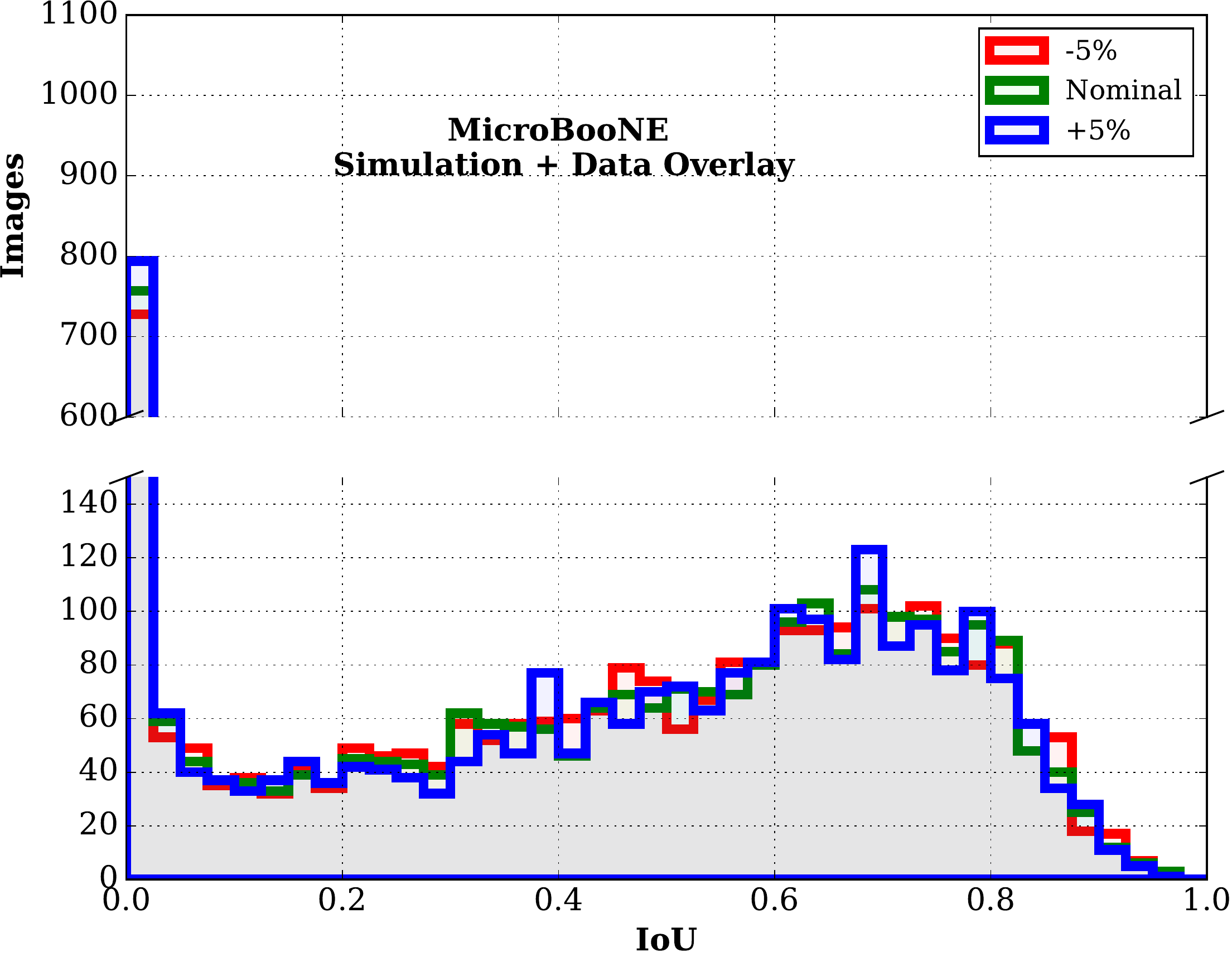}
 \includegraphics[width=0.45\textwidth]{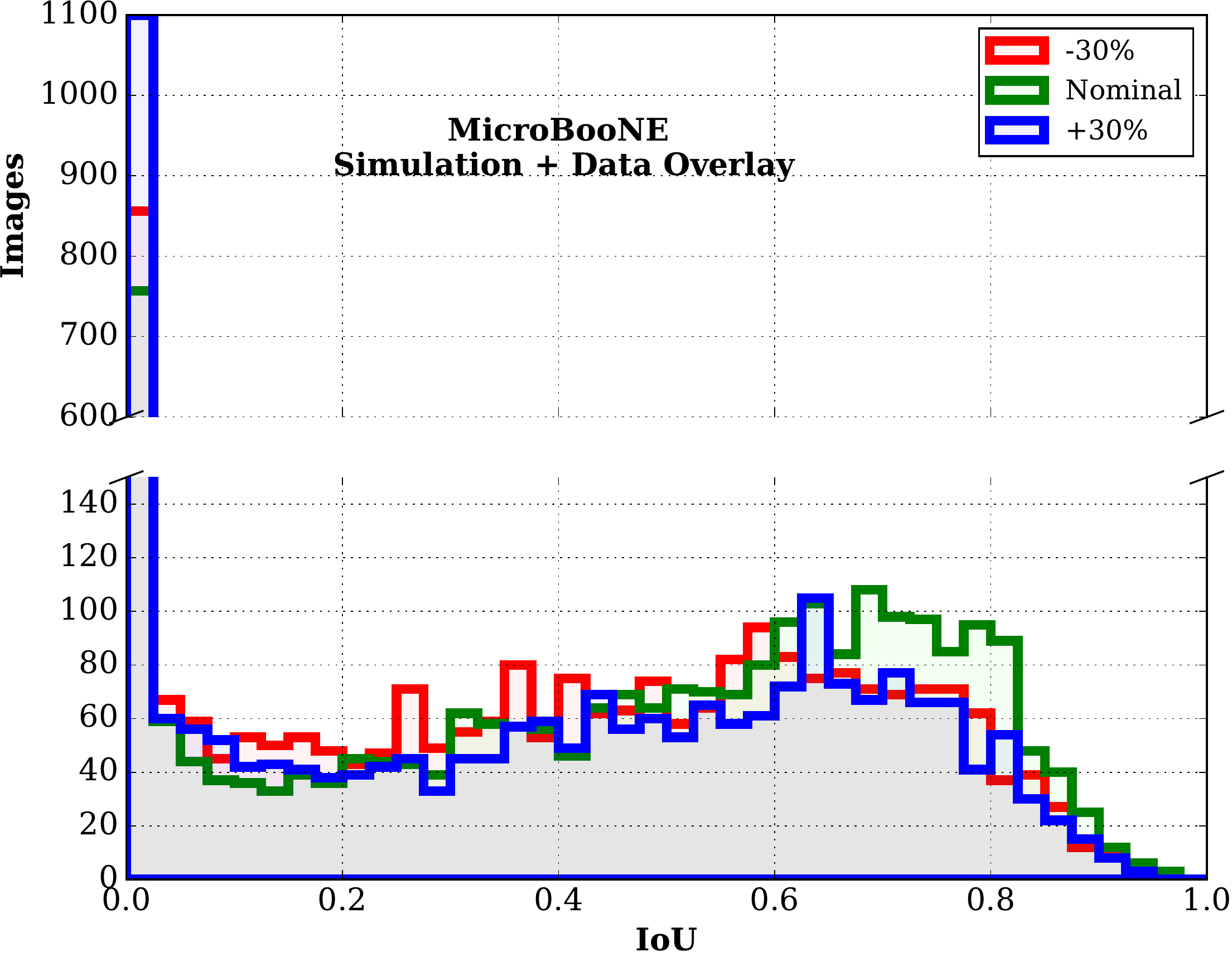}
  \caption{Demonstration 2 distribution of intersection over union. \emph{Left:} Nominal with -5\% and 5\% PI scaling overlaid. 
\emph{Right:} Nominal with -30\% and 30\% pixel scaling overlaid. A break in the y-axis is used in order to be able to look at the zero bin and the rest of the distribution. The zero bin gives the number of events that did not have any overlap with the simulated neutrino interaction.} 
  \label{fig:iou_shift}
\end{figure}

We also observe stability of the IoU distribution for small ($\pm 5\%$) changes in the PI scale, as shown in figure~\ref{fig:iou_shift}. This means the predicted bounding box for the neutrino within the event shifts only slightly in location in the image or varies by a small amount in size. For large shifts in the PI (for example, $\pm30\%$),  we observe the network's bounding box prediction degrades and rapidly populates the zero bin.   

\subsection{Results of Demonstration 3: Three View Neutrino Event Classifier}

\begin{figure}[H]
 \centering
 \includegraphics[width=0.48\textwidth]{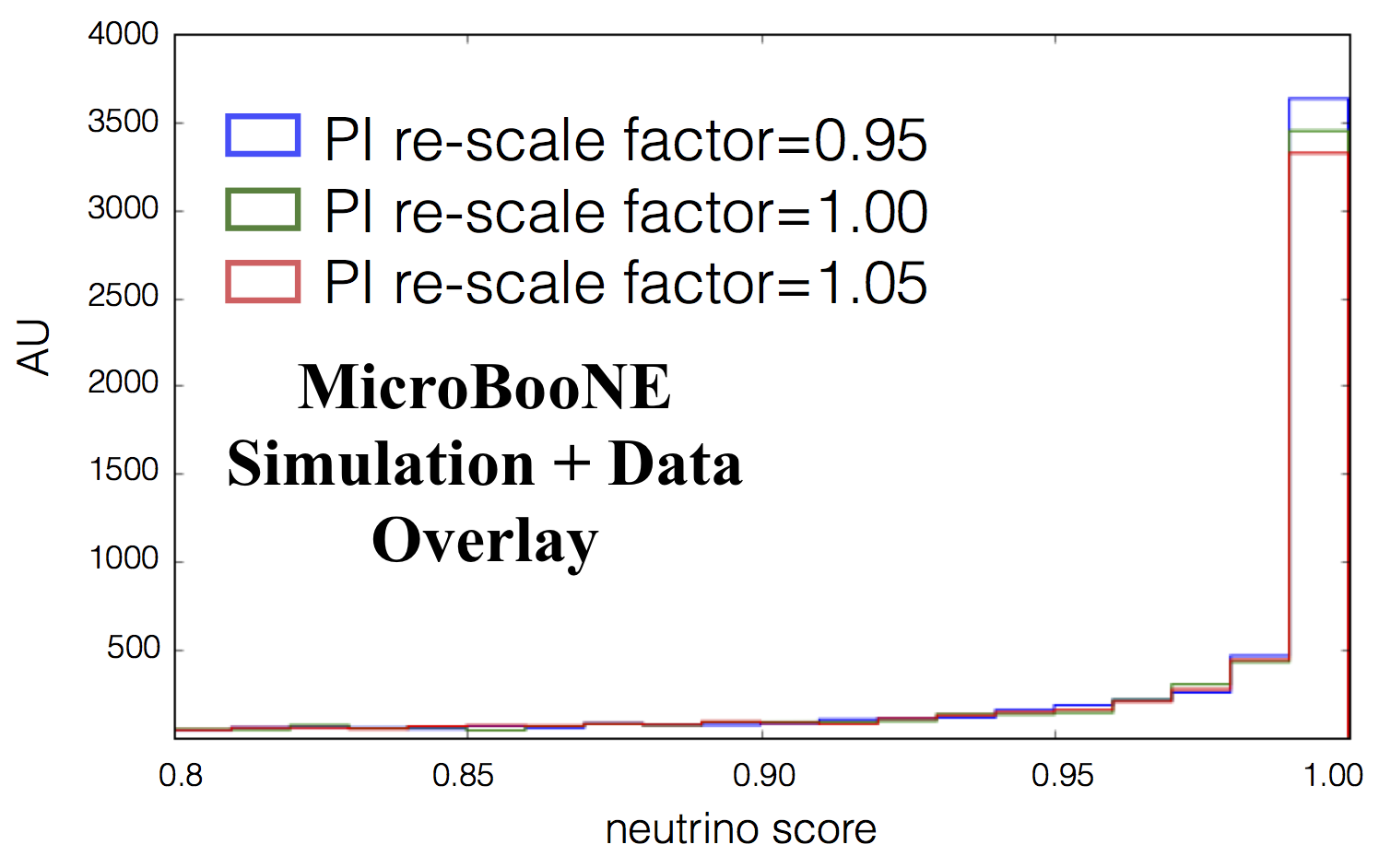}
 \includegraphics[width=0.48\textwidth]{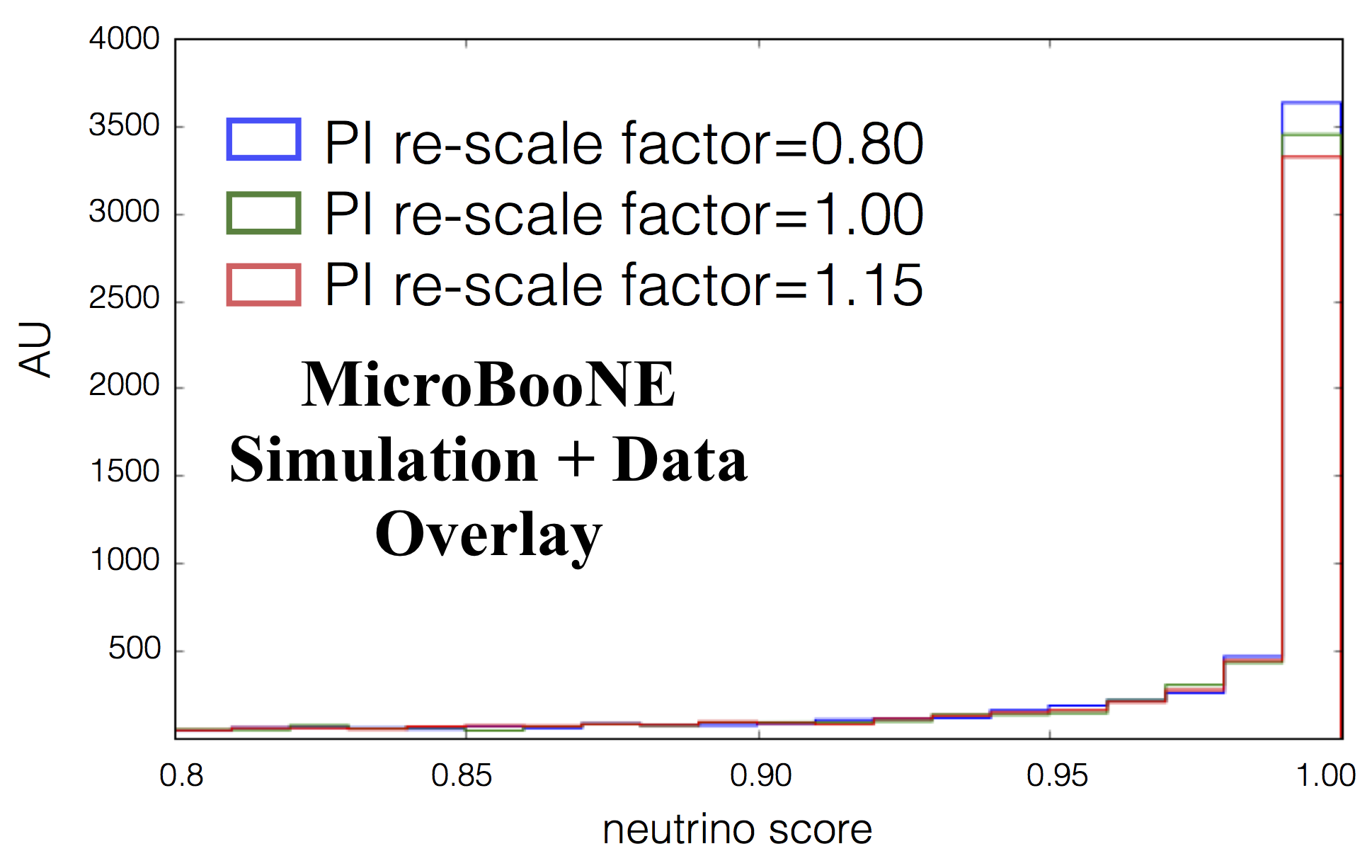}
 \caption{Demonstration 3 comparison of neutrino score distribution between 0.8 and 1.0 between
the nominal (green) with $\pm5$\% (top) and -10\%,+15\% (bottom) variations in the simulated neutrino pixel intensities.}
  \label{fig:demo3_dist_changes}
\end{figure}

We perform the same type of studies using the network in demonstration 3, which is a neutrino event classifier using all three planes of the detector along with PMT-weighted images and images indicating the location of MIP-like and HIP-like energy depositions.  The results are similar to those found in the studies performed with  demonstration 2. Figure~\ref{fig:demo3_dist_changes} shows the variation in the neutrino score from nominal (green) for small variations ($-5\%$,~$+5\%$) (top, blue and red curves) and for larger variations ($-20\%$,~$+15\%$) (bottom, blue and red curves).

\begin{figure}[H]
 \centering
 \includegraphics[width=0.65\textwidth]{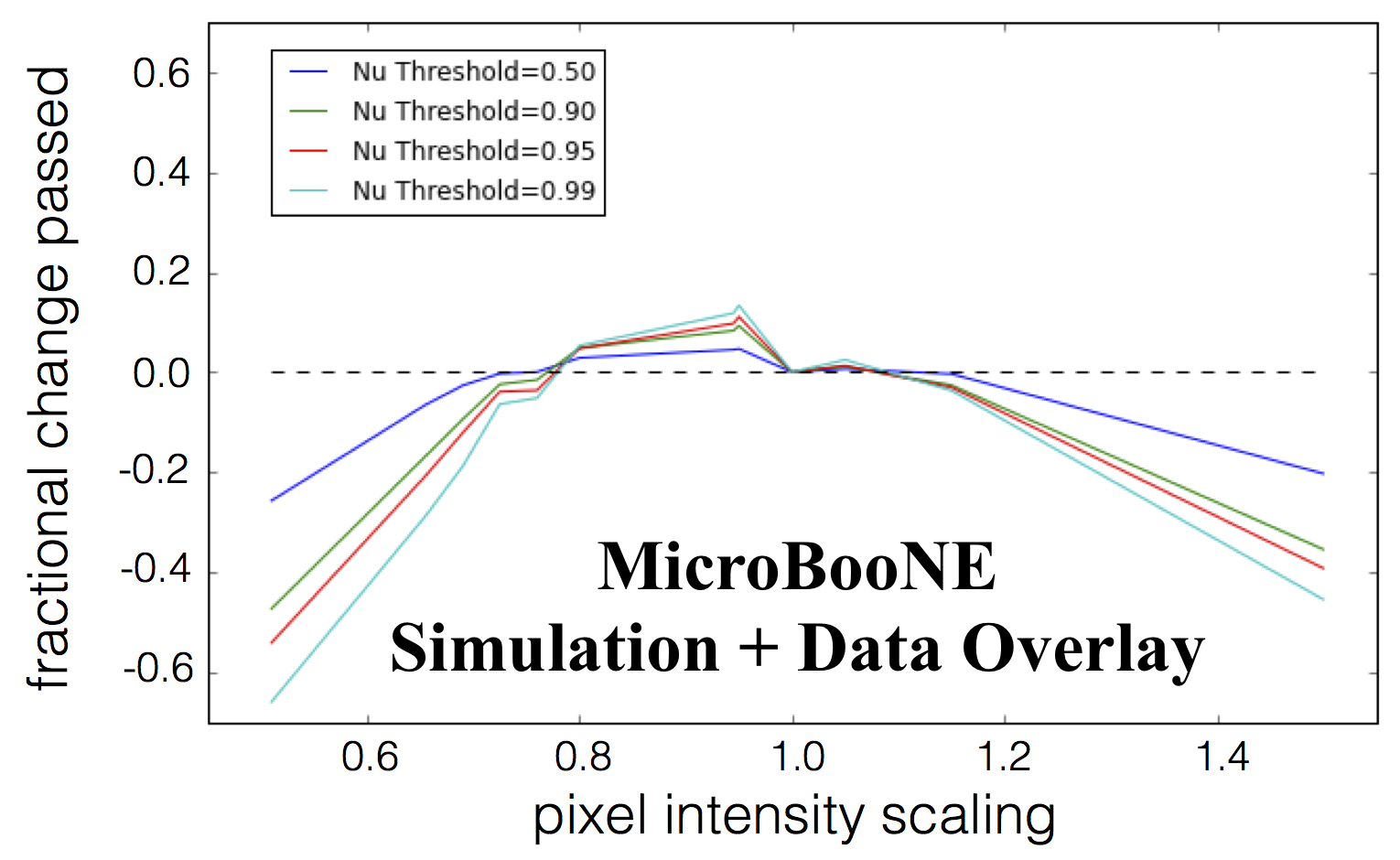}
  \caption{Demonstration 3 fraction of events that pass a given threshold versus the PI rescaling factor.  To pass, an event must have a neutrino score above a certain threshold. Four thresholds are shown: 0.99 (cyan), 0.95 (red), 0.90 (green), and 0.5 (blue). The nominal rescaling factor is defined as 1.0.}
  \label{fig:demo3_change_in_npassed}
\end{figure}

Figure~\ref{fig:demo3_change_in_npassed} shows the impact of such variations on a hypothetical cut based on the neutrino score.  The number of events that pass a threshold neutrino score is plotted versus the size of PI re-scale factor.  A $-5$\% shift causes around 15\% more events to pass a neutrino-like threshold of 0.99.  For very large variations, more neutrinos are classified as not-neutrino-like, indicating that at some point the pixel intensities of the simulated neutrinos fall out of the range that the network expects.

\begin{figure}[H]
 \centering
 \includegraphics[width=0.45\textwidth]{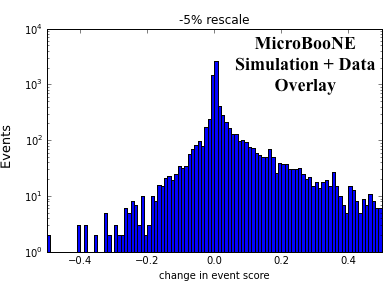}
 \hspace{10pt}
 \includegraphics[width=0.45\textwidth]{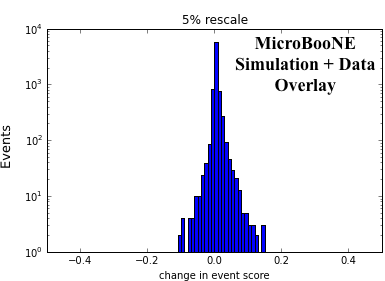}\\
 \includegraphics[width=0.45\textwidth]{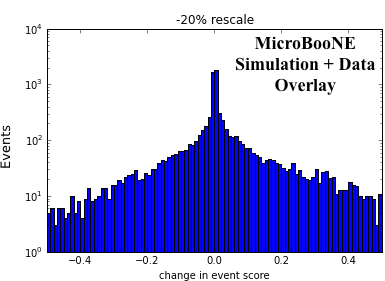}
  \hspace{10pt}
 \includegraphics[width=0.45\textwidth]{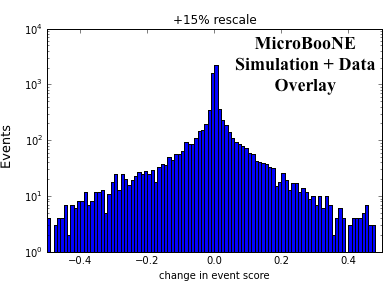}
  \caption{Change in event neutrino score for four different PI rescaling factors. Going clockwise, the top left factor is -5\%, the top right +5\%, the bottom right +15\%, and the bottom left is -20\%.}
  \label{fig:demo3_evtbyevt_dist}
\end{figure}

\begin{figure}[H]
 \centering
 \includegraphics[width=0.65\textwidth]{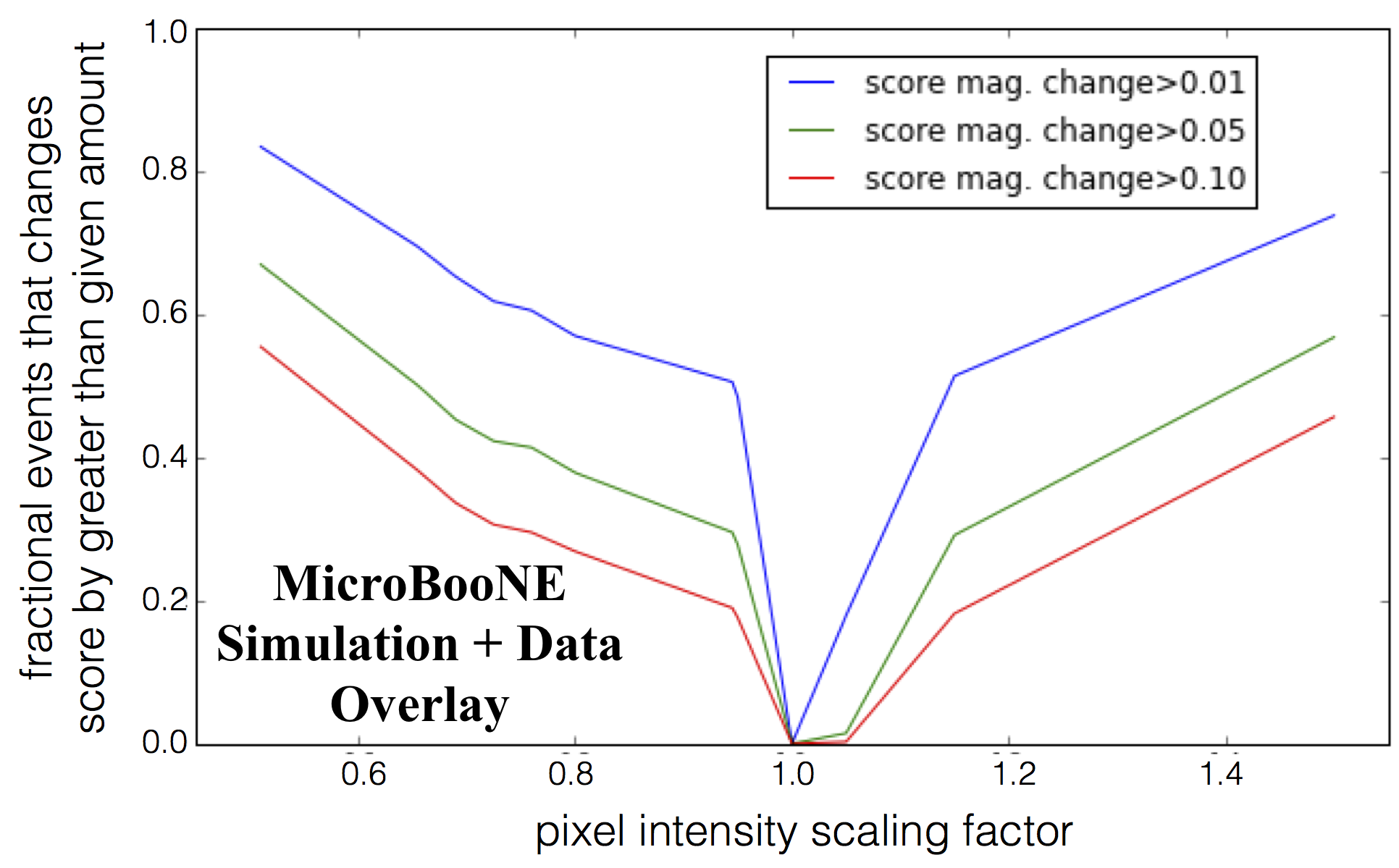}
  \caption{Demonstration 3 fraction of events whose event score changes by a given amount versus the PI rescaling factor.  This plot is meant to quantify the fraction of events that exhibit a certain level of stability. Three change thresholds are shown 1\% (blue), 5\% (green), and 10\% (red). The nominal rescaling factor is defined as 1.0.}
  \label{fig:demo3_evtbyevt_fracchange}
\end{figure}

Finally, we investigate the change in event scores given different PI re-scaling factors. Figure~\ref{fig:demo3_evtbyevt_dist} plots the distribution of score changes observed for 4 different re-scaling factors: -5\%, +5\%, -20\%, +15\%. The plots are in log scale. A large portion of the events have a $<5$\% shift in their scores.  To quantify this better, Figure~\ref{fig:demo3_evtbyevt_fracchange} plots the fraction of events whose scores changed more than a given threshold value. For example, the blue plot in the figure shows the fraction of events that had a score change of $>1\%$.  For a $\pm5$\% variation, more than 40\% of events have a small, $<1$\%, change in their scores.  About 80\% of events have a score change of less than 10\%. About 17\% of events have a score change of $<1$\%, even for a large re-scaling factor.

Overall, we find that the three plane event classifier exhibits the same broad behavior as the single plane detection network.  The class scores vary to a moderate degree for PI rescaling factors of 5\%.  The class scores improve, i.e. up to 15\% more events get classified with a $>99$\% neutrino score, when the PI scaling is slightly reduced.  The variations seen in this network are larger than that for the single plane.  However, as the rescaling factor becomes large, the events become less neutrino-like as a whole.  Currently, we do not know why there is a larger variation in scores output by the network for a given rescaling factor compared to the single plane network.  We can only hypothesize that it has to do with the added information given to the network which is sensitive to the PI scale.  The rescaling factor not only affects the TPC image but affects the PMT-weighted image and how each pixel gets classified as a HIP and MIP image.

\subsection{Discussion of Trends}

\begin{figure}[H]
 \centering
 \includegraphics[width=0.8\textwidth]{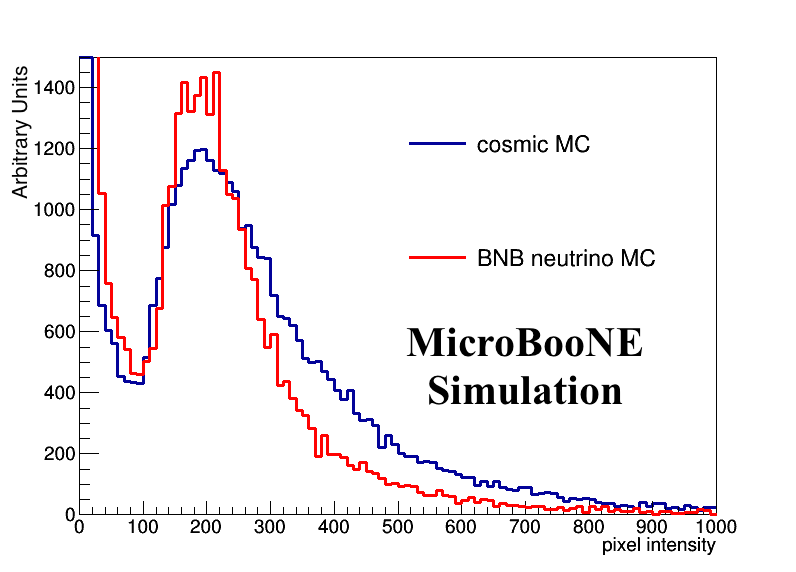}
 \caption{Distribution of peak PI of peaks found by a simple threshold-based peak finding algorithm. The plot compares the distribution from cosmic-only simulation (blue) and BNB-only simulation (red).}
 \label{fig:peak_pi_cosmic_v_bnb}
\end{figure}

We find that the network output depends on the PI scale.  
The dependence that we find seems reasonable. One possible explanation for this dependence is that the network uses the charge per wire pulse for a given track to separate neutrino interactions from cosmic rays. Figure~\ref{fig:peak_pi_cosmic_v_bnb} compares the distribution of maximum PI values from pulses identified by a simple, threshold-based hit finder for simulated cosmic-only images and simulated BNB neutrino-only images.  What we find is that the maximum PI value for cosmic ray tracks is larger than for tracks from BNB neutrino interactions. It would be natural for the networks to learn to associate larger PI values with more cosmic-like tracks.  If so, lowering the overall PI for the entire set of neutrino interactions could cause the network use this feature more strongly, something that is consistent with what we see in both studies of the networks of demonstrations 2 and 3.   
However, as the PI rescaling factors get very large ($\pm$30-50\%), the network begins to have an overall lower neutrino score indicating that the neutrino pixels are beginning to go out of the expect range of values. This occurs for both brighter and dimmer tracks. If only dimmer, one could argue that the neutrino interaction tracks were simply falling below threshold. The fact that these large deviations are associated with less neutrino-like scores is important evidence that the network is not just looking for large discrepancies of the wire pulse amplitudes. This is further supported by that observation that for some events the class scores are stable despite very large PI variations.  


\subsection{Discussion of results}

The stability against changes in PI indicate that the networks are not just looking for a large deviation in the PI to find simulated neutrino interactions.  If they did, then we would expect the neutrino scores to become larger as the variation increases.  This is not observed for reasonable variations.

Looking to the future, we plan to continue working on properly modeling the PI scale in the simulation. We also plan to investigate how we might mitigate the networks' sensitivity to the PI scale, possibly by using things such as hits to make the image or to explicitly train the network to be more stable to modeling errors~\cite{stabilitytraining}. Nevertheless, these studies provide evidence that the PI scale is used by the network in a non-pathological way, i.e. the network is not just responding to very small or very large differences in the PI for tracks coming from neutrino interactions compared to those from cosmics.  However, we do point out that the PI scale is quite a crude handle. In the future, we plan on looking at variations that distort the pulse shapes. This might include variations in the noise filter on the simulated wire signals, modeling of space charge effects, or electron lifetime variations.
    

\end{document}